\documentstyle[aps,preprint]{revtex}           
\begin{document}
\draft

\title{Charge and Spin Currents of the 1D Hubbard Model at Finite 
Energy}

\author{N. M. R. Peres$^1$, P. D. Sacramento$^2$, 
and J. M. P. Carmelo$^1$}
\address{$^1$Department of Physics, University of \'Evora,
Apartado 94, P-7001 \'Evora Codex, Portugal}
\address{$^2$Departamento de F\'{\i}sica and CFIF, Instituto Superior 
T\'ecnico, Av. Rovisco Pais, P-1096 Lisboa Codex, Portugal}

\date{September 1997}

\maketitle

\begin{abstract}
The transport of charge and spin at finite energies is studied
for the Hubbard chain in a magnetic field by means of the 
pseudoparticle perturbation theory. In the general case, this 
involves the solution of an infinite set of Bethe-ansatz equations 
with a flux. Our results refer to all densities and magnetizations. 
We express the charge and the spin-diffusion currents in terms of 
elementary currents associated with the charge and spin carriers. 
We show that these are the $\alpha,\gamma$ pseudoparticles (with
$\alpha =c,s$ and $\gamma =0,1,2,3,...$) and we find their couplings 
to charge and spin. We also study the ratios of the pseudoparticle 
charge and spin transport masses over the corresponding static mass. 
These ratios provide valuable information on the effects of electronic
correlations in the transport properties of the quantum system.
We show that the transport of charge and spin in the Hubbard 
chain can, alternatively, be described by means of pseudoparticle 
kinetic equations. This follows from the occurrence of only 
forward-scattering pseudoparticle interactions at all energies. 
\end{abstract}
\vspace{0.3cm}
\pacs{PACS numbers:71.10. Pm, 05.30. Fk,72.90.+y,03.65. Ca}

\renewcommand{\baselinestretch}{1.656} 
\narrowtext
\section{Introduction}

The transport properties of strongly correlated electron models for 
low-dimensional conductors has been a subject of experimental and 
theoretical interest for over twenty years. Low-dimensional conductors 
show large deviations in their transport properties from the usual 
single-particle description. This suggests that electronic correlations 
might play an important role in these systems \cite{Jacobsen},
even if they are small \cite{Mori}. Solvable 
one-dimensional many-electron models such as the Hubbard chain are 
often used as an approximation for the study of the properties of 
quasi-one-dimensional conductors \cite{Jacobsen,Mori}. Although 
the Hubbard chain has been 
diagonalized long ago \cite{Lieb,Takahashi}, the involved form of 
the Bethe-ansatz (BA) wave function has prevented the calculation of 
dynamic response functions, these including the charge-charge and 
spin-spin response functions and their associate conductivity spectra. 

Information on low-energy expressions for correlation functions
can be obtained by combining BA with conformal-field theory
\cite{Frahm}. On the other hand, several approaches using 
perturbation theory \cite{Maldague}, 
bosonization \cite{Schulz90,Gimarchi}, the pseudoparticle formalism 
\cite{Carm4}, scaling methods \cite{Stafford}, and spin-wave theory 
\cite{Horsch} have been used to investigate the low-energy transport 
properties of the model away from half filling and at the 
metal -- insulator transition \cite{Lieb}. Unfortunately, 
only limited information on the transport properties at finite 
energies has been obtained by numerical methods 
\cite{Loh,Fye,NunoZFPB}.

Recently, a pseudoparticle description of all the BA Hamiltonian 
eigenstates \cite{CarmeloNuno97} has allowed the evaluation of 
analytical expressions for correlation functions at finite energy 
\cite{GCFT}. From these results one can obtain 
expressions for the absorption band edges of the frequency-dependent 
electronic conductivity, $\sigma(\omega)$ \cite{conductivity}. 
The pseudoparticle theory of Ref. \cite{CarmeloNuno97}
introduces new branches of pseudoparticles 
and generalizes for all energy scales previous low-energy studies 
\cite{David}. The new pseudoparticle branches are associated with 
{\it heavy pseudoparticles}. These are the quantum objects needed 
for the description of Hamiltonian eigenstates showing an energy 
gap relatively to the ground state. (This justifies why they are 
called {\it heavy}.) As in the case of the low-energy properties of 
the Hubbard model \cite{David}, it is possible to write the 
Hamiltonian in terms of a set of anticommuting pseudoparticle 
operators. Importantly, all the model eigenstates can be generated
from the $SO(4)$ ground state \cite{Nuno3}.

In this paper we use the above generalized pseudoparticle theory to
study the charge and spin currents of the Hubbard chain at finite
energy. For that we solve the BA equations with a twist angle for 
all densities and magnetizations. Moreover, we express the charge 
and the spin-diffusion currents in terms of the elementary currents 
of the charge and spin carriers. It is shown that the latter 
carriers are the $\alpha,\gamma$ pseudoparticles of the 
pseudoparticle-perturbation theory (PPT) \cite{CarmeloNuno97}. 
We evaluate their couplings to charge and spin and define the 
charge and spin pseudoparticle transport masses. The ratios 
of these masses over the corresponding static mass provide
important information on the role of electronic correlations
in the transport of charge and spin in the 1D quantum liquid.
Furthermore, we find that the transport of charge and spin  
can be described by means of pseudoparticle kinetic equations. 
Our results are a generalization to finite energies of
the low-energy results on transport of charge and spin
presented in Ref. \cite{Carm4}. This is possible by
means of the generalized pseudoparticle representation
introduced in Ref. \cite{CarmeloNuno97} which is a generalization
to finite-energy scales of the usual low-energy operator 
representation \cite{David} in terms of pseudoparticles 
\cite{Carm1,Carm2,Carm3,Carm5,Carm6}.

The paper is organized as follows. In Sec.\,\ref{pppt} we present 
the PPT description of the Hilbert space associated with the
zero-temperature transport of charge and spin. In Sec.\,\ref{baphi} 
we introduce the general BA equations with a spin dependent twist 
angle, $\phi_{\sigma}$, valid for all energy scales, and solve 
these equations by use of the PPT. The pseudoparticle transport
and static masses and the transport of charge and spin are
studied and discussed in Sec.\,\ref{masses}. In Sec.\,\ref{kinetic} 
we show that the transport of charge and spin can be described by 
means of kinetic equations. Finally, the concluding remarks are 
presented in Sec.\,\ref{remarks}.

\section{Pseudoparticle Perturbation Theory}
\label{pppt}

Let us consider the Hubbard-chain Hamiltonian with $N$ electrons in a 
magnetic field $H$ and with chemical potential $\mu$ 

\begin{eqnarray}
\hat{H} &=& -t\sum_{j,\sigma}
[c_{j\sigma}^{\dag}c_{j+1\sigma} + h. c.]
+U\sum_{j} [\hat{n}_{j,\uparrow}- 1/2]
[\hat{n}_{j,\downarrow}-1/2]\nonumber\\
&-&\mu(N_a-\sum_{j,\sigma}\hat{n}_{j,\sigma})
-\mu_0 H\sum_{j,\sigma}\sigma\hat{n}_{j,\sigma}\, ,
\label{hamilt}
\end{eqnarray}
where $c_{j\sigma}^{\dag }(c_{j\sigma})$ creates (annihilates)
one electron with spin $\sigma$ (here and when used as operator 
index, $\sigma = \uparrow,\downarrow$, and $\sigma=\pm1$ otherwise),
$\hat{n}_{j,\sigma}=c_{j\sigma}^{\dag}c_{j\sigma}$ is the number
operator at site $j$, $N_a$ is the number of sites of the chain 
(since we are using periodic boundary conditions it is rather
a ring), and $c_{N_a+1\sigma}=c_{1\sigma}$. In general, we 
use units such that $\hbar=1$, the lattice spacing 
$a=1$, and the electron charge $-e=1$.
The form of the interaction term accounts for the
particle -- hole symmetry of the model at half filling 
\cite{Lieb,Bill}. For simplicity, we consider electronic and 
magnetization densities in the domains $0\leq n \leq 1$ and 
$0\leq m \leq n$, respectively, where 
$n=N/N_a=[N_{\uparrow}+N_{\downarrow}]/N_a$,
$m=[N_{\uparrow}-N_{\downarrow}]/N_a$, and $N_{\sigma}$
is the number of $\sigma$ electrons. 

Although the pseudoparticle description refers to all Hamiltonian
eigenstates \cite{CarmeloNuno97}, in this paper we restrict our 
study to the Hilbert subspace involved in the zero-temperature 
charge and spin frequency-dependent conductivity \cite{conductivity}.
This is spanned by all the Hamiltonian eigenstates contained
in the states $\hat{j}^\zeta|GS\rangle$, where $|GS\rangle$ denotes
the ground state and $\hat{j}^\zeta$ are the charge ($\zeta=\rho$)
and spin ($\zeta=\sigma_z$) current operators (given by Eqs.
(\ref{currrho}) and (\ref{currsigma}) below, respectively).
Since these current operators  
commute with the six generators of the $\eta$-spin and spin algebras 
\cite{Yang}, our Hilbert subspace is in the present parameter 
space spanned only by the lowest-weight states (LWS's) of these 
algebras \cite{Nuno3}. This refers to the Hilbert subspace directly
described by the BA solution \cite{CarmeloNuno97}. [Therefore,
following the studies and notations of Ref. \cite{CarmeloNuno97}, 
we can use the $\alpha,\gamma =0$ pseudoparticles instead of the
$\alpha,\beta$ pseudoholes (with $\beta=\pm {1\over2}$) required 
for the description of the non-LWS's outside the BA solution.]

The PPT introduced in detail in Ref.\,\cite{CarmeloNuno97} involves 
infinite branches of pseudoparticles labelled by the quantum
numbers $\alpha$ and $\gamma$. Here $\alpha=c,s$ and 
$\gamma=0,1,2,3,\ldots\infty$. Fortunately, for most Hamiltonian
eigenstates of physical interest only a small number out of 
the infinite available $\alpha,\gamma $ bands have pseudoparticle 
occupancy. The possible pseudomomentum occupancies correspond 
to the different LWS's of the model. The pseudomomentum values 
$q$ are such that $q^{(-)}_{\alpha ,\gamma}\leq q\leq 
q^{(+)}_{\alpha ,\gamma}$, where the expressions of the
pseudo-Brillouin-zone limits, $q^{(\pm)}_{\alpha ,\gamma}$,
are given in Ref.\,\cite{CarmeloNuno97}. They are of the form 
$q^{(\pm)}_{\alpha ,\gamma} =\pm q_{\alpha ,\gamma}+O(1/N_a)$, where
for the case of a ground state $q_{\alpha ,\gamma}$ is given by

\begin{eqnarray}
q_{c ,0} &=& \pi \, ; \hspace{1cm } q_{c,\gamma } = \pi - 2k_F
\, \hspace{0.5cm} \gamma >0 \, ,\nonumber \\
q_{s ,0} &=& k_{F\uparrow} \, ; \hspace{1cm } q_{s,\gamma } = 
k_{F\uparrow } - k_{F\downarrow } \, \hspace{0.5cm} \gamma >0 \, .
\label{gsbz}
\end{eqnarray} 

The branches $c,0$ and $s,0$ have been called 
in previous low-energy studies $c$ and $s$, 
respectively \cite{Carm4,David}. They were shown to describe the 
low-energy excitations of the Hubbard chain and to determine the 
low-energy behavior of its charge and spin transport properties 
\cite{Carm4,David}. (In the limit of low energy the 
heavy-pseudoparticle branches are empty.) On the other hand, description 
of the LWS's of the model that have a finite-energy gap, $\omega_0$, 
relatively to the ground state involves the heavy pseudoparticle 
branches $c,\gamma>0$ and $s,\gamma>0$ \cite{CarmeloNuno97}.

A very useful concept in this theory is that of generalized ground 
state (GGS). In Ref. \cite{CarmeloNuno97} it was defined as the 
Hamiltonian eigenstate(s) of lowest energy in the Hilbert subspace 
associated with a given sub-canonical ensemble. The concept of 
sub-canonical ensemble follows from the conservation laws of the 
$\alpha ,\gamma $ pseudoparticle numbers, $N_{\alpha ,\gamma}$. 
Each Hamiltonian eigenstate has constant values for these numbers 
and a sub-canonical ensemble refers to a given choice of constant 
$N_{\alpha ,\gamma}$ numbers. 

On the other hand, in Ref. \cite{GCFT} the concept of GGS was 
extended to (i) filled $\alpha ,\gamma$ pseudoparticle seas with 
compact occupations around $q=0$, i.e. for 
$q^{(-)}_{F\alpha ,\gamma,+1}\leq q\leq q^{(+)}_{F\alpha ,\gamma,+1}$,
where the pseudo-Fermi points are given by $q^{(\pm)}_{F\alpha 
,\gamma,+1}=\pm {\pi N_{\alpha ,\gamma}\over N_a}+O(1/N_a)$, and (ii)
filled $\alpha ,\gamma$ pseudoparticle seas with compact occupations 
for $q^{(-)}_{\alpha ,\gamma}\leq q\leq q^{(-)}_{F\alpha ,\gamma,-1}$
and for $q^{(+)}_{F\alpha ,\gamma,-1}\leq 
q\leq q^{(-)}_{\alpha ,\gamma}$,
where the pseudo-Fermi points are given by 
$q^{(\pm)}_{F\alpha ,\gamma,-1}=\pm [q_{\alpha ,\gamma} - 
{\pi N_{\alpha ,\gamma}\over N_a}]+O(1/N_a)$. 
From the studies of Refs. \cite{GCFT,conductivity}, it will
be shown elsewhere that the creation of one $\alpha,\gamma$
pseudoparticle from the ground state involves, to
leading order, a number $2\gamma$ of electrons. Since the
currents are two-electron operators, it follows that
the creation of single $\alpha,1$ pseudoparticles from
the ground state are the most important contributions to the
transport of charge and spin at finite energies. On the other
hand, since the states $\hat{j}^\zeta|GS\rangle$ which
define our Hilbert space have zero momentum
(relatively to the ground state) and the creation 
from the ground state of single $\alpha,1$ pseudoparticles of type
(ii) is a finite-momentum excitation, for simplicity
in this paper we restrict our study to GGS's 
of type (i). Therefore, in order to simplify our notation
we denote the pseudo-Fermi points $q^{(\pm)}_{F\alpha ,\gamma,+1}$
simply by $q^{(\pm)}_{F\alpha ,\gamma}$. These are given by
$q^{(\pm)}_{F\alpha ,\gamma}=\pm q_{F\alpha ,\gamma}+O(1/N_a)$
where the pseudo-Fermi momentum \cite{CarmeloNuno97} 

\begin{equation}
q_{F\alpha ,\gamma} = {\pi N_{\alpha ,\gamma}\over N_a} \, ,
\label{qF}
\end{equation}
appears in several expressions below. Note, however, that the generalization
of our results to GGS's of type (ii) is straightfoward. We emphasize 
that Ref. \cite{CarmeloNuno97} definition of GGS refers to the
above choice (i) for the $c,0$ and $s,\gamma$ branches
and to the choice (ii) for the $c,\gamma$ branch with $\gamma >0$.
Therefore, in the case of the $c,\gamma >0$ pseudoparticles
our GGS choice differs from the choice of that reference.

The ground state is a special case of a GGS where there is no 
$\alpha,\gamma >0$ heavy-pseudoparticle occupancy 
\cite{CarmeloNuno97} and the pseudo-Fermi points (\ref{qF})
are of the form

\begin{equation}
q_{Fc ,0} = 2k_F \, ; \hspace{1cm } q_{Fs ,0} = k_{F\downarrow} 
\, ; \hspace{1cm } q_{F\alpha,\gamma } = 0 \hspace{0.5cm} 
\gamma >0 \, .
\label{gspfs}
\end{equation} 

The PPT consists in expanding the Hamiltonian\,(\ref{hamilt}) in the 
density of excited pseudoparticles relatively to the initial ground state.
This allows us to write Hamiltonian\,(\ref{hamilt}) in normal 
order relatively to that ground state as \cite{CarmeloNuno97}

\begin{equation}
:{\hat{H}}: = {\hat{H}}_{0} + {\hat{H}}_{Landau} \, , 
\label{hamiltnorm}
\end{equation}
where up to second order in the density of excited pseudoparticles, 
${\hat{H}}_{Landau}$ is of the form

\begin{equation}
{\hat{H}}_{Landau} = \hat{H}^{(1)} + \hat{H}^{(2)} \, , 
\label{H12}
\end{equation}
with

\begin{equation}
\hat{H}^{(1)} = \sum_{q,\alpha,\gamma}
\epsilon_{\alpha ,\gamma}(q):\hat{N}_{\alpha ,\gamma}(q): \, ,
\end{equation}

\begin{equation}
\epsilon_{\alpha ,0}(q) = \epsilon^{(0)}_{\alpha ,0}(q) 
- \epsilon^{(0)}_{\alpha ,0}(q_{F\alpha,0}) 
\, , \hspace{1cm}
\epsilon_{\alpha ,\gamma}(q) = 
\epsilon^{(0)}_{\alpha ,\gamma}(q) 
- \epsilon^{(0)}_{\alpha ,\gamma}(0) \, ,
\label{bands}
\end{equation}
and $\hat{H}^{(2)}$ is given by 

\begin{equation}
\hat{H}^{(2)}={1\over {N_a}}\sum_{q,\alpha ,\gamma} 
\sum_{q',\alpha',\gamma'}{1\over 2}f_{\alpha,\gamma;
\alpha',\gamma'}(q,q') :\hat{N}_{\alpha ,\gamma}(q):
:\hat{N}_{\alpha' ,\gamma'}(q'): \, .
\label{H2}
\end{equation}
Here $\hat{N}_{\alpha,\gamma}(q)=b_{q,\alpha,\gamma}^{\dag}
b_{q,\alpha,\gamma}$ is the $\alpha ,\gamma$-pseudoparticle 
operator number at pseudomomentum $q$ and the operators
$b_{q,\alpha,\gamma}^{\dag}$ and $b_{q,\alpha,\gamma}$ obey 
the usual anticommuting algebra \cite{CarmeloNuno97}. The 
Hamiltonian eigenstates decribed by the BA are also eigenstates 
of $\hat{N}_{\alpha,\gamma}(q)$ with eigenvalue $N_{\alpha,\gamma}(q)$
and eigenstates of $:\hat{N}_{\alpha,\gamma}(q):$ with eigenvalue
$\delta N_{\alpha,\gamma}(q)\equiv N_{\alpha,\gamma}(q)-
N^0_{\alpha,\gamma}(q)$, where $N^0_{\alpha,\gamma}(q)$ is 
the ground-state pseudomometum distribution \cite{CarmeloNuno97}.
These distributions characterize the occupancy configurations 
of the pseudomomenta in the $\alpha,\gamma$-pseudoparticle bands.

The physical meaning of the Hamiltonian terms
${\hat{H}}_{0}$ and ${\hat{H}}_{Landau}$ is explained
in Ref. \cite{CarmeloNuno97}. These Hamiltonian terms are
such that $[{\hat{H}}_{0},{\hat{H}}_{Landau}]=0$ and 
${\hat{H}}_{0}$ has eigenvalue $\omega_0$ given by 
\cite{CarmeloNuno97,GCFT}

\begin{equation}
\omega_0=2\mu\sum_{\gamma >0} \gamma N_{c,\gamma}
+ 2\mu_0H\sum_{\gamma >0}(1+\gamma) N_{s,\gamma}+
\sum_{\alpha,\gamma>0}\epsilon^0_{\alpha,\gamma}(0)
N_{\alpha,\gamma} \, ,
\label{gap}
\end{equation}
where $N_{\alpha,\gamma}$ are the numbers of $\alpha,\gamma$ 
heavy pseudoparticles created in the transition from the ground 
state to the GGS. 

The set of energies $\omega_0$, Eq. (\ref{gap}), play
a central role in the theory. This is because for a given
initial ground state the PPT is associated with a final Hilbert
subspace characterized by a set of finite $N_{\alpha,\gamma}$
numbers. The states which span such subspace differ from the
initial ground state by a small density of pseudoparticles
and have small positive $(\omega -\omega_0)$ energy. 
The main point of the PPT is that for
low-excitation energy, $(\omega -\omega_0)$, only the first
two Hamiltonian terms, Eq. (\ref{H12}), are relevant
\cite{CarmeloNuno97,GCFT}. Therefore, the truncated
Hamiltonian (\ref{hamiltnorm}) - (\ref{H12}) describes the
physics for energies just above the set of finite-energy
values $\omega_0$ of the form (\ref{gap}). [This includes
$\omega_0 =0$.]

Since the conservation of the electron numbers imposes the following
sum rules on the numbers $N_{\alpha,\gamma}$ \cite{CarmeloNuno97}

\begin{equation}
N_\downarrow = \sum_{\gamma >0}\gamma N_{c,\gamma} 
+ \sum_{\gamma}(1+\gamma)N_{s,\gamma}\, ,
\label{nbaixo}
\end{equation}
and

\begin{equation}
N = N_{c,0} + 2\sum_{\gamma>0}\gamma N{c,\gamma}\, ,
\label{nc}
\end{equation}
the creation of heavy pseudoparticles from the ground state
at constant electron numbers requires the annihilation
of $\alpha,0$ pseudoparticles. It follows from Eqs.
(\ref{nbaixo}) and (\ref{nc}) that the changes 
$\Delta N_{\alpha,0}$ associated with a 
corresponding ground-state -- GGS transition read

\begin{equation}
\Delta N_{s,0} = -\sum_{\gamma >0}\gamma N_{c,\gamma} 
- \sum_{\gamma >0}(1+\gamma)N_{s,\gamma}\, ,
\label{DNs0}
\end{equation}
and

\begin{equation}
\Delta N_{c,0} = - 2\sum_{\gamma>0}\gamma N{c,\gamma}\, .
\label{DNc0}
\end{equation}
For instance, the creation of one $c,\gamma $ heavy 
pseudoparticle from the ground state requires the annihilation
of a number $2\gamma $ of $c,0$ pseudoparticles and of a
number $\gamma $ of $s,0$ pseudoparticles, whereas the
creation of one $s,\gamma$ pseudoparticle involves the
annihilation of a number $1+\gamma$ of $s,0$ pseudoparticles
and conserves the number of $c,0$ pseudoparticles.

Although, following Eq.\,(\ref{gap}), $\omega_0$ can be large, 
we emphasize that the PPT is always a low $(\omega-\omega_0)$ energy 
theory. This is because within the PPT the densities of removed
$\alpha,0$ pseudoparticles, $-\Delta N_{\alpha,0}/N_a$, of added
$\alpha,\gamma >0$ heavy pseudoparticles, 
$\Delta N_{\alpha,\gamma}/N_a=N_{\alpha,\gamma}/N_a$, and
of their pseudoparticle -- pseudohole processes 
are always kept small. Moreover, for each set of finite 
$N_{\alpha,\gamma >0}$ numbers there is one PPT and one value of 
energy (\ref{gap}).

The energy bands $\epsilon^0_{\alpha,\gamma}(q)$ and the $f$-functions
$f_{\alpha,\gamma;\alpha',\gamma'}(q,q')$ are given, respectively, 
by \cite{CarmeloNuno97}

\begin{equation}
\epsilon_{c,0}^0(q) = -{U\over 2} - 2t\cos K^{(0)}(q) +
2t\int_{-Q}^{Q}dk\widetilde{\Phi }_{c,0;c,0}
\left(k,K^{(0)}(q)\right)\sin k \, ,
\label{e0c0q}
\end{equation}

\begin{equation}
\epsilon_{c,\gamma}^0(q) = - \gamma U
+ 4t Re \sqrt{1 - u^2[R^{(0)}_{c,\gamma}(q) - i\gamma]^2}
+ 2t\int_{-Q}^{Q}dk\widetilde{\Phi }_{c,0;c,\gamma}
\left(k,R^{(0)}_{c,\gamma}(q)\right)\sin k \, ,
\label{e0cgq}
\end{equation}

\begin{equation}
\epsilon_{s,\gamma}^0(q) = 2t\int_{-Q}^{Q}dk
\widetilde{\Phi }_{c,0;s,\gamma}
\left(k,R^{(0)}_{s,\gamma} (q)\right)\sin k \, ,
\label{e0sgq}
\end{equation}
and

\begin{eqnarray}
&\frac 1{2\pi}&
f_{\alpha,\gamma;\alpha',\gamma'}(q,q') =  v_{\alpha,\gamma}(q)
\Phi_{\alpha,\gamma;\alpha',\gamma'}(q,q')+v_{\alpha',\gamma'}(q')
\Phi_{\alpha',\gamma';\alpha,\gamma}(q',q) \nonumber \\
& + & \sum_{j=\pm 1} 
\sum_{\alpha''}\sum_{\gamma''}^{\infty}
\theta(N_{\alpha'',\gamma''})v_{\alpha'',\gamma''}
\Phi_{\alpha'',\gamma'';\alpha,\gamma}(jq_{F\alpha''
,\gamma''},q)\Phi_{\alpha'',\gamma'';\alpha',\gamma'}
(jq_{F\alpha'',\gamma''},q') \, .
\end{eqnarray} 
Here $\theta(x)=1$ for $x>0$ and $\theta(x)=0$ otherwise is the
usual theta function and 
$v_{\alpha,\gamma}=v_{\alpha,\gamma}(q_{F\alpha ,\gamma})$ is 
the velocity at the pseudo-Fermi point. The phase-shift functions 
$\tilde{\Phi }_{\alpha,\gamma;\alpha',\gamma'}$ and
the phase shifts $\Phi_{\alpha,\gamma;\alpha',\gamma'}$
are defined in Ref.\,\cite{CarmeloNuno97}. 
$\Phi_{\alpha,\gamma;\alpha ',\gamma '}(q,q')$ 
measures the shift in the phase of the $\alpha',\gamma'$ 
pseudoparticle of pseudomomentum $q'$ due to a zero-momentum 
forward-scattering collision with the $\alpha ,\gamma$ pseudoparticle
of pseudomomentum $q$. It is useful to introduce
the function $W^0(q)$ such that $W=K,R_{c,\gamma},R_{s,\gamma}$.
(Here, $\gamma=1,2,\ldots,\infty$ for $R_{c,\gamma}$ and 
$\gamma=0,1,2,\ldots,\infty$ for $R_{s,\gamma}$.)   
It represents any of the three ground-state rapidity functions
$K^{(0)}(q)$, $R^{(0)}_{c,\gamma}(q)$, and $R^{(0)}_{s,\gamma}(q)$,
whereas the functional $W(q)$ represents any of the three general
functional rapidity functions $K(q)$, $R_{c,\gamma}(q)$, 
and $R_{s,\gamma}(q)$ \cite{CarmeloNuno97}. These functionals are 
obtained from Eqs.\,(\ref{int1}), (\ref{int2}), and (\ref{int3}) 
with $\phi=0$. (In that Appendix we solve the BA equations
with twist angles.) The ground-state functions $W^0(q)$ are 
obtained by taking the particular choice $N_{\alpha,\gamma}(q)=
N_{\alpha,\gamma}^0(q)$ in the latter equations. It is useful
to introduce the pseudo-Fermi rapidity parameters
\cite{CarmeloNuno97}

\begin{equation}
Q = K^{(0)}(q_{Fc,0}) \, ; \hspace{1cm } r_{c ,0} =
{\sin Q\over u} \, ; \hspace{1cm } r_{\alpha,\gamma } = 
R^{(0)}_{\alpha,\gamma}(q_{F\alpha ,\gamma}) \, ,
\label{gsfs}
\end{equation} 
where $Q$ appears in the integrals of Eqs. (\ref{e0c0q}) -
(\ref{e0sgq}) and in the last expression $r_{\alpha,\gamma }$ refers
to all $\alpha,\gamma$ branches except $c,0$.

\section{Charge and Spin Currents: Solution of the BA
Equations}
\label{baphi}

Within linear response theory the charge and spin currents 
of the 1D Hubbard model can be computed by performing
a spin-dependent Peierls-phase substitution in the
hopping integral of Hamiltonian (\ref{hamilt}), 
$t \rightarrow te^{\pm i\phi_{\sigma}/N_a}$
\cite{Maldague,Shastry}.

It has been possible to solve the Hamiltonian (\ref{hamilt}) 
with the additional hopping phase $e^{\pm i\phi_{\sigma}/N_a}$
by means of the coordinate BA both with twisted and toroidal 
boundary conditions, both approaches giving essentially the 
same results \cite{Shastry,Martins}. One obtains the energy 
spectrum of the model parameterized by a set a numbers 
$\{k_j,\Lambda_\delta\}$ which are solution of the BA
interaction equations given by

\begin{equation}
e^{ik_j N_a}=e^{i\phi_{\uparrow}}\prod_{\delta=1}^{N_\downarrow}
\frac{\sin(k_j)-\Lambda_\delta+iU/4}{\sin(k_j)-\Lambda_\delta-
iU/4}\, ,
\hspace{1cm} (j=1,\ldots,N)\, ,
\label{inter1}
\end{equation}
and
\begin{equation}
\prod_{j=1}^{N}
\frac{\sin(k_j)-\Lambda_\delta+iU/4}{\sin(k_j)-\Lambda_\delta-iU/4}
=e^{i(\phi_{\downarrow}-\phi_{\uparrow})}
\prod_{\beta=1,\neq\delta}^{N_\downarrow}
\frac{\Lambda_\beta-\Lambda_\delta+iU/2}
{\Lambda_\beta-\Lambda_\delta-iU/2}\, ,
\hspace{1cm} (\delta=1,\ldots,N_\downarrow)\, .
\label{inter2}
\end{equation}

However, previous studies of the $\phi_\sigma\neq 0$ problem 
\cite{Shastry,Martins} have only considered the real
BA rapidities of Eqs. (\ref{inter1}) and (\ref{inter2})
which refer to low energy. Here we follow the same steps as 
Takahashi \cite{Takahashi} for the $\phi_\sigma= 0$ interaction 
Eqs.\,(\ref{inter1}) and (\ref{inter2}) and consider both
real and complex rapidities. We then arrive to the 
following $\phi_\sigma\neq 0$ equations which refer to
the real part of these rapidities 

\begin{eqnarray}
k_jN_a &=&
2\pi I_j^c+\phi_\uparrow
-\sum_{\gamma }\sum_{j'=1}^{N_{s,\gamma}}
2\tan^{-1}\left(\frac{\sin(k_j)/u-R_{s,\gamma,j'}}{(\gamma+1)} \right)
\nonumber \\
&-& \sum_{\gamma >0}\sum_{j'=1}^{N_{c,\gamma}}
2\tan^{-1}\left(\frac{\sin(k_j)/u-R_{c,\gamma,j'}}{\gamma}\right)\, ,
\label{tak1}
\end{eqnarray}

\begin{eqnarray} 
2N_a Re \, \sin^{-1}([R_{c,\gamma,j}+i\gamma]u)
&=& 2\pi I^{c,\gamma}_j
+ \gamma (\phi_\uparrow+\phi_\downarrow)
-\sum_{j'=1}^{N_c}2\tan^{-1}\left(\frac{\sin(k_{j'})/u-R_{c,\gamma,j}}
{\gamma}\right)\nonumber\\
&+&\sum_{\gamma'>0}\sum_{j'=1}^{N_{c,\gamma'}}
\Theta_{\gamma,\gamma'}(R_{c,\gamma,j}-R_{c,\gamma',j'})\, ,
\label{tak2}
\end{eqnarray}
and

\begin{eqnarray}
\sum_{j'=1}^{N_c}2\tan^{-1}\left(\frac{R_{s,\gamma,j}-\sin(k_j')/u}
{(1+\gamma)}\right) &+& (1+\gamma)(\phi_\uparrow-\phi_\downarrow)
\nonumber \\
&=& 2\pi I^{s,\gamma}_j+
\sum_{\gamma'}\sum_{j'=1}^{N_{s,\gamma'}}
\Theta_{\gamma+1,\gamma'+1}(R_{s,\gamma,j}-R_{s,\gamma',j'})\, .
\label{tak3}
\end{eqnarray}
The functions $\Theta_{\gamma,\gamma'}(x)$ [and 
$\Theta_{\gamma+1,\gamma'+1}(x)$] of Eqs.
\,(\ref{tak1}), (\ref{tak2}), and (\ref{tak3}) are
defined in Ref. \cite{CarmeloNuno97}. The following definitions
for the real part of the rapidities, 
$\Lambda_\alpha^{n+1}/u=R_{s,\gamma,j}$ (with $n+1=\gamma$ and 
$\alpha=j$), $\Lambda_\alpha^{' \, n}/u=R_{c,\gamma,j}$ 
(with $n=\gamma$ and $\alpha=j$), and $\gamma=1,2,\ldots,\infty$ 
for the $N_{c,\gamma}$ sums and $\gamma=0,1,2,\ldots,\infty$ for 
the $N_{s,\gamma}$ sums, allows us to recover Takahashi's 
formulae for $\phi=0$ \cite{Takahashi}. Here and often below
we use the notation $c\equiv c,0$, which allows the $c,\gamma$
sums to run over $1,2,3,\ldots,\infty$. Whether we are using this 
notation or the previous one will be obvious from the context.

The important numbers $I_j^c$, $I^{c,\gamma}_j$, and 
$I^{s,\gamma}_j$ which appear in going from Eqs.\,(\ref{inter1}) 
and (\ref{inter2}) to Eqs.\,(\ref{tak1}), (\ref{tak2}), and 
(\ref{tak3}) are the quantum numbers which describe
the Hamiltonian eigenstates. Depending on the parity of the numbers 
$[\sum_{\gamma=0}N_{s,\gamma}+\sum_{\gamma=1}
N_{c,\gamma}]$, $[N_a-N+N_{c,\gamma}]$, and $[N-N_{s,\gamma}]$, 
respectively, they are consecutive integers or half-odd integers   
\cite{CarmeloNuno97}. All the LWS's of the model are described
by the different occupancies of these quantum numbers. For example,
the ground state is described by a compact symmetric occupancy around 
the origin of the numbers $I_j^c$ and $I_j^{s,0}$, and by zero 
occupancy for the numbers $I^{c,\gamma}_j$ and 
$I^{s,\gamma>0}_j$ \cite{CarmeloNuno97}.
It is convenient to describe the eigenstates of the model  
in terms of pseudomomentum $\{q_j^{\alpha,\gamma}=
2\pi I_j^{\alpha,\gamma}/N_a\}$ distributions, where 
$I_j^{c,0} \equiv I_j^c$.

The energy and momentum eigenvalues are given by \cite{CarmeloNuno97}

\begin{eqnarray}
E&=&-2t\sum_{j=1}^{N_c}\cos(k_j)+
\sum_{\gamma >0}\sum_{j=1}^{N_{c,\gamma}}4t Re
\sqrt{1-u^2[R_{c,\gamma,j}-i\gamma]^2}\nonumber\\
&+&N_a(U/4-\mu)+N(\mu-U/2)-\mu_0 H(N_\uparrow-N_\downarrow),
\label{energy}
\end{eqnarray}
and

\begin{equation}
P = \frac{2\pi}{N_a}\left[\sum_{j=1}^{N_c}I_j^c+
\sum_{\gamma}\sum_{j=1}^{N_{s,\gamma}}I_j^{s,\gamma}
-\sum_{\gamma >0}\sum_{j=1}^{N_{c,\gamma}}I_j^{c,\gamma}\right]
+\pi \sum_{\gamma >0} N_{c,\gamma} \, ,
\label{momentum}
\end{equation}
respectively. We emphasize that the rapidity dependence on
$\phi_{\sigma}$ is defined by Eqs. (\ref{tak1})-(\ref{tak3}) 
and determines the energy-functional (\ref{energy}) dependence on 
$\phi_{\sigma}$. The corresponding $\phi_{\sigma}=0$ expressions 
recover the rapidity and energy expressions of 
Ref \cite{CarmeloNuno97}.

In the limit of a large system ($N_a\rightarrow \infty$, $N/N_a$ fixed)
we can develop a generalization of the low-energy 
pseudoparticle-Landau-liquid description of the Hubbard model 
\cite{Carm1,Carm2,Carm3} and of its operational representation 
\cite{David}. This generalization refers to energies just 
above the set of energies $\omega_0$, Eq. (\ref{gap}), where the
Hamiltonian (\ref{hamiltnorm}) describes the quantum-problem
physics. Note that the choice $\omega_0=0$, which refers 
to $N_{\alpha,\gamma}=0$
for $\gamma >0$, recovers the usual low-energy theory
of Refs. \cite{David,Carm1,Carm2,Carm3,Carm5,Carm6}. On the other hand,
when $\omega_0>0$, in addition to finite occupancy of 
the usual $c,0\equiv c$ and $s,0\equiv s$ pseudoparticle bands, 
there is finite occupancy for some of the branches of the heavy 
$c,\gamma$ and $s,\gamma$ pseudoparticles \cite{CarmeloNuno97}. 

In the above thermodynamic limit the rapidity real parts
$k_j=k_j(q_j)$, $R_{s,\gamma,j}=R_{s,\gamma,j}(q_j)$, and 
$R_{c,\gamma,j}=R_{c,\gamma,j}(q_j)$ proliferate on the 
real axis. As in Refs. \cite{Carm1,Carm2,Carm3},
equations \,(\ref{tak1}), (\ref{tak2}), and (\ref{tak3}) can be 
rewritten as integral equations with an explicit dependence on the 
pseudomomentum distribution functions $N_{\alpha,\gamma}(q)$. These 
are Eqs.\, (\ref{int1}), (\ref{int2}), and (\ref{int3}) of
Appendix\,\ref{normalOrder} which refer to the case
$\phi_{\sigma}\neq 0$. In that Appendix we derive ground-state 
normal-ordered expressions for the rapidities and charge
and spin currents. 

The combination of Eqs.\,(\ref{energy}), (\ref{int1}), (\ref{int2}), 
and (\ref{int3}) allows the evaluation of several interesting 
transport quantities. This includes the charge and spin currents 
and the charge and spin pseudoparticle transport masses. The 
charge and spin current 
operators $\hat{j}^\zeta$ (with $\zeta=\rho$ for charge, and 
$\zeta=\sigma_z$ for spin) are for the 1D Hubbard model
given by \cite{Carm4}

\begin{equation}
\hat{j}^\rho=-eit\sum_{\sigma}\sum_{j=1}^{N_a}
(c_{j\sigma}^{\dag}c_{j+1\sigma}-
c_{j+1\sigma}^{\dag }c_{j\sigma})\, ,
\label{currrho}
\end{equation}
and
\begin{equation}
\hat{j}^{\sigma_z}=-(1/2)it\sum_{\sigma}\sum_{j=1}^{N_a}
\sigma(c_{j\sigma}^{\dag}c_{j+1\sigma}-
c_{j+1\sigma}^{\dag}c_{j\sigma})\, .
\label{currsigma}
\end{equation}
Importantly, the discrete nature of the model 
implies that the commutators of the Hamiltonian (\ref{hamilt}) 
and of the current operators $\hat{j}^\zeta$,
Eqs. (\ref{currrho}) and (\ref{currsigma}), are non zero. It 
follows that the BA wave function 
does not diagonalizes simultaneously the Hamiltonian 
(\ref{hamilt}) and the current operators (\ref{currrho}) 
and (\ref{currsigma}). Since the BA solution alone only provides 
the diagonal part in the Hamiltonian-eigenstate basis
of the physical operators \cite{CarmeloNuno97},
we can only evaluate expressions for the diagonal part
of the currents which provide the mean values of the 
charge and spin currents. These refer to all LWS's and are important 
quantities for they allow us to compute the transport masses of 
the charge and spin carriers of the system. In addition, our 
formalism defines the charge and spin carriers. These are found 
to be the $c$ (i.e., $c,0$) and $c,\gamma$ pseudoparticles for charge, and 
the $c$ and $s,\gamma$ pseudoparticles for spin. 
This follows from Eqs.\,(\ref{int1})-(\ref{int3})
and also from the Boltzmann transport analysis of 
Sec.\,\ref{kinetic}. 

We emphasize that combining the generalized pseudoparticle
representation \cite{CarmeloNuno97} with a low-energy 
$(\omega -\omega_0)$ conformal-field theory \cite{GCFT},
leads to finite-energy current -- current correlation function 
expressions which are determined by the non-diagonal terms 
(in the Hamiltonian-eigenstate basis) of the current operators.
This is a generalization of the low-energy correlation-function
studies of Refs. \cite{Frahm,Carm5}. 
However, these expressions cannot be derived within the BA solution
alone. Therefore, these studies go beyond the scope of the 
present paper and here we consider the diagonal part of the 
charge and spin current operators only.

The mean value of the current operator $\hat{j}^\zeta$
in a given LWS, $\vert m\rangle$, is given by

\begin{equation}
\langle m\vert \, \hat{j}^\zeta\vert m\rangle =-\left.
\frac{d(E_m/N_a)}{d(\phi/N_a)} \right\vert_{\phi=0} \, ,
\label{mean}
\end{equation}
where $E_m$ is the energy of the Hamiltonian eigenstate 
$\vert m\rangle$ and \cite{Shastry}

\begin{eqnarray}
\phi & = & \phi_\uparrow=\phi_\downarrow \, ,
\hspace{1cm}  \zeta=\rho \, ,\nonumber \\
\phi & = & \phi_\uparrow=-\phi_\downarrow \, ,
\hspace{1cm}  \zeta=\sigma_z \, .
\label{phis}
\end{eqnarray}
 
In our basis the LWS's are simply obtained 
by considering all the possible occupation distributions of the 
pseudomomenta $q_j =2\pi I_j^{\alpha,\gamma}/N_a$. Therefore, it is 
convenient to describe the matrix elements 
$\langle m\vert \, \hat{j}^\zeta\vert 
m\rangle $ in terms of the pseudomomentum occupation
distributions $N_{\alpha,\gamma}(q)$. This leads
to a functional form for the current mean values. 
The computation of $\langle m\vert \, \hat{j}^\zeta\vert m\rangle$ 
involves the expansion of Eqs. (\ref{energy}) and 
(\ref{int1})-(\ref{int3}) up to first order in the flux
$\phi$. Writing Eq.\,(\ref{energy})
in the limit of $N_a\rightarrow\infty$, expanding it up to first order
in the flux $\phi$, and using Eq.\,(\ref{mean}) we obtain

\begin{eqnarray}
\langle m\vert \, \hat{j}^\zeta\vert m\rangle &=& -2t\frac 1{2\pi}
\int_{-q_c}^{q_c}dqN_c(q)K^{\phi}(q)\sin(K(q))\nonumber\\
&+&\sum_{\gamma >0}4t\frac 1{2\pi}
\int_{-q_{c,\gamma}}^{q_{c,\gamma}}dq N_{c,\gamma}(q)
Re\, \frac{u^2[R_{c,\gamma}(q)-i\gamma]}
{\sqrt{1-u^2[R_{c,\gamma}(q)-i\gamma]^2}}
R_{c,\gamma}^{\phi}(q)\, ,
\label{jmean}
\end{eqnarray}
where the important functions $W^{\phi}(q)$ (with $W=K, R_{s,\gamma}$, 
and $R_{c,\gamma}$) are the derivatives of the rapidity functions
defined by Eqs. (\ref{int1}) - (\ref{int3}) in order to the flux 
$\phi$ at $\phi=0$. They obey a set of integral equations obtained 
from differentiation of Eqs. (\ref{int1}) - (\ref{int3}).

It is convenient to write $\langle m\vert \, 
\hat{j}^\zeta\vert m\rangle$ in normal order relatively to 
the ground state. To achieve this goal we expand 
all the rapidities $W(q)$ and the functions $W^{\phi}(q)$ as

\begin{eqnarray}
&W&(q)=W^0(q)+W^1(q)+\ldots\, ,
\label{exp1}\\
&W&^{\phi}(q)=W^{0,\phi}(q)+W^{1,\phi}(q)+\ldots\, ,
\label{exp2}
\end{eqnarray}
respectively. In these equations the functions $W^0(q)$ and 
$W^{0,\phi}(q)$ are both referred to the ground state, and 
the functions $W^1(q)$ and $W^{1,\phi}(q)$ are first-order 
functionals of the deviations $\delta N_{\alpha,\gamma}(q)$.
In Appendix \ref{normalOrder} we show that the above expansions lead
to a ground-state normal-ordered representation. To first order in
the deviations the normal-ordered expression for the matrix element 
(\ref{jmean}) simply reads

\begin{equation}
\langle m\vert \, \hat{j}^\zeta\vert m\rangle=\sum_{\alpha}
\sum_{\gamma}
\int_{-q_{\alpha,\gamma}}^{q_{\alpha,\gamma}}dq
\delta N_{\alpha,\gamma}(q)
j^\zeta_{\alpha,\gamma}(q)\, ,
\label{deltaj}
\end{equation}
where the elementary-current spectrum $j_{\alpha,\gamma}^\zeta(q)$ is 
given by

\begin{equation}
j_{\alpha,\gamma}^\zeta(q)=
\sum_{\alpha'}\sum_{\gamma'}
\theta(N_{\alpha',\gamma'})
{\cal C}^{\zeta}_{\alpha',\gamma'}\left[
v_{\alpha,\gamma}(q)\delta_{\alpha ,\alpha'}
\delta_{\gamma ,\gamma'} +
F^1_{\alpha,\gamma;\alpha',\gamma'}(q)\right]\, .
\label{jq}
\end{equation}
Here

\begin{equation}
F^1_{\alpha,\gamma;\alpha',\gamma'}(q)={1\over 2\pi}
\sum_{j=\pm 1}jf_{\alpha,\gamma;\alpha'
,\gamma'}(q,jq_{F\alpha',\gamma'})\, ,
\label{Fq}
\end{equation}
and ${\cal C}^{\zeta}_{\alpha,\gamma}$ are the coupling constants 
of the pseudoparticles to charge and spin given by

\begin{equation}
{\cal C}^{\zeta}_{\alpha,\gamma} = \delta_{\alpha,c}\delta_{\gamma,0}
+ {\cal K}^{\zeta}_{\alpha,\gamma} \, ,
\label{coupling}
\end{equation}
where

\begin{equation}
{\cal K}_{\alpha,\gamma}^{\rho}=\delta_{\alpha,c}2\gamma \, ;
\hspace{1cm}
{\cal K}_{\alpha,\gamma}^{\sigma_z}=
-\delta_{\alpha,s}2(1+\gamma) \, .
\label{coupling2}
\end{equation}

As in a Fermi liquid \cite{Pine,Baym}, the expressions of the 
elementary currents (\ref{jq}) involve the velocities 
$v_{\alpha,\gamma}(q)$ and the interactions [or $f$-functions] 
$f_{\alpha,\gamma;\alpha',\gamma'}(q,q')$. However, the 
pseudoparticle coupling contants to charge and spin, Eqs. 
(\ref{coupling}) and (\ref{coupling2}), are very different
from the corresponding couplings of the Fermi-liquid 
quasiparticles. We emphasize that at low energy
Eq. (\ref{deltaj}) recovers the expression already obtained
in Ref. \cite{Carm4} which only contains the $c\equiv c,0$ and
$s\equiv s,0$ elementary currents. The coupling constants 
(\ref{coupling})-(\ref{coupling2})
play an important role in the description of charge and
spin transport and are a generalization for
$\gamma >0$ of the couplings introduced in Ref.
\cite{Carm4}. They define the $\alpha,\gamma$ pseudoparticles 
as charge and spin carriers. We emphasize that
when ${\cal C}^{\zeta}_{\alpha,\gamma}=0$ the corresponding 
$\alpha,\gamma$ pseudoparticles do not couple to $\zeta $
(i.e. charge or spin). Therefore, for $\gamma >0$ the
$c,\gamma$ and $s,\gamma$ pseudoparticles do not couple to 
spin and charge, respectively. This is related to the
charge and spin separation of one-dimensional quantum
liquids which in the case of the present model was
studied in Refs. \cite{David,Carm6}. Importantly, when 
${\cal C}^{\zeta}_{\alpha,\gamma}=0$ the $\alpha,\gamma$
pseudoparticle -- pseudohole processes do not contribute
to the $\zeta $ correlation functions.
It will be shown elsewhere that this provides a
powerful selection rule which implies that some
of the terms obtained from the small $(\omega-\omega_0)$
conformal-field theory \cite{GCFT} for the correlation
functions vanish.

In contrast to the general current expression
(\ref{jmean}), expression (\ref{deltaj}) is only
valid for Hamiltonian eigenstates which differ from
the ground-state pseudoparticle occupancy by a small
density of psudoparticles. This is because in expression
(\ref{deltaj}) we are only considering the first-order 
deviation term.

The velocity term of current-spectrum expression \,(\ref{jq}) 
is what we would expect for a non 
interacting gas of pseudoparticles and the 
second term takes account for the dragging effect on a single 
pseudoparticle due to its interactions with the other
pseudoparticles. (This is similar to the Fermi-liquid 
quasiparticle elementary currents \cite{Pine,Baym}.) 
We remind that Eq.\,(\ref{jq}) is valid for 
finite energies $\omega $ just above the energy $\omega_0$ 
corresponding to the suitable set of finite $N_{\alpha,\gamma}$ 
numbers. These numbers characterize the state 
$\vert m\rangle$. Therefore, the sum over $\gamma$ is
in Eq.\,(\ref{jq}) restricted to the $\alpha,\gamma$ bands that have 
non-zero occupancy of pseudoparticles, as is imposed by the
presence of the step-function. Within the PPT, the deviation
second-order pseudoparticle energy expansion corresponds
to the deviation first-order current expansion (\ref{deltaj})
which refers to small positive values $(\omega -\omega_0)$ 
of the excitation energy. In contrast to Fermi liquid theory, 
our PPT is valid for finite energies [just above the 
energy values $\omega_0$, Eq. (\ref{gap})] because
(i) there is only forward scattering among the pseudoparticles 
at all energy scales and (ii) at small $(\omega-\omega_0)$ 
energy values only two-pseudoparticle forward-scattering 
interactions are relevant. (In a Fermi liquid this
is only true for $\omega_0=0$ and $\omega\rightarrow 0$
\cite{Pine,Baym}.) We emphasize that the current expression
(\ref{jmean}) includes all orders of scattering and, therefore,
applies to all energies without restrictions. Elsewhere
it will be shown to be useful in the study of charge and
spin transport at finite temperatures.

\section{PSEUDOPARTICLE TRANSPORT AND STATIC MASSES}
\label{masses}

In reference \cite{Carm4} the charge and spin transport
masses of the $c,0$ and $s,0$ pseudoparticles were defined
and were shown to play an important role in the transport
of charge and spin.
For instance, they were shown to fully determine
the charge and spin stiffnesses
\cite{Carm4,Stafford,Fye,Shastry,Millis,Scalapino}.
Here we generalize the mass definitions of Ref. \cite{Carm4} to 
$\gamma >0$ and define the charge and spin transport masses, 
$m_{\alpha,\gamma}^{\zeta}$, as

\begin{equation} 
m^{\zeta}_{\alpha,\gamma}=
\frac {q_{F\alpha,\gamma}} {{\cal C}^{\zeta}_{\alpha,\gamma}
j^\zeta_{\alpha,\gamma}}\, ,
\label{tmass}
\end{equation}
where $j^\zeta_{\alpha,\gamma}=j^\zeta_{\alpha,\gamma}
(q_{F\alpha,\gamma})$. They contain important physical information. 
As in a Fermi liquid \cite{Pine,Baym}, the ratio

\begin{equation} 
r^{\zeta}_{\alpha,\gamma} = 
m^{\zeta}_{\alpha,\gamma}/m^{\ast}_{\alpha,\gamma} \, ,
\label{ratio}
\end{equation}
of the transport mass over the static mass provides a measure 
of the correlations importance in transport. Similarly to the 
$\gamma =0$ case \cite{Carm4}, the latter is in general defined as

\begin{equation} 
m^{\ast}_{\alpha,\gamma}=
\frac {q_{F\alpha,\gamma}} {v_{\alpha,\gamma}}\, .
\label{smass}
\end{equation}
In Appendix \ref{static} we define the mass (\ref{smass})
in terms of suitable functions and find some limiting
expressions.

It can be shown from the transport- and static-mass
expressions that the ratio $m^{\zeta}_{\alpha,\gamma}
/m^{\ast}_{\alpha,\gamma}$ involves the Landau parameters

\begin{equation}
F^i_{\alpha,\gamma;\alpha',\gamma'}\equiv 
F^i_{\alpha,\gamma;\alpha',\gamma'}(q_{F\alpha,\gamma}) \, ;
\hspace{1cm} i = 0,1 \, ,
\label{Fi}
\end{equation}
with $F^i_{\alpha,\gamma;\alpha',\gamma'}(q)$ given by Eq.
(\ref{Fq}). These parameters can be written as follows

\begin{equation}
F^i_{\alpha,\gamma;\alpha',\gamma'} =
-\delta_{\alpha,\alpha'}\delta_{\gamma,\gamma'}v_{\alpha,\gamma}+
\sum_{\alpha'',\gamma''}\theta(N_{\alpha'',\gamma''}) 
v_{\alpha'',\gamma''}
[\xi^i_{\alpha'',\gamma'';\alpha,\gamma}
\xi^i_{\alpha'',\gamma'';\alpha',\gamma'}]\, ,
\end{equation}
where the quantities $\xi^i_{\alpha,\gamma;\alpha',\gamma'}$ are
given by

\begin{equation}
\xi^i_{\alpha,\gamma;\alpha',\gamma'}=
\delta_{\alpha,\alpha'}\delta_{\gamma,\gamma'}+
\sum_{l=\pm 1}l^i \Phi_{\alpha,\gamma;\alpha',\gamma'}
(q_{F\alpha,\gamma},lq_{F\alpha',\gamma'})\, .
\label{xi}
\end{equation}
We find for the ratios $m^{\zeta}_{\alpha,\gamma}
/m^{\ast}_{\alpha,\gamma}$ the following expressions

\begin{equation}
\frac{m^{\zeta}_{\alpha,0}}{m^{\ast}_{\alpha,0}} =
\frac {v_{\alpha,0}}{{\cal C}^{\zeta}_{\alpha,\gamma}
(\sum_{\alpha',\alpha''}{\cal C}^{\zeta}_{\alpha',0}
v_{\alpha'',0}\xi^1_{\alpha'',0;\alpha,0}
\xi^1_{\alpha'',0;\alpha',0})} \, ,
\hspace{1cm}\gamma=0\, , 
\label{ratio0}
\end{equation}
and

\begin{equation}
\frac{m^{\zeta}_{\alpha,\gamma}}{m^{\ast}_{\alpha,\gamma}} = \frac 1{
{\cal C}^{\zeta}_{\alpha,\gamma}({\cal C}^{\zeta}_{\alpha,\gamma}+
\sum_{\alpha'}{\cal C}^{\zeta}_{\alpha',0}\xi^1_{\alpha,\gamma;
\alpha',0})}\,, \hspace {1cm} \gamma > 0 \, . 
\label{ratios}
\end{equation}
In the Table analytical limiting values for the mass ratios of form
(\ref{ratio}) are listed. Obviously, since for $\gamma >0$ the $c,\gamma$ 
and $s,\gamma$ pseudoparticles do not couple to spin and charge,
respectively, the ratios $m_{c,\gamma}^{\sigma_z}/m^{\ast}_{c,\gamma}$ 
and $m_{s,\gamma}^\rho/m^{\ast}_{s,\gamma}$ are infinite. 

As was referred previously, it can be shown from the results
of Refs. \cite{CarmeloNuno97,GCFT,conductivity} that the
creation of one $\alpha ,\gamma$ pseudoparticle from
the ground state is a finite-energy excitation which,
to leading order, involves a number $2\gamma $ of electrons.
Therefore, and since the current operators are
of two-electron character and couple to charge and spin
according to the values of the constants
(\ref{coupling}) and (\ref{coupling2}), at finite energies 
the $c,1$ and $s,1$ heavy pseudoparticles play the major role in 
charge and spin transport, respectively. On the other hand,
the $\alpha,\gamma >1$ heavy pseudoparticles contribute
very little to charge and spin transport.
It follows that in the present section we restrict our study to the
ratios (\ref{ratio}) involving $\gamma =1 $ heavy pseudoparticles.
We consider the ratios $m_{c,1}^{\rho}/m^{\ast}_{c,1}$ 
and $m_{s,1}^{\sigma_z}/m^{\ast}_{s,1}$. (Note that
$m_{c,1}^{\sigma_z}/m^{\ast}_{c,1}=m_{s,1}^{\rho}/
m^{\ast}_{s,1}=\infty$.) We also consider the case of the 
$c,0$ charge-mass ratio which is closely related to the charge 
stiffness studied in detail in Ref. \cite{Carm4}. 
In Figs.\,\ref{f1}-\ref{f12} these ratios 
are plotted as functions of the onsite repulsion 
$U$ in units of $t$, electronic density $n$, and magnetic 
field $h=H/H_c$. Note that the ratios of the figures are 
smaller than one. 
Moreover, the $\alpha ,1$ mass ratios never achieve the value 
$1$ whereas the $\alpha ,0$ mass ratios tend to one in some 
limits because of the generalized adiabatic principle of Ref. 
\cite{Carm4}. 

Combined analysis of Figs. \ref{f1} and \ref{f2} reveals that the 
charge-mass ratio for the $c,1$ pseudoparticle is, for large $U$, 
fairly independent both of the band filling $n$ and magnetic field 
$h$. It is a decreasing function of $U$ and as a function of the
density, $n$, goes through a maximum for a density which is
a decreasing function of $U$. Moreover, figures \ref{f3} and \ref{f4}
show that this ratio is a decreasing function of the magnetic
field. 

In contrast, Figs. \ref{f5} - \ref{f8} reveal that the 
charge-mass ratio for the $c,0$ pseudoparticle is an increasing 
function of $U$ and of $h$ and as a function of the density, $n$, 
goes through a minimum for a density which is a decreasing 
function of $U$. Note that from Fig. \ref{f5} the evolution of 
the $c,0$ pseudoparticles to free spinless fermions as $U$ 
increases is clear. This is signaled by the ratio going to one
as $U\rightarrow\infty$. This behavior follows from the
generalized adiabatic principle of Ref. \cite{Carm4} and
agrees with the well known decoupling of the BA wave 
function in free spinless 
fermions (in the low-energy sector \cite{Ricardo97}) and 
localized antiferromagnetic spins \cite{Ogata90}. 
Figures \ref{f7} and \ref{f8} also reveal that
in the fully-polarized ferromagnetic limit, $h\rightarrow 1$, 
the ratio goes to one. This mass-ratio behavior also
follows from the generalized adiabatic principle \cite{Carm4}
and confirms that in that limit the onsite Coulomb interations 
play no role in charge transport (they are froozen by the Pauli 
principle) .
 
Note that in the large-$U$ Figs. \ref{f2} - (c) and \ref{f6} - 
(c) the ratios $m_{c,1}^\rho/m^{\ast}_{c,1}$ and 
$m_{c,0}^\rho/m^{\ast}_{c,0}$, respectively, are almost symmetric 
around the density $n=0.5$. This implies that for large $U$ 
the charge transport properties show similarities in the
cases of vanishing densities and of densities 
closed to one. 

Figure \ref{f9} shows that the spin-mass ratio of the $s,1$
pseudoparticles is a decreasing function of $U$ but that it
depends little on $U$ for $U>6$. For large $U$ this ratio
almost does not depend on the density $n$, as revealed by
Fig. \ref{f10} - (c). Figures \ref{f10} show that, in general,
it is an increasing function of $n$ but that for $h\rightarrow
1$ it has a maximum for an intermediate density. In figures
\ref{f11} and \ref{f12} this spin-mass ratio is plotted as a 
function of $h$. It is a decreasing function of $h$ 

The transport masses are very sensitive to the effects of 
electronic correlations, as for instance to the metal-insulator 
transition which occurs at zero temperature 
when $n\rightarrow 1$ \cite{Lieb}. As a direct result of
this transition, $m_{c,0}^{\rho}\rightarrow\infty$ as
$n\rightarrow 1$, as was shown and discussed in Ref.
\cite{Carm4}. Moreover, the zero-temperature charge and
spin stiffnesses, $D^{\zeta}$, 
\cite{Carm4,Stafford,Fye,Shastry,Millis,Scalapino}, defined
as
 
\begin{equation}
D^{\zeta}= \left. \frac 1{2}
\frac{d^2(E_0/N_a)}{d(\phi/N_a)^2}\right\vert_{\phi=0} \, ,
\label{drude}
\end{equation}
where $\phi $ is defined for charge $\zeta =\rho$ and
spin $\zeta =\sigma_z$ in Eq. (\ref{phis}),
are such that $2\pi D^{\zeta}=\sum_{\alpha}q_{F\alpha,0}
/m^{\zeta}_{\alpha,0}$ \cite{Carm4}. For charge,
$m^\rho_{s,0}=\infty$, and the latter expression reads  
$2\pi D^{\rho}=q_{Fc,0}/m^{\rho}_{c,0}$ and is such that
$D^{\rho}\rightarrow 0$ as $n\rightarrow 1$, satisfying Kohn 
criterion \cite{Kohn}. These quantities can be computed 
within the pseudoparticle formalism by direct evaluation 
of Eq.\,(\ref{drude}). They can also be obtained by combining a 
pseudoparticle Boltzmann transport description with linear response 
theory, as in Ref. \cite{Carm4}. In order to confirm the validity and
correctness of our formalism, in Appendix \ref{stiffness} we have 
recovered the charge and spin stiffness expressions $(135)$ - $(137)$ of 
Ref. \cite{Carm4} by direct use of Eq. (\ref{drude}). 

Equations (\ref{coupling}) and (\ref{coupling2}) show 
that the Hubbard-chain charge carriers are the $c,\gamma$ 
pseudoparticles. In contrast to the zero-temperature limit 
where the $c,0$ pseudoparticles fully determine the charge
stiffness, we expect that the $c,\gamma$ heavy pseudoparticles 
play an important role in the charge-transport properties at finite 
temperatures \cite{Castella,Zotos}. Moreover, elsewhere it 
will be shown that the limiting behavior of the 
$s,\gamma$ and $c,\gamma$ heavy-pseudoparticle bands as 
$H\rightarrow 0$ and $n\rightarrow 1$, respectively, will
have effects on the charge- and spin-transport properties at
finite temperatures. In order to obtain some information on 
that behavior, it is useful to consider limiting values for 
the quantities whose general expressions we have introduced in 
previous sections. 

In Appendix \ref{zeroH} we present simpler equations to define 
the pseudoparticle bands and phase shifts in the limit of zero 
magnetic field. These results show that for $H\rightarrow 0$ and
$\gamma >0$ the bands $\epsilon_{s,\gamma}^0(q)$ collapse to a point. 
This is because both the bandwidth [see Eq. (\ref{esgh0})] 
and the momentum pseudo-Brillouin zone width [see Eq. 
(\ref{gsbz})] go to zero as $H\rightarrow 0$. This behavior is
also present in the Heisenberg chain and, therefore, in that 
model the triplet and singlet exitations are degenerated at 
zero magnetic field \cite{Faddeev81}. This also holds true for 
the Hubbard chain at $H=0$ and in the limit $U\gg t$, where the BA wave 
function factorizes in a spinless-fermion Slater determinant 
and in the BA wave function for the 1D anti-ferromagnetic 
Heisenberg chain. On the other hand, in the limit $n\rightarrow 1$ 
the bands $\epsilon_{c,\gamma}^0(q)$ (for $\gamma >0$) collapse to a 
point also because
both the bandwidth and the momentum pseudo-Brillouin zone width
[see Eq. (\ref{gsbz})] go to zero in that limit. 

\section{Kinetic Equations for the Pseudoparticles}
\label{kinetic}

In the previous sections the quantum-liquid physics for energies 
just above the $\omega_0$ values, Eq. (\ref{gap}), was described 
in terms of homogeneous pseudoparticle distributions.  The 
pseudoparticles experience only zero-momentum forward-scattering 
interactions at all energy scales. This property is 
absent in Fermi-liquid theory where it holds true only
at low excitation energy when the quasiparticles are well
defined quantum objects \cite{Pine,Baym,Landau}. This unconventional 
character of integrable models \cite{Shastry86} allows us to 
extend the use of the kinetic equations to energy scales just above 
the $\omega_0$ energy values, Eq. (\ref{gap}), and not only to 
low energies \cite{Carm4}. The results presented
in this section are a generalization of the kinetic-equation
low-energy study presented in Ref. \cite{Carm4}. 

In the final Hilbert subspace of energy $\omega $ relative to 
the initial ground state the Hubbard model can be mapped onto a 
continuum field theory of small energy $(\omega -\omega_0)$
\cite{GCFT}. The time coordinate $t$ of such theory is the 
Fourier transform of the small energy $(\omega-\omega_0)$ which 
corresponds to a finite energy $\omega $ in the original Hubbard model.
The validity of this approach is confirmed by the fact that it fully
reproduces the rigorous results of Section \ref{baphi}. 

Let us consider excitations described by space and time dependent
pseudoparticle distribution functions, $N_{\alpha,\gamma}(q,x,t)$,
given by

\begin{equation}
N_{\alpha,\gamma}(q,x,t)=N_{\alpha,\gamma}^0(q)+
\delta N_{\alpha,\gamma}(q,x,t)\, ,
\label{nqxt}
\end{equation}
where $N_{\alpha,\gamma}^0(q)$ is the ground-state distribution.
It follows from the PPT introduced in Ref. \cite{CarmeloNuno97} 
and discussed in the previous sections that the
single-pseudoparticle local energy is given, to first order in the 
deviations $\delta N_{\alpha,\gamma}(q,x,t)$, by

\begin{equation}
\check {\varepsilon}_{\alpha,\gamma} (q,x,t)=
\epsilon_{\alpha,\gamma} (q)+ \frac 1{2 \pi}
\sum_{\alpha',\gamma'}
\int_{-q_{\alpha',\gamma'}}^{q_{\alpha',\gamma'}}dq' 
\delta N_{\alpha',\gamma'}(q',x,t)
f_{\alpha,\gamma;\alpha',\gamma'}(q,q')\, .
\label{classicale}
\end{equation}

Let ${\cal A}^{\zeta}$ represent the total charge, $\zeta=\rho$, 
or spin, $\zeta=\sigma_z$. It follows from the relations 
(\ref{nbaixo}) and (\ref{nc}) involving the pseudoparticle and 
electron numbers that ${\cal A}^{\zeta}$ depends linearly on the 
pseudoparticle deviation numbers. Thus, in the case of 
inhomogeneous excitations described by Eq.\,(\ref{nqxt}) the 
corresponding expectation value at point $x$ and time $t$, 
$\langle {\cal A}^{\zeta} (x,t)\rangle$, can be written as

\begin{equation}
\langle {\cal A}^{\zeta}(x,t) \rangle=
\langle {\cal A}^{\zeta} \rangle_0+
\frac {N_a}{2 \pi}
\sum_{\alpha',\gamma'}
\int_{-q_{\alpha',\gamma'}}^{q_{\alpha',\gamma'}}dq' 
\delta N_{\alpha',\gamma'}(q',x,t)
{\cal C}_{\alpha',\gamma'}^\zeta \times a^\zeta\, ,
\label{conserved}
\end{equation}
where $a^\rho=-e$ and $a^{\sigma_z}=1/2$. 

In this``semi-classical'' approach the response to a scalar field, 
$V^{\zeta}(x,t)$, is proportional to the conserved quantity 
${\cal A}^{\zeta}$. As for low energy \cite{Carm4}, in the presence 
of the inhomogeneous potential the force 
${\cal F}^{\zeta}(x,t)_{\alpha,\gamma}$ that acts upon 
the $\alpha,\gamma$ pseudoparticle is given by 
${\cal F}^{\zeta}_{\alpha,\gamma}(x,t)=
-[\partial V^{\zeta}(x,t)/ \partial x]
{\cal C}_{\alpha,\gamma}^\zeta \times a^\zeta$. It
follows that the deviations $\delta N_{\alpha,\gamma}(q,x,t)$ are 
determined by the solution of a system of kinetic equations 
(one equation for each occupied $\alpha,\gamma $
branch) which reads

\begin{eqnarray}
0&=&\frac{\partial N_{\alpha,\gamma}(q,x,t)}{\partial t}+
\frac{\partial N_{\alpha,\gamma}(q,x,t)}{\partial x}
\frac{\partial \check {\varepsilon}_{\alpha,\gamma} (q,x,t)}
{\partial q}-\frac{\partial N_{\alpha,\gamma}(q,x,t)}{\partial q}
\frac{\partial \check {\varepsilon}_{\alpha,\gamma} (q,x,t)}
{\partial x} \nonumber\\
&-&\frac{\partial N_{\alpha,\gamma}(q,x,t)}{\partial q}
\frac {\partial V^{\zeta}(x,t)}{\partial x}
{\cal C}_{\alpha,\gamma}^\zeta \times a^\zeta \, .
\label{kin}
\end{eqnarray}

Introducing Eq.\,(\ref{nqxt}) in Eq.\,(\ref{kin}), expanding
to first order in the deviations $\delta N_{\alpha,\gamma}(q,x,t)$, and
using Eq.\,(\ref{classicale}) we obtain the following set of 
linearized kinetic equations

\begin{eqnarray}
0&=&\frac{\partial \delta N_{\alpha,\gamma}(q,x,t)}{\partial t}
+v_{\alpha,\gamma}(q) 
\frac{\partial \delta N_{\alpha,\gamma}(q,x,t)}{\partial x}
\nonumber\\
&-&\frac{\partial \delta N_{\alpha,\gamma}(q,x,t)}{\partial q}
\left \{\frac {\partial V^{\zeta}(x,t)}{\partial x}
{\cal C}_{\alpha,\gamma}^\zeta \times a^\zeta 
\right . \nonumber\\
&+&\left . \sum_{\alpha',\gamma'} \frac 1{2\pi}
\int_{-q_{\alpha',\gamma'}}^{q_{\alpha',\gamma'}} dq'
\frac{\partial \delta N_{\alpha',\gamma'}(q',x,t)}{\partial x}
f_{\alpha,\gamma;\alpha',\gamma'}(q,q')
\right \}\, .
\label{linear}
\end{eqnarray}

The conservation law for $\langle {\cal A}^{\zeta}(x,t) \rangle$ 
leads in one dimension to

\begin{equation}
\frac{\partial \langle {\cal A}^{\zeta}(x,t) \rangle}
{\partial t}+\frac{\langle {\cal J}^{\zeta}(x,t) \rangle}
{\partial x}=0\, ,
\label{conservation}
\end{equation}
where $\langle {\cal A}^{\zeta}(x,t) \rangle$ is given by 
Eq.\,(\ref{conserved}) and $\langle {\cal J}^{\zeta}(x,t) 
\rangle$ is the associate current. Multiplying Eq.\,(\ref{linear}) 
by ${\cal C}_{\alpha,\gamma}^\zeta \times a^\zeta$, summing 
over $\alpha$ and $\gamma$, and integrating over $q$ we find for
$V^{\zeta}(x,t)=0$ and by comparing
the result with Eq.\,(\ref{conservation}) that the current 
spectrum $j^\zeta_{\alpha,\gamma}(q)$ is given by $a^\zeta$ times 
expression\,(\ref{jq}). (This expression has been derived from the 
solution of the BA equations with $a^\zeta=1$.) 

This agreement confirms the validity of the above 
low-$(\omega-\omega_0)$ continuum-field theory.
The unusual spectral properties associated with 
the zero-momentum forward-scattering character of the pseudoparticle
interactions follow from the integrability of the Hubbard
chain \cite{CarmeloNuno97,Shastry86}. 

\section{Concluding Remarks}
\label{remarks}

In this paper we have generalized the finite-energy PPT 
\cite{CarmeloNuno97} to the case of the Hubbard chain with 
a spin dependent Peierls substitution. This has allowed
the evaluation of the charge and spin currents in terms of the 
elementary currents of the charge and spin carriers. 
We have shown that at all energy scales these carriers are the 
$\alpha,\gamma$ pseudoparticles of the PPT. We have evaluated 
their couplings to charge and spin and introduced the associate 
charge and spin transport masses. Our results are also a
generalization for finite energies of the low-energy studies of 
Ref. \cite{Carm4}, our charge and spin current expressions recovering 
the expressions already obtained in that reference in the limit 
of low energy.

The obtained heavy-pseudoparticle transport masses are important quantities. 
They are believed to control the charge and spin stiffnesses at 
finite temperatures. Moreover, the pseudoparticle couplings 
to charge and spin obtained in the present paper provide important 
selection rules concerning the ground-state transitions which 
contribute to the finite-energy charge -- charge and 
spin -- spin correlation functions.
Expressions for these functions can be derived by combining the BA 
solution with a low-energy $(\omega-\omega_0)$ generalized 
conformal-field theory \cite{GCFT}. The above
selection rules will be shown elsewhere to provide 
important information on the
finite-energy correlation functions which
cannot be extracted from conformal-field theory alone.   

Finally, the $\phi_{\sigma}$-dependent charge and spin 
current expressions of general form (\ref{jmean}) will
be used elsewhere to find out whether the half-filling 
Hubbard model is or is not an insulator at all temperatures 
\cite{Castella,Zotos}. 

\section*{ACKNOWLEDGMENTS}

We thank David K. Campbell, Carlos Lobo, and Ivo Souza, from
the University of Illinois at Urbana Champaign, for stimulating 
discussions. N.M.R.P. has been partially supported by PRODEP 
$5.2$, REF. N.193.007.


\appendix
\section{NORMAL-ORDERED SOLUTION OF THE BA EQUATIONS 
WITH THE FLUX $\phi$}
\label{normalOrder}

In this appendix we derive the normal-ordered BA equations required
for the evaluation of Eqs.\,(\ref{deltaj}) and (\ref{jq}). Writing 
$W^{\phi}(q)$ from Eq.\,(\ref{exp2}) as

\begin{equation}
W^{\phi}(q)=\frac{dW(q)}{dq}L^{\phi}(q)\, ,
\end{equation}
where $L$ equals $L_{c,0}$, $L_{c,\gamma}$, or $L_{s,\gamma}$ 
when $W$ equals $K$, $R_{s,\gamma}$, or $R_{c,\gamma}$, respectively, 
we find that $W^{1,\phi}(q)$ obeys the following equality

\begin{equation}
W^{1,\phi}(q)=\frac {dW^0}{dq}L^{1,\phi}(q)+
\frac {dW^1}{dq}L^{0,\phi}(q) \, .
\end{equation}
Introducing the above equation in Eq.\,(\ref{jmean}) 
and writing the distributions functions $N_{\alpha,\gamma}(q)$ 
as $N_{\alpha,\gamma}^0(q)+\delta N_{\alpha,\gamma}(q)$, 
we can expand $J\equiv\langle m\vert \, 
\hat{j}^\zeta\vert m\rangle$ in terms of the pseudomomentum
deviations as

\begin{equation}
J = J^0 + J^1 + J^2 ... \, ,
\label{Jexp}
\end{equation}
where the first term, $J^1$, of the current normal-ordered 
expansion (\ref{Jexp}) can after some algebra be written as

\begin{eqnarray}
J^1&=& J^1_0
-2t\sum_{j=\pm 1} {L^{0,\phi}_{c,0}(jq_{Fc})L_{c,0}^1(jq_{Fc})
\over 2\pi\rho_{c,0}(Q)} \sin (Q)
+\nonumber\\
&+&\sum_{\gamma >0} \theta(N_{c,\gamma})Re\, 4t \sum_{j=\pm 1}
\frac{u^2[jr_{c,\gamma} -i\gamma]}
{\sqrt{1-u^2[jr_{c,\gamma} -i\gamma]^2}} 
{L_{c,\gamma}^{0,\phi}(jq_{Fc,\gamma}) 
L^1_{c,\gamma}(jq_{Fc,\gamma})\over 2\pi\rho_{c,\gamma}(r_{c,\gamma})}
\nonumber\\
&-&2t\int_{-q_{Fc}}^{q_{Fc}} dq 
\frac{dK^{(0)}(q)}{dq}\sin(K^{(0)}(q))
{\cal L}^{1,\phi}_{c,0}(q) \, ,  
\label{jnorm}
\end{eqnarray}
where the functions $2\pi\rho_{c,0}(k)$ and $2\pi\rho_{\alpha,\gamma}(r)$
were defined in Ref. \cite{CarmeloNuno97} and

\begin{eqnarray}
J^1_0&=&-2t\int_{-q_{c}}^{q_{c}} dq \delta N_c(q)
\frac{dK^{(0)}(q)}{dq}\sin(K^{(0)}(q)) L^{0,\phi}_{c,0}(q) \nonumber\\
&+& \sum_{\gamma >0} Re\, 4t\int_{-q_{c,\gamma}}
^{q_{c,\gamma}} dq \delta N_{c,\gamma}(q)
\frac{u^2[R^{(0)}_{c,\gamma}(q) -i\gamma]}
{\sqrt{1-u^2[R^{(0)}_{c,\gamma}(q) -i\gamma]^2}}
\frac{dR^{(0)}_{c,\gamma}(q)}{dq}
L^{0,\phi}_{c,\gamma}(q) \, .
\end{eqnarray}
The function ${\cal L}^{1,\phi}(q)$ is defined as

\begin{equation}
{\cal L}^{1,\phi}(q)=L^{1,\phi}(q)-W^1(q)
\frac{L^{0,\phi}(q)}{dq}\,.
\end{equation} 

In order to obtain the integral equations for $L^{0,\phi}(q)$ and 
${\cal L}^{1,\phi}(q)$ (with ${\cal L}={\cal L}_{c,0}, 
{\cal L}_{c,\gamma}$, and ${\cal L}_{s,\gamma}$), we start from 
the continuum limit of Eqs. (\ref{tak1}), (\ref{tak2}), and 
(\ref{tak3}) which reads

\begin{eqnarray}
K(q) &=& q
+\phi_\uparrow/N_a-\sum_{\gamma'}\frac 1{2\pi}
\int_{-q_{s,\gamma'}}^{q_{s,\gamma'}}
dq\,'N_{s,\gamma'}(q\,')\, 2\tan^{-1}\left(\frac{\sin(K(q))/u-
R_{s,\gamma'}(q\,')}
{(\gamma'+1)} \right)\nonumber\\
&-&\sum_{\gamma' >0} \frac 1{2\pi}
\int_{-q_{c,\gamma'}}^{q_{c,\gamma'}}
dq\,'N_{c,\gamma'}(q\,')\, 2\tan^{-1}\left(\frac{\sin(K(q))/u-
R_{c,\gamma'}(q\,')}
{\gamma'}\right)\, ,
\label{int1}
\end{eqnarray}

\begin{eqnarray}
2 Re \, \sin^{-1}[(R_{c,\gamma}(q)-i\gamma)u] &=& q
+ \gamma (\phi_\uparrow+\phi_\downarrow)/N_a-\nonumber\\
&-&\frac 1{2\pi} \int_{-q_c}^{q_c}
dq\,'N_c(q\,')\, 2\tan^{-1}\left(\frac{\sin(K(q\,'))/u-R_{c,\gamma}(q)}
{\gamma}\right)\nonumber\\
&+&\sum_{\gamma'>0} \frac 1{2\pi}
\int_{-q_{c,\gamma'}}^{q_{c,\gamma'}}
dq\,'N_{c,\gamma'}(q\,')\Theta_{\gamma,\gamma'}
(R_{c,\gamma}(q)-R_{c,\gamma'}(q\,')\, ,
\label{int2}
\end{eqnarray}
and

\begin{eqnarray}
q&=& (\gamma+1)(\phi_\uparrow-\phi_\downarrow)/N_a+
\frac 1{2\pi} \int_{-q_c}^{q_c} dq\,'N_c(q\,')\,
2\tan^{-1}\left(\frac{R_{s,\gamma}(q)-\sin(K(q\,'))/u}
{(\gamma+1)}\right)\nonumber\\
&-&
\sum_{\gamma'}\frac
1{2\pi}\int_{-q_{s,\gamma}}^{q_{s,\gamma}}
dq\,'N_{s,\gamma'}(q\,')\Theta_{\gamma+1,\gamma'+1}
(R_{s,\gamma} (q)-R_{s,\gamma'}(q\,'))\, .
\label{int3}
\end{eqnarray}

It is convenient to write the function 
$\Theta^{[1]}_{\gamma,\gamma'}(x)$,
defined by Eq. (B7) of Ref. \cite{CarmeloNuno97}, as follows

\begin{equation}
\Theta^{[1]}_{\gamma,\gamma'}(x) =
\sum_{l}{2b_{l}^{\gamma ,\gamma'}\over 1 +
[x/l]^2} \, .
\label{T1}
\end{equation} 
We emphasize that comparision term by term of expression
(B7) of Ref. \cite{CarmeloNuno97} with expression 
(\ref{T1}) fully defines the coefficients
$b_{l}^{\gamma ,\gamma'}$ and the corresponding 
set of integer numbers $l$.  

Following equation (\ref{phis}), we have 
$\phi_\uparrow=\phi_\downarrow$ for a charge-probe
current and $\phi_\uparrow=-\phi_\downarrow$ for a spin probe. 
With the above equations written in terms of
$\phi_\uparrow$ and $\phi_\downarrow$, Eq.\,(\ref{jnorm})
provides both the charge and spin currents. In what follows,
we introduce in the functions $L^{\phi}(q)$ the index $\zeta=
\rho,\sigma_z$
to label the equations for either the charge or the spin current,
respectively. We start by expanding 
Eqs.\,(\ref{int1}), (\ref{int2}), and (\ref{int3}) up
to first order in $\phi$. This procedure reveals that the functions 
$L^{\phi,\zeta}(q)$ obey the following integral equations

\begin{eqnarray}
L^{\phi,\zeta}_{c,0} (q) &=& {\cal C}^\zeta_{c,0}
+\sum_{\gamma'}\frac 1{(\gamma'+1)\pi }
\int_{-q_{s,\gamma'}}^{q_{s,\gamma'}}
dq\,'{N_{s,\gamma'}(q\,')\over {1 + 
[\frac {\sin(K(q))/u-R_{s,\gamma'}(q\,')}{\gamma' +1}]^2}}
\frac{dR_{s,\gamma'}(q\,')}{dq\,'}L_{s,\gamma'}^{\phi,\zeta}(q\,')
\nonumber\\
&+&\sum_{\gamma' >0}\frac 1{\pi \gamma'}
\int_{-q_{c,\gamma'}}^{q_{c,\gamma'}}
dq\,'{N_{c,\gamma'}(q\,')\over {1 +  
[\frac {\sin(K(q))/u-R_{c,\gamma'}(q\,')}{ \gamma'}]^2}}
\frac{dR_{c,\gamma'}(q\,')}{dq\,'}
L_{c,\gamma'}^{\phi(q'),\zeta}(q\,') \, ,
\label{lint1}
\end{eqnarray}

\begin{eqnarray}
L^{\phi,\zeta}_{c,\gamma} (q) &=& {\cal C}^\zeta_{c,\gamma}+
\frac 1{\pi u \gamma}\int_{-q_{c}}^{q_{c}}
dq\,'{N_c(q\,')\over {1 + 
[\frac{\sin(K(q\,'))/u-R_{c,\gamma}(q)}{ \gamma}]^2}}
\frac{dK(q\,')}{dq\,'}\cos(K(q\,'))L_{c,0}^{\phi,\zeta}(q\,')\nonumber\\
&+&\sum_{\gamma'>0}\sum_l\frac 1{\pi l}
\int_{-q_{c,\gamma'}}^{q_{c,\gamma'}}
dq\,'{N_{c,\gamma'}(q\,') b^{\gamma ,\gamma'}_l\over {1 +
[\frac{R_{c,\gamma}(q)-R_{c,\gamma'}(q\,')}{l}]^2}}
\frac{dR_{c,\gamma'}(q\,')}{dq\,'}L_{c,\gamma'}^{\phi,\zeta}(q')\, ,
\label{lint2}
\end{eqnarray}
and

\begin{eqnarray}
L^{\phi}_{s,\gamma} (q) &=& {\cal C}^\zeta_{s,\gamma}+\nonumber\\
&+&\frac 1{u(\gamma+1) \pi}
\int_{-q_{c}}^{q_{c}}dq\,'{N_c(q\,')\over {1 + 
[\frac {\sin(K(q\,'))/u-R_{s,\gamma}(q)}{
\gamma+1}]^2}} \frac{dK(q\,')}{dq\,'}\cos(K(q\,'))
L_{c,0}^{\phi,\zeta}(q\,')\nonumber\\
&-&\sum_{\gamma'}\sum_l\frac 1{\pi l}
\int_{-q_{s,\gamma'}}^{q_{s,\gamma'}}
dq\,'{N_{s,\gamma'}(q\,') b^{\gamma +1,\gamma' +1}_l\over {1 +
[\frac{R_{s,\gamma}(q)-R_{s,\gamma'}(q\,')}{l}]^2}}
\frac{dR_{s,\gamma'}(q\,')}{dq\,'}L_{s,\gamma'}^{\phi,\zeta}(q') \, , 
\label{lint3}
\end{eqnarray}
where the coupling constants ${\cal C}^{\zeta}_{\alpha,\gamma}$
are defined by Eqs. (\ref{coupling}) and (\ref{coupling2}).
We again write the distributions functions $N_{\alpha,\gamma}(q)$ of
Eqs.\,(\ref{lint1}), (\ref{lint2}), and (\ref{lint3}) as 
$N_{\alpha,\gamma}^0(q)+\delta N_{\alpha,\gamma}(q)$. This allows 
us to obtain integral equations
for $L^{0,\phi,\zeta}(q)$ and ${\cal L}^{1,\phi}(q)$
(we remark that the functions ${\cal L}^{1,\phi}(q)$
are the same both for $\zeta=\rho,\sigma_z$). It is then
straighforward to find the integral equations obeyed 
by $L^{0,\phi,\zeta}(q)$ and show that $L^{0,\phi,\zeta}(q)$  
can be simply expressed in terms of linear combinations of 
phase shifts. The final result is

\begin{equation}
L_{\alpha,\gamma}^{0,\phi,\zeta}(q) =
{\cal C}^\zeta_{\alpha,\gamma}+
\sum_{\alpha',\gamma'}\sum_{j=\pm 1}j\theta(N_{\alpha',\gamma'}) 
{\cal C}^\zeta_{\alpha',\gamma'}
\Phi_{\alpha,\gamma;\alpha',\gamma'}(q,jq_{F\alpha',\gamma'})\, .
\label{l0zeta}
\end{equation}
The integral equations obeyed by ${\cal L}^{1,\phi}(q)$ are related
to the integral equations obeyed by $\tilde{{\cal L}}^{1,\phi}(r)$, 
where $r$ equals $\sin(K^{(0)}(q))/u$, $R^{(0)}_{c,\gamma}(q)$, and 
$R^{(0)}_{s,\gamma }(q)$ for ${\cal L}={\cal L}_{c,0}$, 
${\cal L}_{c,\gamma}$, and ${\cal L}_{s,\gamma}$, respectively. 
The functions $\tilde{{\cal L}}^{1,\phi}(r)$ obey the 
following integral equations

\begin{equation}
\tilde{{\cal L}}_{c,0}^{1,\phi}(r)=
\tilde{{\cal L}}_{c,0}^{1,\phi,0}(r)+
\frac 1{\pi} \int_{-r_{s,0}}^{r_{s,0}} dr'
\frac{ \tilde{{\cal L}}_{s,0}^{1,\phi}(r')}
{1+(r-r')^2}\, ,        
\label{qtilde}
\end{equation}

\begin{equation}
\tilde{{\cal L}}^{1,\phi}_{c,\gamma}(r)=
\tilde{{\cal L}}^{1,\phi,0}_{c,\gamma}(r)-
\frac 1{\pi\gamma u} 
\int_{-r_{c}}^{r_{c}} dr'
\frac{ \tilde{{\cal L}}_{c,0}^{1,\phi}(r')}
{1+(\frac{r-r'}{\gamma})^2} \, ,        
\label{qgtilde}
\end{equation}
and

\begin{eqnarray}
\tilde{{\cal L}}^{1,\phi}_{s,\gamma} (r)&=&
\tilde{{\cal L}}^{1,\phi,0}_{s,\gamma}(r)-
\frac 1{\pi(\gamma+1)u} \int_{-r_{c}}^{r_{c}} dr'
\frac{ \tilde{{\cal L}}_{c,0}^{1,\phi}(r')}
{1+(\frac{r-r'}{\gamma})^2} \nonumber\\
&-&\sum_l \frac 1{\pi l} \int_{-r_{s,0}}^{r_{s,0}} 
dr'{b^{\gamma +1,1}_l\tilde{{\cal L}}_{s,0}^{1,\phi}(r')\over {1 +
[\frac{r-r'}{l}]^2}} \, ,       
\label{pgtilde}
\end{eqnarray}
where the free terms $\tilde{{\cal L}}_{c,0}^{1,\phi,0}(r)$,
$\tilde{{\cal L}}^{1,\phi,0}_{c,\gamma}(r)$, and 
$\tilde{{\cal L}}^{1,\phi,0}_{s,\gamma}(r)$ are, respectively, 
given by

\begin{eqnarray}
\tilde{{\cal L}}_{c,0}^{1,\phi,0}(r)&=&
\sum_{\gamma'}\frac1{\pi \gamma'} \int_{-q_{c,\gamma'}}^{q_{c,\gamma'}}
dq' \delta N_{c,\gamma'}(q'){L_{c,\gamma'}^{0,\phi}(q')\over 
{1+[\frac{r-R^{(0)}_{c,\gamma'}(q')}{\gamma'}]^2}}
\frac{dR^{(0)}_{c,\gamma'}(q')}{dq'}\nonumber\\ 
&+&\sum_{\gamma'}\frac1{\pi (\gamma'+1)} 
\int_{-q_{s,\gamma'}}^{q_{s,\gamma'}}
dq' \delta N_{s,\gamma'}(q')\frac{L_{s,\gamma'}^{0,\phi}(q')}
{(1+[\frac{r-R^{(0)}_{s,\gamma' }(q')}{\gamma'+1}]^2)}
\frac{dR^{(0)}_{s,\gamma' }(q')}{dq'}\nonumber\\
&+& \sum_{\gamma'} \theta (N_{c,\gamma'})\frac 1{\gamma' \pi}
\sum_{j=\pm 1}{jL^1_{c,\gamma'}(jq_{Fc,\gamma'})\over
{2\pi\rho_{c,\gamma'}(r_{c,\gamma'})}}
\frac {L_{c,\gamma'}^{0,\phi}(jq_{Fc,\gamma'})}
{(1+[\frac{r-jr_{c,\gamma'}}{\gamma'}]^2)}\nonumber\\
&+&\sum_{\gamma'} \theta (N_{s,\gamma'})\frac 1{(\gamma'+1) \pi}
\sum_{j=\pm 1}{jL^1_{s,\gamma'}(jq_{Fs,\gamma'})\over {
2\pi\rho_{s,\gamma'}(r_{s,\gamma'})}}
\frac {L_{s,\gamma'}^{0,\phi}(jq_{Fs,\gamma'})}
{(1+[\frac{r-jr_{s,\gamma' }}{\gamma'+1}]^2)}\, ,
\end{eqnarray}

\begin{eqnarray}
\tilde{{\cal L}}^{1,\phi,0}_{c,\gamma}(r)&=&
-\frac1{\pi u \gamma} \int_{-q_c}^{q_c}
dq' \delta N_c(q')
\frac{L_{c,0}^{0,\phi}(q')}
{1+[\frac{\sin(K^{(0)}(q'))/u-r}{\gamma}]^2}\cos(K^{(0)}(q'))
\frac{dK^{(0)}(q')}{dq'}\nonumber\\ 
&-&\sum_{\gamma'} \sum_l \frac1{\pi l} 
\int_{-q_{c,\gamma'}}^{q_{c,\gamma'}}
dq' \delta N_{c,\gamma'}(q'){b^{\gamma ,\gamma'}_l
L_{c,\gamma'}^{0,\phi}(q')\over {1 +
[\frac{r-R^{(0)}_{c,\gamma'}(q')}{l}]^2}}
\frac{dR^{(0)}_{c,\gamma'}(q')}{dq'}
\nonumber\\
&-& \frac1{\gamma u \pi}
\sum_{j=\pm 1}{jL^1_{c,0}(jq_{Fc})\over
{2\pi\rho_{c,0}(Q)}}
\frac{L_{c,0}^{0,\phi}(jq_{Fc})}
{(1+[\frac{r-jr_{c,0}}{\gamma}]^2)}\cos (Q)\nonumber\\
&-&\sum_{\gamma'} \theta (N_{c,\gamma'})\sum_l \frac1{\pi l} 
\sum_{j=\pm 1}{j b^{\gamma ,\gamma'}_lL^1_{c,\gamma'}(jq_{Fc,\gamma'})\over
{2\pi\rho_{c,\gamma'}(r_{c,\gamma'})}}
{L_{c,\gamma'}^{0,\phi}(jq_{Fc,\gamma'})
\over {(1 + [\frac{r-jr_{c,\gamma'}}{l}]^2)}} \, ,
\end{eqnarray}
and

\begin{eqnarray}
\tilde{{\cal L}}^{1,\phi,0}_{s,\gamma}(r)&=&
\frac1{\pi u (\gamma +1)} \int_{-q_c}^{q_c}
dq' \delta N_c(q')
\frac{L_{c,0}^{0,\phi}(q')}
{1+[\frac{\sin(K^{(0)}(q'))/u-r}{\gamma +1}]^2}\cos(K^{(0)}(q'))
\frac{dK^{(0)}(q')}{dq'}\nonumber\\
&-&\sum_{\gamma'} \sum_l \frac1{\pi l}
\int_{-q_{s,\gamma'}}^{q_{s,\gamma'}}
dq' \delta N_{s,\gamma'}(q'){b^{\gamma +1 ,\gamma' +1}_l
L_{s,\gamma'}^{0,\phi}(q')\over {1 +
[\frac{r-R^{(0)}_{s,\gamma'}(q')}{l}]^2}}
\frac{dR^{(0)}_{s,\gamma'}(q')}{dq'}
\nonumber\\
&+& \frac1{(\gamma +1) u \pi}
\sum_{j=\pm 1}{jL^1_{c,0}(jq_{Fc})\over
{2\pi\rho_{c,0}(Q)}}
\frac{L_{c,0}^{0,\phi}(jq_{Fc})}
{(1+[\frac{r-jr_{c,0}}{\gamma +1}]^2)}\cos (Q)\nonumber\\
&-&\sum_{\gamma'} \theta (N_{s,\gamma'})\sum_l \frac1{\pi l}
\sum_{j=\pm 1}{j b^{\gamma +1 ,\gamma' +1}_l
L^1_{s,\gamma'}(jq_{Fs,\gamma'})\over
{2\pi\rho_{s,\gamma'}(r_{s,\gamma'})}}
{L_{s,\gamma'}^{0,\phi}(jq_{Fs,\gamma'})
\over {(1 + [\frac{r-jr_{s,\gamma'}}{l}]^2)}} \, .
\end{eqnarray}
Introducing the functions ${\cal L}^{1,\phi}(q)$ obtained 
from Eqs.\,(\ref{qtilde}), (\ref{qgtilde}), and (\ref{pgtilde}), in 
Eq.\,(\ref{jnorm}) and
keeping terms only up to second order in the density of 
heavy pseudoparticles, we obtain Eq.\,(\ref{jmean}) with 
$j_{\alpha,\gamma}^\zeta(q)$ given by

\begin{equation}
j_{\alpha,\gamma}^\zeta(q)=v_{\alpha,\gamma}(q)
L^{0,\phi,\zeta}_{\alpha,\gamma}(q)+\sum_{\alpha',\gamma'}
\sum_{j=\pm 1}j \theta (N_{\alpha',\gamma'})
v_{\alpha',\gamma'}
L^{0,\phi,\zeta}_{\alpha',\gamma'}(jq_{F\alpha',\gamma'})
\Phi_{\alpha',\gamma';\alpha,\gamma}(jq_{F\alpha',\gamma'},q)\, .
\label{jq2}     
\label{jnorm2}
\end{equation}
Inserting  Eq.\,(\ref{l0zeta}) in Eq.\,(\ref{jq2}) we obtain
Eq.\,(\ref{jq}).

\section{Static Masses for the Heavy Pseudoparticles}
\label{static}

The static mass $m^{\ast}_{\alpha,\gamma}$ is defined in 
Ref. \cite{CarmeloNuno97} as

\begin{equation}
\frac 1 {m^{\ast}_{\alpha,\gamma}}= \left .
\frac{2t\,d\eta_{\alpha,\gamma}(r)/dr}
{(2\pi \rho_{\alpha,\gamma}(r))^2}\right \vert_{r=r^0}
-\left . \frac{2t\eta_{\alpha,\gamma}(r)
(2\pi\,d\rho_{\alpha,\gamma}(r)/dr)}
{(2\pi \rho_{\alpha,\gamma}(r))^3}\right \vert_{r=r^0} \, ,
\label{staticdef}
\end{equation}
where the functions $2t\eta_{\alpha,\gamma}(r)$ and 
$2\pi\rho_{\alpha,\gamma}(r)$ are defined in Ref.\, 
\cite{CarmeloNuno97} and $r^0$ is 
$W^0_{\alpha,\gamma}(q_{F\alpha,\gamma})$ which
represents $Q$, $r_{c,\gamma}$, and $r_{s,\gamma}$,

After some straightforward algebra, the general expressions 
(\ref{staticdef}) lead to the following simple expressions 
for $1/m^{\ast}_{\alpha,\gamma}$

\begin{equation}
\frac 1 {m^{\ast}_{c,\gamma}}=\frac 
{-4tu^2/(1+u^2\gamma^2)^{3/2}+\Lambda^{\eta}_{c,\gamma}}
{(2u/\sqrt{1+u^2\gamma^2}-\Lambda^{\rho}_{c,\gamma})^2}\,,
\hspace{1cm}\gamma>0 \,,
\label{1overmc}
\end{equation}
and

\begin{equation}
\frac 1 {m^{\ast}_{s,\gamma}}=\frac 
{\Lambda^{\eta}_{c,\gamma+1}-
\Lambda^{\eta}_{s,\gamma}-\Lambda^{\eta}_{s,\gamma+2}}
{(\Lambda^{\rho}_{c,\gamma+1}-\Lambda^{\rho}_{s,\gamma}-
\Lambda^{\rho}_{s,\gamma+2})^2}\,,\hspace{1cm}
\gamma>0 \,.
\label{1overms}
\end{equation}
In Eqs.\,(\ref{1overmc}) and (\ref{1overms}) the functions
$\Lambda^{\eta}_{\alpha,x}$ and
$\Lambda^{\rho}_{\alpha,x}$ read

\begin{equation}
\Lambda^{\eta}_{\alpha,x}=2\int_{-q_{F\alpha,0}}
^{q_{F\alpha,0}} \frac {dq}{\pi x^3}
\frac{v_{\alpha,0}(q)R^{(0)}_{\alpha,0}(q)}
{[1+(R^{(0)}_{\alpha,0}(q)/x)^2]^2} \, ,
\end{equation}
and 

\begin{equation}
\Lambda^{\rho}_{\alpha,x}=\int_{-q_{F\alpha,0}}
^{q_{F\alpha,0}} \frac {dq}{\pi x}
\frac{1}
{1+[R^0_{\alpha,0}(q)/x]^2} \, ,
\end{equation}
with

\begin{equation}
R^{(0)}_{c,0}(q)=\frac {\sin (K^{(0)}(q))}{u} \, .
\end{equation}

In the limit of fully polarized ferromagnetism, these expressions 
lead to the following closed-form expressions for the static masses

\begin{equation}
\frac 1{m_{c,\gamma}^{\ast}}=\frac {t\pi}
{8[\eta_{1,\gamma}]^2}\left ( -\frac {\pi
+2[\eta_{1,\gamma}]}{\sqrt {1+[u\gamma]^2}}
-\frac {u\gamma\sin(2n\pi)}{[u\gamma]^2+\sin^2(n\pi)}
\right)\, ,
\label{mcast}
\end{equation} 

\begin{equation}
\frac 1{m_{s,\gamma}^{\ast}}=\frac {t\pi}
{[\eta_{2,\gamma+1}]}\left ( \frac {1}{\sqrt {1+[u(\gamma+1)]^2}}
-\frac{u(\gamma+1)}{2[\eta_{2,\gamma+1}]}
\frac {\sin(2n\pi)}{[u(\gamma+1)]^2+\sin^2(n\pi)}
\right)\, ,
\label{msast}
\end{equation}
where

\begin{equation}
\eta_{1,x}=\tan^{-1}[\cot (n\pi)\frac {x u}
{\sqrt{1+u^2 x^2}}]\, ,
\end{equation}
and $\eta_{2,x}=\pi/2-\eta_{1,x}$. 
We remark that the static masses of the $c,\gamma$ pseudoparticles 
are, in general, negative. The static masses of the $\alpha,0$ 
pseudoparticles have been studied in Ref. \cite{Carm3}.

\section{Charge and Spin Stiffnesses at Zero Temperature}
\label{stiffness}

In this Appendix we show that the direct use of Eq. (\ref{drude})
leads to the stiffness expressions $(135)$ - $(137)$ of
Ref. \cite{Carm4}.

The calculation of the charge and spin stiffnesses (\ref{drude})
requires the expansion of Eq.\,(\ref{energy}) and of 
Eqs.\,(\ref{int1}), (\ref{int2}), and (\ref{int3}) up 
to second order in $\phi$. As in the case of the charge and spin 
current, both he charge and spin stiffnesses can be computed from 
Eq.\,(\ref{drude}), and we obtain one or the other depending on the 
coupling constants we choose in Eqs.\, (\ref{int1}), (\ref{int2}), 
and (\ref{int3}). Expanding the ground-state 
energy up to second order in $\phi$, we obtain

\begin{equation}
\left . \frac{d^2(E_0/N_a)}{d(\phi/N_a)^2}
\right \vert_{\phi=0}=
\frac 1{2\pi} \int_{-q_{Fc}}^{q_{Fc}} dq \left[
2t K^{0,\phi \phi}(q) \sin(K^{(0)}(q)) +
2t [K^{0,\phi}(q)]^2 \cos(K^{(0)}(q))\right]\, ,
\label{e2fi}
\end{equation}
where the function $K^{0,\phi \phi}(q)$ is the second 
derivative of the rapidity function defined by Eq. \,(\ref{int1}) 
in order to $\phi/N_a$ at $\phi=0$. The functions $K^{0,\phi}(q)$ 
and $K^{0,\phi \phi}(q)$ can be written as

\begin{equation}
K^{0,\phi}(q)=\frac {dK^{(0)}(q)}{dq} L_{c,0}^{0,\phi}(q)\, ,
\label{k0phi}
\end{equation}
and

\begin{equation}
K^{0,\phi \phi}(q)=\frac d{dq}
\left(\frac {dK^{(0)}(q)}{dq}[L_{c,0}^{0,\phi}(q)]^2\right)+
2\frac {dK^{(0)}(q)}{dq}L^{0,\phi \phi}_{c,0,\ast}\, ,
\label{k02phi}
\end{equation}
respectively. The use of Eqs.\,(\ref{k0phi}) and (\ref{k02phi}) in 
Eq.\,(\ref{e2fi}) leads then to

\begin{eqnarray}
\left . \frac{d^2(E_0/N_a)}{d(\phi/N_a)^2}
\right \vert_{\phi=0}&=&
\frac1{2\pi}\int_{-q_{Fc}}^{q_{Fc}} dq
2t\sin(K^{(0)}(q))             
2\frac{dK^{(0)}(q)}{dq}L^{0,\phi\phi}_{c,0,\ast}(q)
\nonumber\\
&+&\frac 1{2\pi}\sum_{j=\pm 1} {2t\sin(Q)
[L_{c,0}^{0,\phi}(jq_{Fc})]^2\over
{2\pi\rho_{c,0}(Q)}} \, ,
\label{e2phi}
\end{eqnarray}
where the function $L_{c,0}^{0,\phi}(jq_{Fc})$ is defined in Appendix 
\ref{normalOrder}. The function $L^{0,\phi\phi}_{c,0,\ast}(q)$ 
obeys the following integral equation 

\begin{eqnarray}
L^{0,\phi \phi}_{c,0,\ast} (q) &=& \frac 1{2\pi} 
\sum_{j=\pm 1} \frac {j[L_{s,0}^{0,\phi}(jq_{Fs,0})]^2}
{2\pi\rho_{s,0}(r_{s,0})(1+[\sin(K^{(0)}(q))/u-jr_{s,0}]^2)}\nonumber\\
&+&\frac 1{\pi}\int_{-q_{Fs,0}}^{q_{Fs,0}}dq' \frac 
{dR^{(0)}_{s,0}(q')}{dq'}
\frac {L^{0,\phi \phi}_{s,0,\ast} (q')}
{1+[\sin(K^{(0)}(q))/u-R^0_{s,0}(q')]^2} \, ,
\label{Qast}
\end{eqnarray}
which was obtained by performing the type of expansions developed 
in Appendix \ref{normalOrder}.
Moreover, $L^{0,\phi \phi}_{s,0,\ast} (q)$ is given by

\begin{eqnarray}
L^{0,\phi \phi}_{s,0,\ast} (q)&=&\frac 1{2 u \pi} \sum_{j=\pm 1}
{j[\cos(Q) L_{c,0}^{0,\phi}(jq_{Fc})]^2\over 
{2\pi\rho_{c,0}(Q)(1+[R^{(0)}_{s,0}(q)-jr_{c,0}]^2)}}\nonumber\\
&-&\frac 1{4\pi} \sum_{j=\pm 1}{j[L_{s,0}^{0,\phi}(jq_{Fs,0})]^2
\over {2\pi\rho_{s,0}(r_{s,0})
(1+[(R^{(0)}_{s,0}(q)-jr_{s,0})/2]^2)}}\nonumber\\
&-&\frac 1{2\pi}\int_{-q_{Fs,0}}^{q_{Fs,0}}dq' 
\frac {dR^{(0)}_{s,0}(q')}{dq'}
\frac {L^{0,\phi \phi}_{s,0,\ast} (q')}
{1+[(R^{(0)}_{s,0}(q)-R^{(0)}_{s,0}(q'))/2]^2}\nonumber\\
&+&\frac 1{\pi u}\int_{-q_{Fc}}^{q_{Fc}}dq' \frac {dK^{(0)}(q')}{dq'}
\frac {\cos (K^{(0)}(q'))L^{0,\phi \phi}_{c,0,\ast} (q')}
{1+(\sin(K^{(0)}(q'))/u-R^{(0)}_{s,0}(q))^2} \, .
\label{Past}
\end{eqnarray}
Introducing Eqs.\,(\ref{Qast}) and (\ref{Past}) in Eq.\,(\ref{e2phi}) 
we obtain, after some algebra, the following expression
for the (charge and spin) stiffness $D^\zeta$

\begin{equation}
4\pi D^\zeta = \sum_{j=\pm 1}v_{c,0}
[L_{c,0}^{0,\phi,\zeta}(jq_{Fc,0})]^2
+\sum_{j=\pm 1}v_{s,0}[L_{s,0}^{0,\phi,\zeta}(jq_{Fs,0})]^2\, , 
\label{dzeta}
\end{equation}
where the functions $L^{0,\phi,\zeta}(jq_{F\alpha,0})$ are 
defined by Eq.\,(\ref{l0zeta}). After some simple
algebra, expression\,(\ref{dzeta})
can be shown to be the same as expressions $(135)$ - $(137)$
of Ref. \cite{Carm4}.

\section{the zero magnetic-field case}
\label{zeroH}

For the case of zero magnetic field it is 
possible to cast the equations for the energy bands, phase shifts, 
and rapidities in a simpler form. After some algebra, the 
$\epsilon_{s,\gamma}(q)$ band (with $\gamma=0,1,2,\ldots,\infty$) 
and the $\epsilon_{c,\gamma}(q)$ band with 
($\gamma=1,2,\ldots,\infty$) can at zero magnetic field
be rewritten as

\begin{equation}
\epsilon_{s,\gamma}^0(q) =
-\delta_{\gamma ,0} \left[2t\int_0^{\infty}d \omega
{\cos(\omega R^{(0)}_{s,0}(q))\over \omega\cosh (\omega)}
\Upsilon_1(\omega)\right] \, ,
\label{esgh0}
\end{equation}

and
\begin{eqnarray}
\epsilon_{c,\gamma}^0(q)&=&
Re\,4t\sqrt{1-u^2(R_{c,\gamma}^{(0)}(q)+i\gamma)^2}-\nonumber\\
&-&4t\int_0^{\infty}d \omega
\frac{e^{-\gamma \omega}}{\omega}\cos(\omega R^{(0)}_{c,\gamma}(q))
\Upsilon_1(\omega) \, ,
\label{ecgh0}
\end{eqnarray}
where $\Upsilon_1(\omega)$ obeys the integral equation

\begin{eqnarray}
\Upsilon_1(\omega)=\Upsilon_1^0(\omega)+\int_{-\infty}^{\infty}
d \omega' \Upsilon_1(\omega') \Gamma(\omega',\omega) \, .
\label{up1}
\end{eqnarray}
Here the free term and the kernel read

\begin{equation}
\Upsilon_1^0(\omega)=\frac 1 {2 \pi} \int_{-Q}^{Q}dk 
\sin(k)\sin(\omega \sin(k)/u)\, ,
\label{up01}
\end{equation}
and

\begin{equation}
\Gamma(\omega',\omega)=\frac {\sin((\omega-\omega')r_{c,0})}
{\pi(\omega-\omega')(1+e^{2\vert \omega' \vert})}\, ,
\label{gama1}
\end{equation}
respectively. The kernel (\ref{gama1}) was already obtained
in Ref. \cite{Carm88} (see Eq. (A5) of that reference). 
Equation \,(\ref{esgh0}), together with the fact
that in the limit of zero magnetic field the width of the
$s,\gamma >0$ momentum pseudo-Brillouin zone vanishes
[see Eq. (\ref{gsbz})], shows that the $s,\gamma $ bands collapse for 
$\gamma>0$ and all values of $U$ and $n$ to the point 
zero. 

In this limit it is also possible to cast the integral
equations for the phase shifts, whose expressions are given in Ref.\,
\cite{CarmeloNuno97}, in the following alternative form

\begin{equation} 
\bar {\Phi}_{c,0;c,0}(r,r')=-B(r-r')+\int_{-r_{c,0}}^{r_{c,0}}dr''
\bar {\Phi}_{c,0;c,0}(r'',r')A(r-r'') \, ,
\label{c0c0}
\end{equation}

\begin{equation}
\bar {\Phi}_{c,0;c,\gamma'}(r,r')=-\frac 1{\pi}
\tan^{-1}(\frac {r-r'}{\gamma'})+\int_{-r_{c,0}}^{r_{c,0}}dr''
\bar {\Phi}_{c,0;c,\gamma'}(r'',r') A(r-r'') \, ,
\label{c0cg}
\end{equation}

\begin{eqnarray}
\bar {\Phi}_{c,0;s,\gamma'}(r,r')&=&-\delta_{0,\gamma'}
\frac 1{2\pi}\tan^{-1}[\sinh(\pi /2(r-r'))]
+\nonumber\\
&+&\int_{-r_{c,0}}^{r_{c,0}}\frac {dr''}{\gamma'+1}
\bar {\Phi}_{c,0;s,\gamma'}(r'',r') A(r-r'') \, ,
\label{c0sg}
\end{eqnarray}

\begin{equation}
\bar {\Phi}_{c,\gamma;c,0}(r,r')=\frac 1{\pi}
\tan^{-1}(\frac {r-r'}{\gamma})-\int_{-r_{c,0}}^{r_{c,0}}
\frac {dr''}{\pi \gamma}
\frac {\bar {\Phi}_{c,0;c,0}(r'',r')}
{1+(\frac {r-r''}{\gamma})^2} \, ,
\label{cgc0}
\end{equation}

\begin{equation}
\bar {\Phi}_{c,\gamma;c,\gamma'}(r,r')=
\frac 1{2\pi}
\Theta_{\gamma,\gamma'}(r-r')
-\int_{-r_{c,0}}^{r_{c,0}}
\frac {dr''}{\pi \gamma}
\frac {\bar {\Phi}_{c,0;c,\gamma'}(r'',r')}
{1+(\frac {r-r''}{\gamma})^2} \, ,
\label{cgcg}
\end{equation}

\begin{equation}
\bar {\Phi}_{c,\gamma;s,\gamma'}(r,r')=
-\int_{-r_{c,0}}^{r_{c,0}}
\frac {dr''}{\pi \gamma}
\frac {\bar {\Phi}_{c,0;s,\gamma'}(r'',r')}
{1+(\frac {r-r''}{\gamma})^2} \, ,
\label{cgsg}
\end{equation}

\begin{equation}
\bar {\Phi}_{s,\gamma;c,0}(r,r')=0 \, ,
\label{sgc0}
\end{equation}

\begin{equation}
\bar {\Phi}_{s,\gamma;c,\gamma'}(r,r')=
\delta_{0,\gamma}\frac 1{4}
\int_{-r_{c,0}}^{r_{c,0}}dx''
\frac {\bar {\Phi}_{c,0;c,\gamma'}(r'',r')}
{\cosh(\pi /2(r''-r))} \, ,
\label{sgcg}
\end{equation}

\begin{eqnarray}
\bar {\Phi}_{s,0;s,\gamma'}(r,r')&=& 
\frac 1{\pi} \int_0^\infty d\omega {\sin(\omega[r-r'])
\over \omega(1+e^{2\omega})}
[e^{-\gamma' \omega}+(1-\delta_{0,\gamma'})
e^{-(\gamma'-2)\omega}]\nonumber\\
&+&\frac 1{4}
\int_{-r_{c,0}}^{r_{c,0}}
\frac {dr''}{\gamma'+1}
\frac {\bar {\Phi}_{c,0;s,\gamma'}(r'',r')}
{\cosh(\pi /2(r-r''))} \, ,
\label{s0sg}
\end{eqnarray}
and

\begin{eqnarray}
\bar {\Phi}_{s,\gamma >0;s,\gamma'}(r,r')&=&
\frac 1{2\pi}
\Theta_{\gamma+1,\gamma'+1}(r-r')
-\nonumber\\
&-& \frac 1{\pi} \int_0^\infty d\omega\frac 
{\sin[\omega(r-r')]}
{\omega(1+e^{2\omega})}e^{-(\gamma'+\gamma)\omega}
[2-\delta_{0,\gamma'}+e^{-2 \omega}+(1-\delta_{0,\gamma'})
e^{2\omega}] \, .
\label{sgsg}
\end{eqnarray}
The functions $A(r)$ and $B(r)$ are defined as

\begin{equation}
A(r)=\frac 1{\pi} \int_0^{\infty} d\omega \frac {\cos(r\omega)}
{1+e^{2|\omega|}}\, ,
\end{equation}
and

\begin{equation}
B(r)=\frac 1{\pi} \int_0^{\infty} d\omega \frac {\sin(r\omega)}
{\omega (1+e^{2|\omega|)}}\, ,
\end{equation}
respectively.

At $n=1$ we have that $\Upsilon_1(x)=\Upsilon^0_1(x)
=J_1(x/u)$, where $J_1(x/u)$ is the Bessel function 
of order one, and the bands (\ref{esgh0}) and (\ref{ecgh0}) are 
obtained in closed form. 


\begin{figure}
\caption{The ratio $m_{c,1}^{\rho}/m_{c,1}^{\ast}$ as function of $U$ 
at electronic density $n=0.7$ and for values of the magnetic field 
$h=H/H_c=0.1$ (full line),
$h=0.5$ (dashed line) and $h=0.9$ (dashed-dotted line). 
For other electronic densities, the plots follow the same trends as for
$n=0.7$.}
\label{f1}
\end{figure}

\begin{figure}
\caption{The ratio $m_{c,1}^{\rho}/m_{c,1}^{\ast}$ as function of
the electronic density $n$ and for values of the magnetic field
$h=0.1$, $h=0.3$, $h=0.5$, $h=0.7$, and $h=0.9$. The onsite
Coulomb interaction is (a) $U=1$, (b) $U=5$, and (c) $U=20$.}
\label{f2}
\end{figure}

\begin{figure}
\caption{The ratio $m_{c,1}^{\rho}/m_{c,1}^{\ast}$ as function of 
the the magnetic field $h$ and for values of the electronic density 
$n=0.1$, $n=0.3$, $n=0.5$, $n=0.7$, and $n=0.9$. The onsite
Coulomb interaction is (a) $U=1$, (b) $U=5$, and (c) $U=20$.}
\label{f3}
\end{figure}

\begin{figure}
\caption{The ratio $m_{c,1}^{\rho}/m_{c,1}^{\ast}$ as function of the 
the magnetic field $h$ and for values of the onsite Coulomb 
interaction $U=1$, $U=2$, $U=3$, $U=5$, $U=10$, and $U=20$. The 
electronic density is (a) $n=0.5$ and (b) $n=0.9$.}
\label{f4}
\end{figure}

\begin{figure}
\caption{The ratio $m_{c,0}^{\rho}/m_{c,0}^{\ast}$ as function of 
$U$, for electronic density $n=0.7$, and for values of the 
magnetic field $h=0.1$, $h=0.3$, $h=0.5$, $h=0.7$, and 
$h=0.9$. For other electronic densities, the plots follow the same 
trends as for $n=0.7$.}
\label{f5}
\end{figure}

\begin{figure}
\caption{The ratio $m_{c,0}^{\rho}/m_{c,0}^{\ast}$ as 
function of the electronic density $n$ and for 
values of the magnetic field
$h=0.1$, $h=0.3$, $h=0.5$, $h=0.7$, and $h=0.9$. The onsite
Coulomb interaction is (a) $U=1$, (b) $U=5$, and (c) $U=20$.}
\label{f6}
\end{figure}

\begin{figure}
\caption{The ratio $m_{c,0}^{\rho}/m_{c,0}^{\ast}$ as function 
of the magnetic field $h$ and for values of the electronic 
density $n=0.1$, $n=0.3$, $n=0.5$, $n=0.7$, and $n=0.9$. The onsite
Coulomb interaction is (a) $U=1$, (b) $U=5$, and (c) $U=20$.}
\label{f7}
\end{figure}

\begin{figure}
\caption{The ratio $m_{c,0}^{\rho}/m_{c,0}^{\ast}$ as function 
of the the magnetic field $h$ and for values of the 
onsite Coulomb interaction $U=1$, $U=2$, $U=3$, $U=5$, $U=10$, and 
$U=20$. The electronic density is (a) $n=0.5$ and (b) $U=0.9$.}
\label{f8}
\end{figure}

\begin{figure}
\caption{The ratio $m_{s,1}^{\sigma_z}/m_{s,1}^{\ast}$ as function 
of $U$, for electronic density $n=0.7$, and for values 
of the magnetic field $h=0.1$, $h=0.3$, $h=0.5$, $h=0.7$, 
and $h=0.9$. For other electronic densities, the plots follow the 
same trends as for $n=0.7$.}
\label{f9}
\end{figure}

\begin{figure}
\caption{The ratio $m_{s,1}^{\sigma_z}/m_{s,1}^{\ast}$ as 
function of the electronic density $n$ and for values of the 
magnetic field $h=0.1$, $h=0.3$, $h=0.5$, $h=0.7$, and 
$h=0.9$. The onsite Coulomb interaction is (a) $U=1$, (b) $U=5$, 
and (c) $U=20$.}
\label{f10}
\end{figure}

\begin{figure}
\caption{The ratio $m_{s,1}^{\sigma_z}/m_{s,1}^{\ast}$ as function 
of the magnetic field $h$ and for values of the electronic 
density $n=0.1$, $n=0.3$, $n=0.5$, $n=0.7$, and $n=0.9$. The onsite
Coulomb interaction is (a) $U=1$, (b) $U=5$, and (c) $U=20$.}
\label{f11}
\end{figure}

\begin{figure}
\caption{The ratio $m_{s,1}^{\sigma_z}/m_{s,1}^{\ast}$ as function 
of the the magnetic field $h$ and for values of the onsite 
Coulomb interaction $U=1$, $U=2$, $U=3$, $U=5$, $U=10$, and $U=20$. 
The electronic density is (a) $n=0.3$, (b) $n=0.5$, and (c) $U=0.9$.}
\label{f12}
\end{figure}
\mediumtext
\begin{table}
\label{tab1}
\begin{tabular}{cccc}
 & $H\rightarrow H_c$ & $H\rightarrow 0$ & $n\rightarrow 1$\\
\tableline
$m_{c,\gamma}^\rho/m^{\ast}_{c,\gamma}$ & $\frac 1
{2\gamma(2\gamma-\eta_{\gamma-1})}$ & 
$\frac 1
{2{\gamma}(2\gamma+\xi_{c,\gamma;c,0}^1)}$ & $\frac {1}{4{\gamma}^2}$\\ 
$m_{c,0}^\rho/m^{\ast}_{c,0}$ & $\frac {v_c}
{2t\sin(\pi n_{\uparrow})}$&
$\frac 1{(\xi_0)^2}$& $\infty$\\
$m_{s,\gamma}^{\sigma_z}/m^{\ast}_{s,\gamma}$& $\frac 1
{2(\gamma +1)(2\gamma +2-\eta_{\gamma})}$ & $\frac 1 
{2(\gamma +1)(2\gamma+2+2\xi^1_{s,\gamma;s,0})}$& 
$\frac 1 {2(\gamma +1)(2\gamma+2+2\xi^1_{s,\gamma;s,0}
-\xi^1_{s,\gamma;c,0})}$\\ 
\end{tabular}
\end{table}
Mass ratios in several limits of physical interest. 
The function $\eta_{\gamma }$ is defined as 
$\eta_{\gamma }=2/(\pi)\tan^{-1}[(\sin(n\pi))/(u[\gamma +1])]$. 
The equations for the static 
masses $m^{\ast}_{\alpha,\gamma}$ are given in Appendix 
\ref{static}. In the case $H\rightarrow 0$, simple expressions
for the parameters $\xi_{\alpha,\gamma;\alpha',0}^1$, Eq. (\ref{xi}), 
can be obtained from the results of Appendix \ref{zeroH}. 
The ratios $m_{c,\gamma}^{\sigma_z}/m^{\ast}_{c,\gamma}$ 
and $m_{s,\gamma}^\rho/m^{\ast}_{s,\gamma}$ are infinite. The 
dependence on $U$ and $n$ of the parameter $\xi_0$ has been studied 
in Refs. \cite{Frahm,Woy89}.
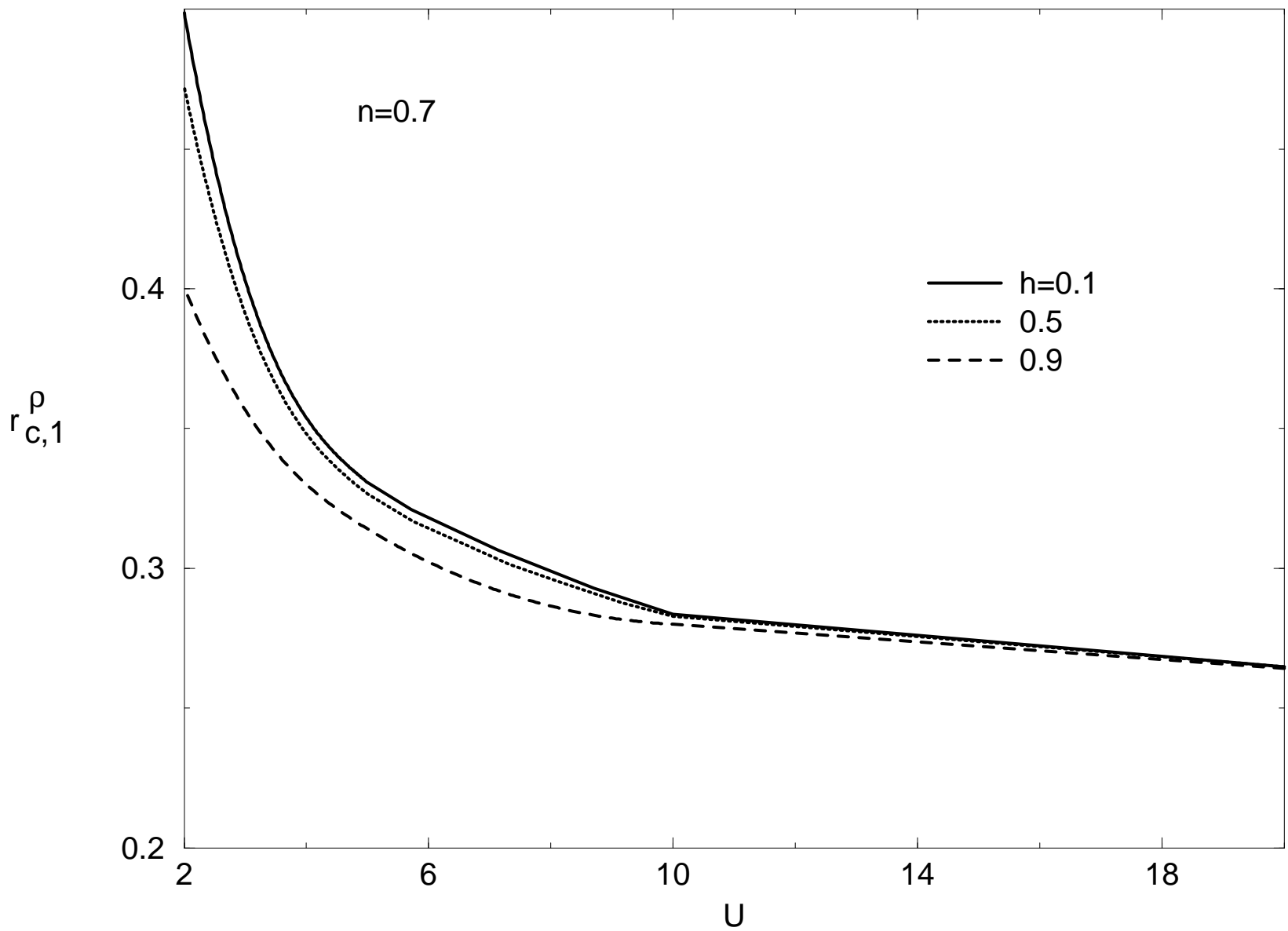
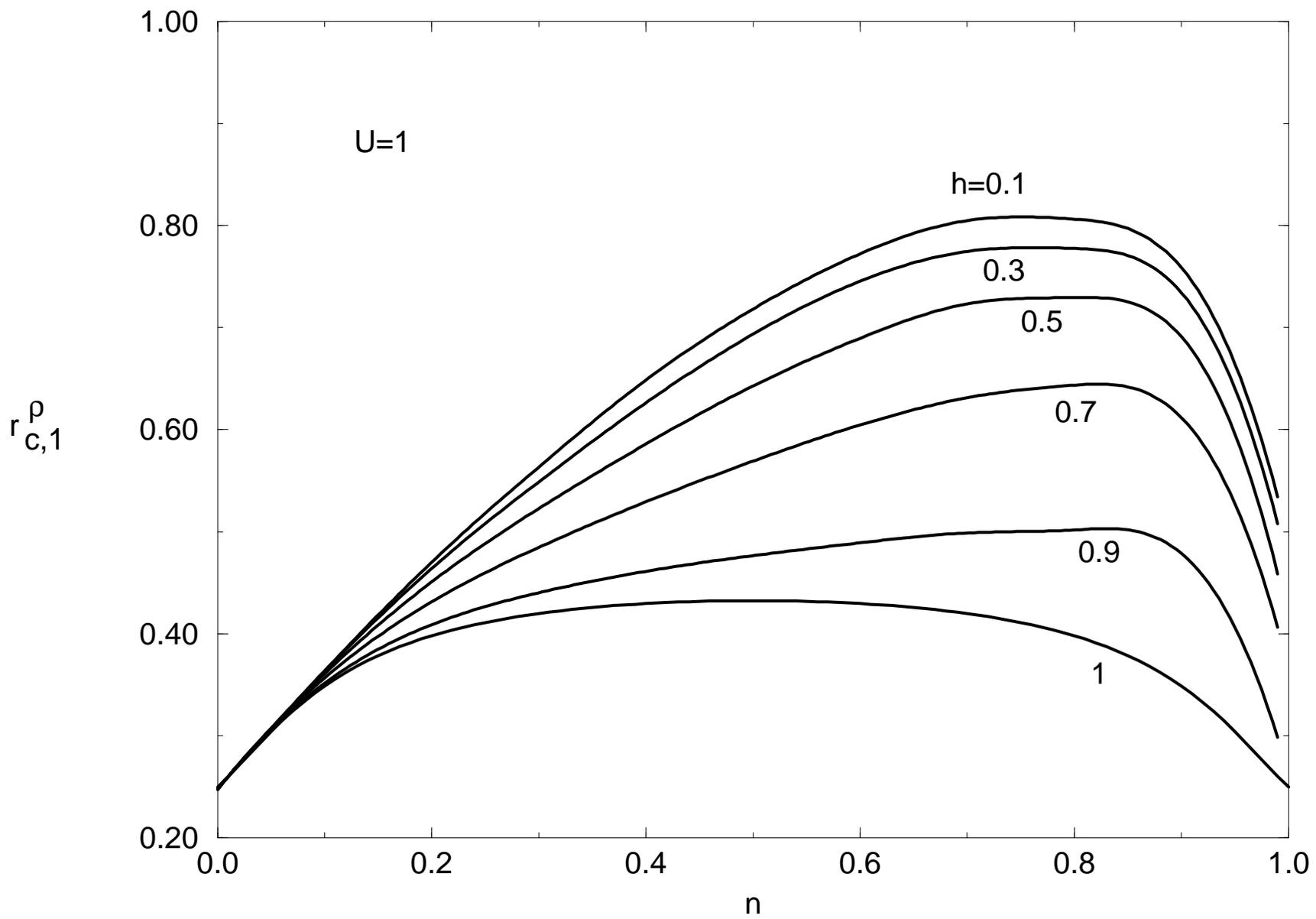
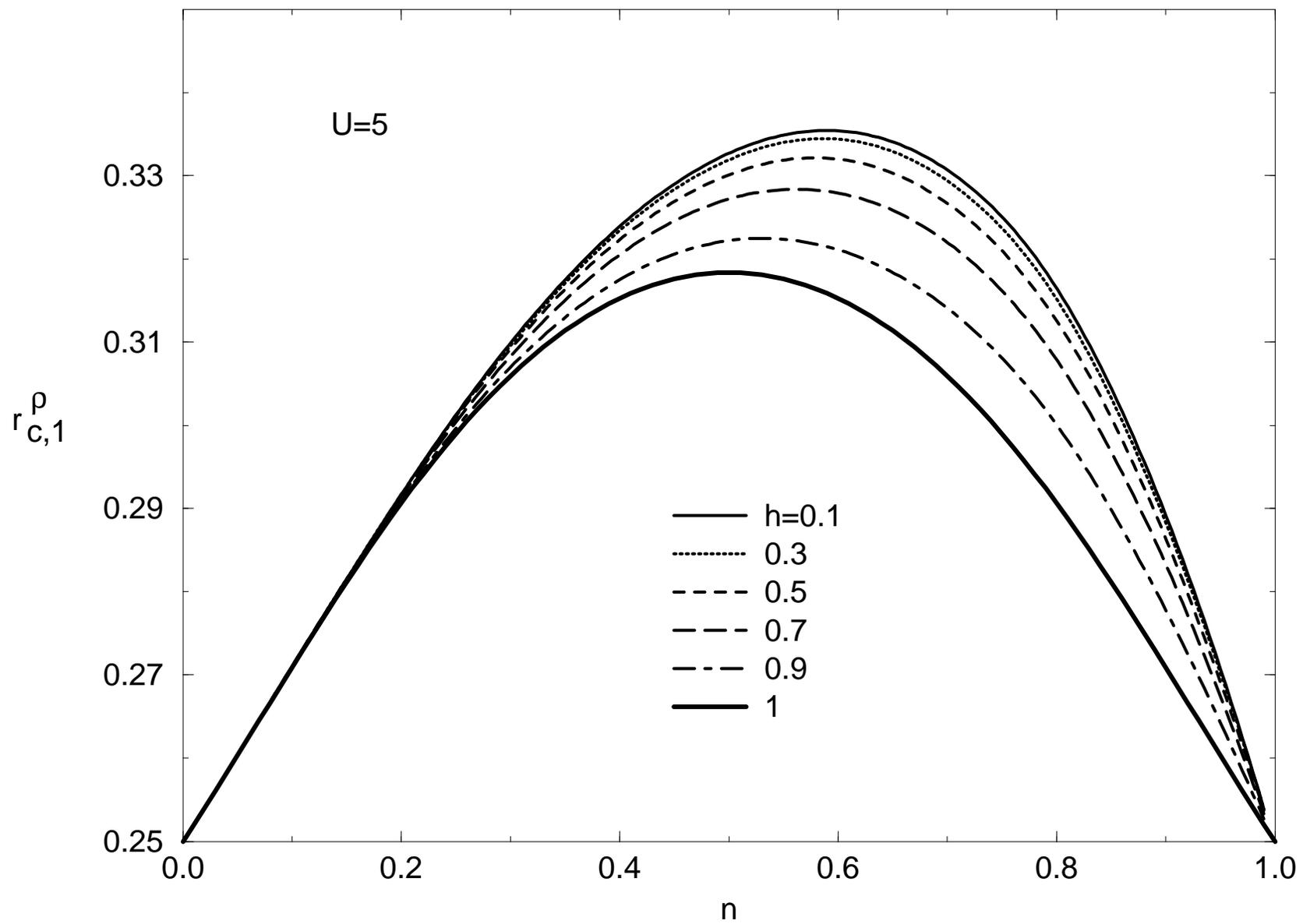
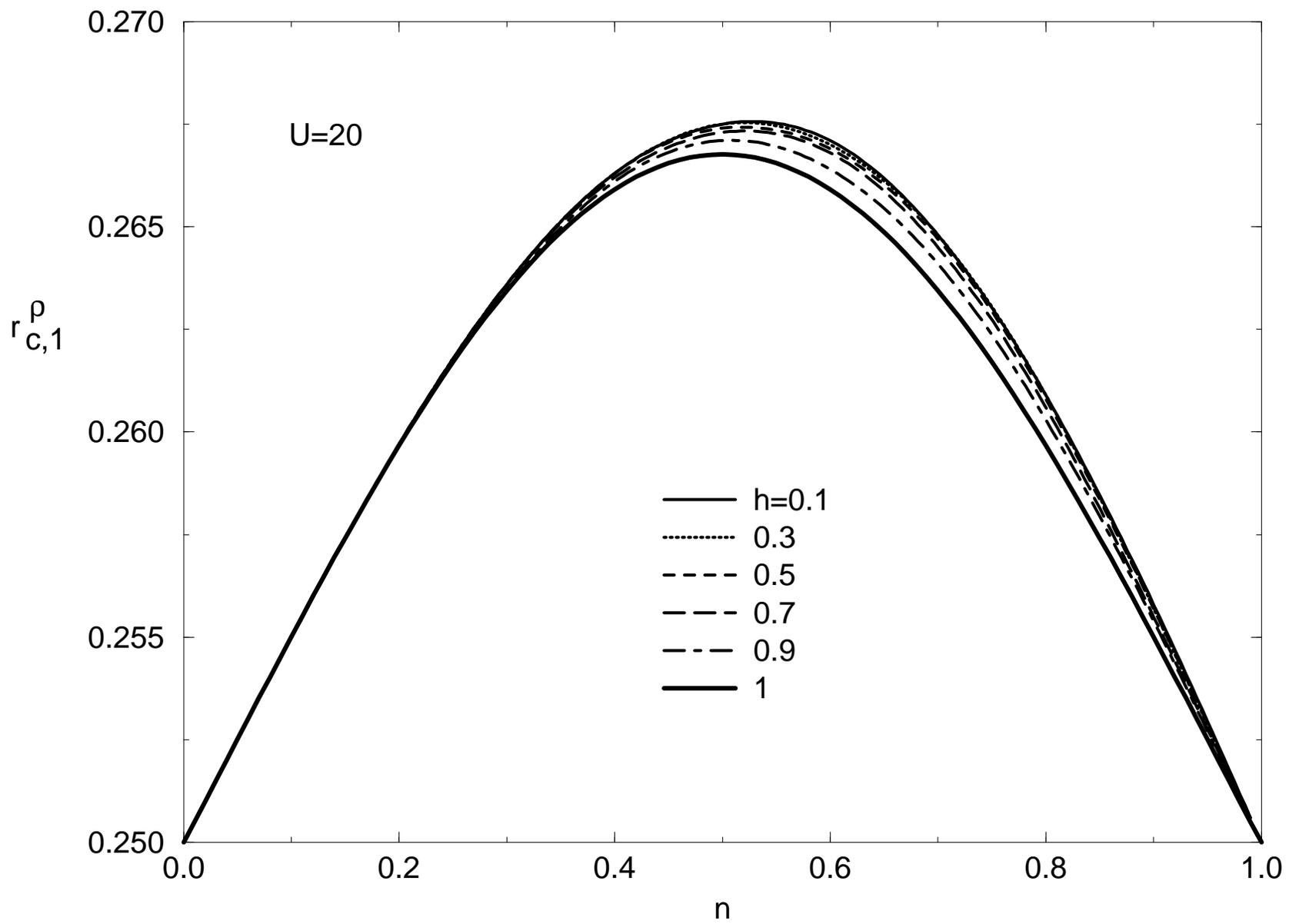
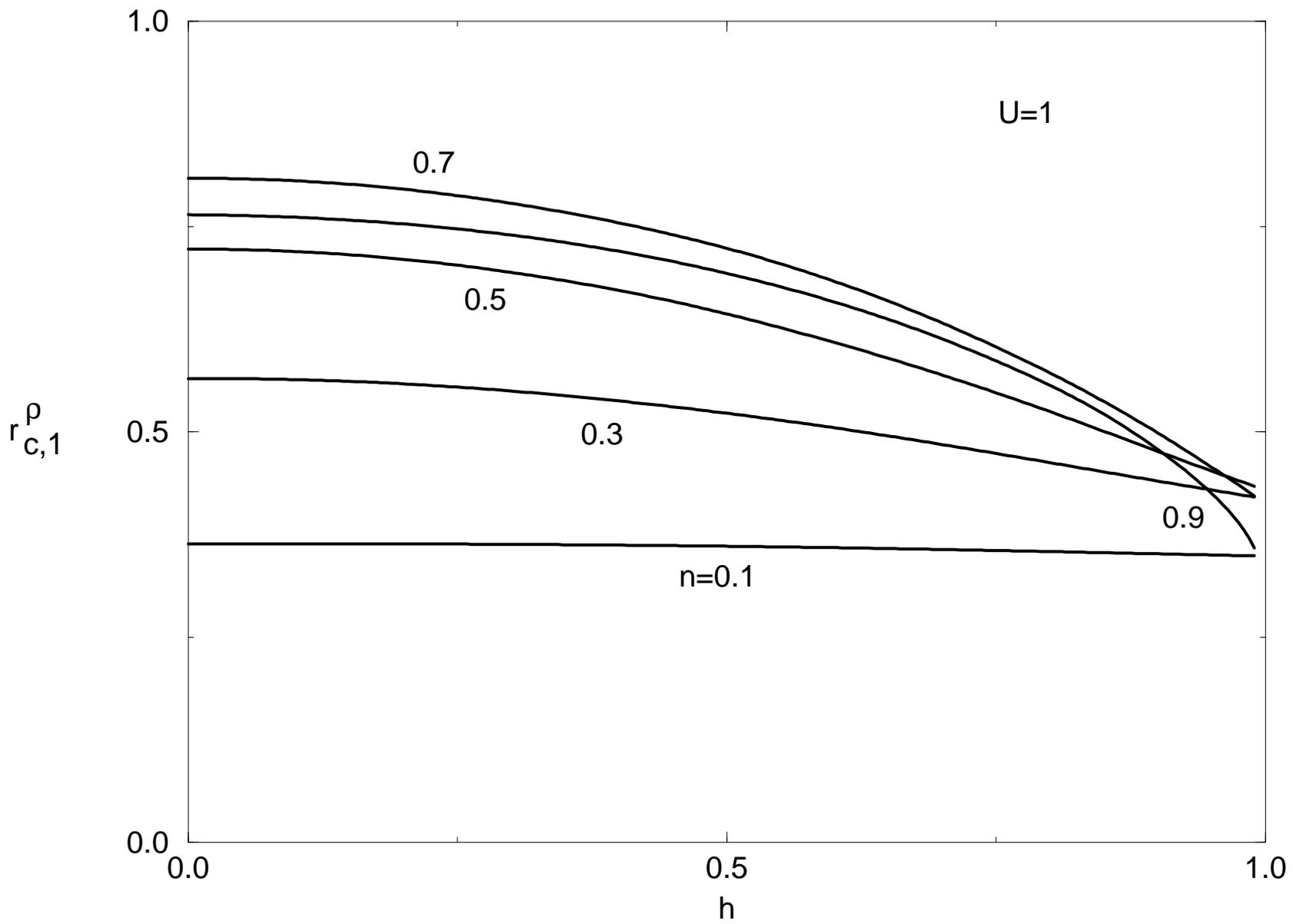
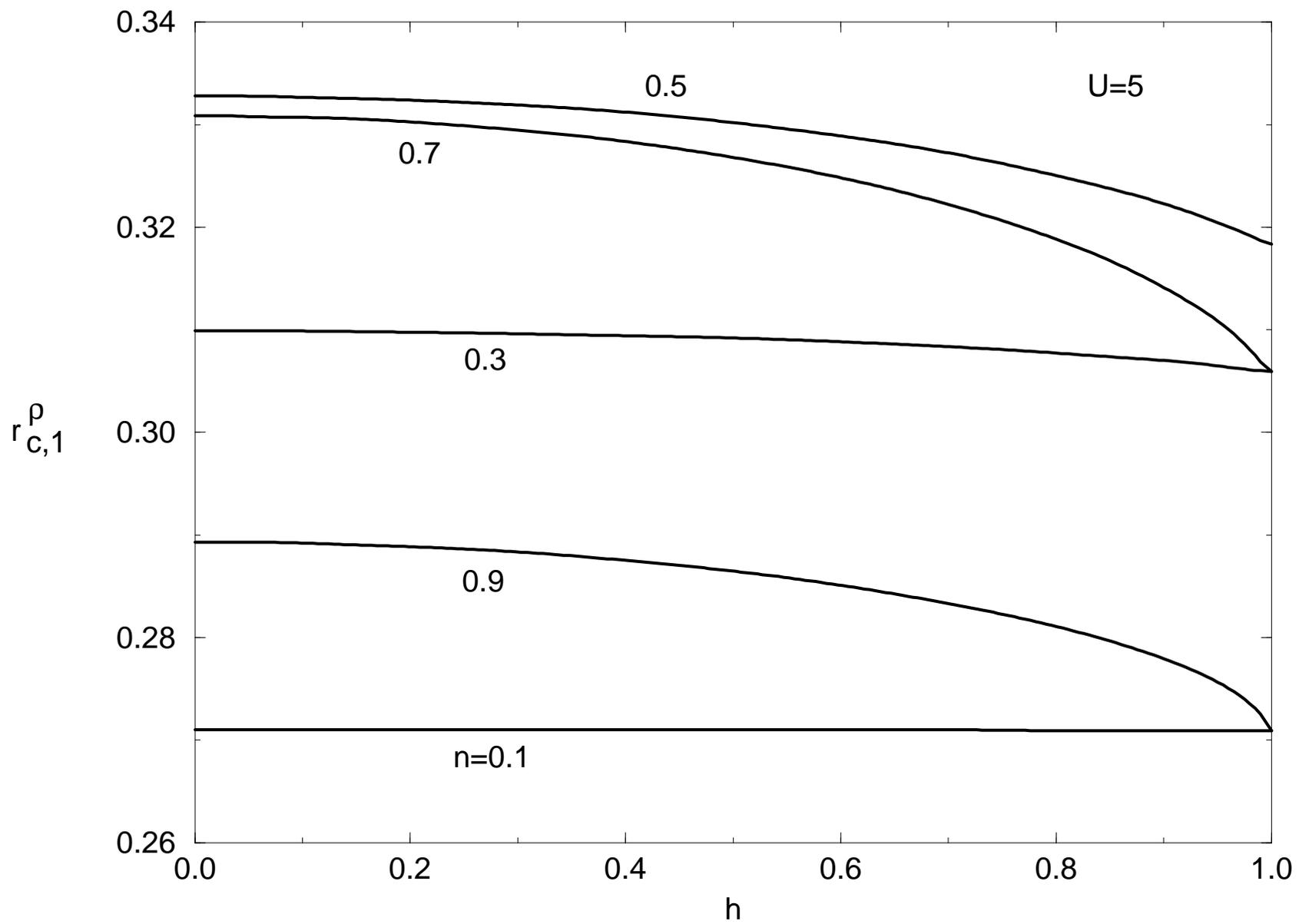
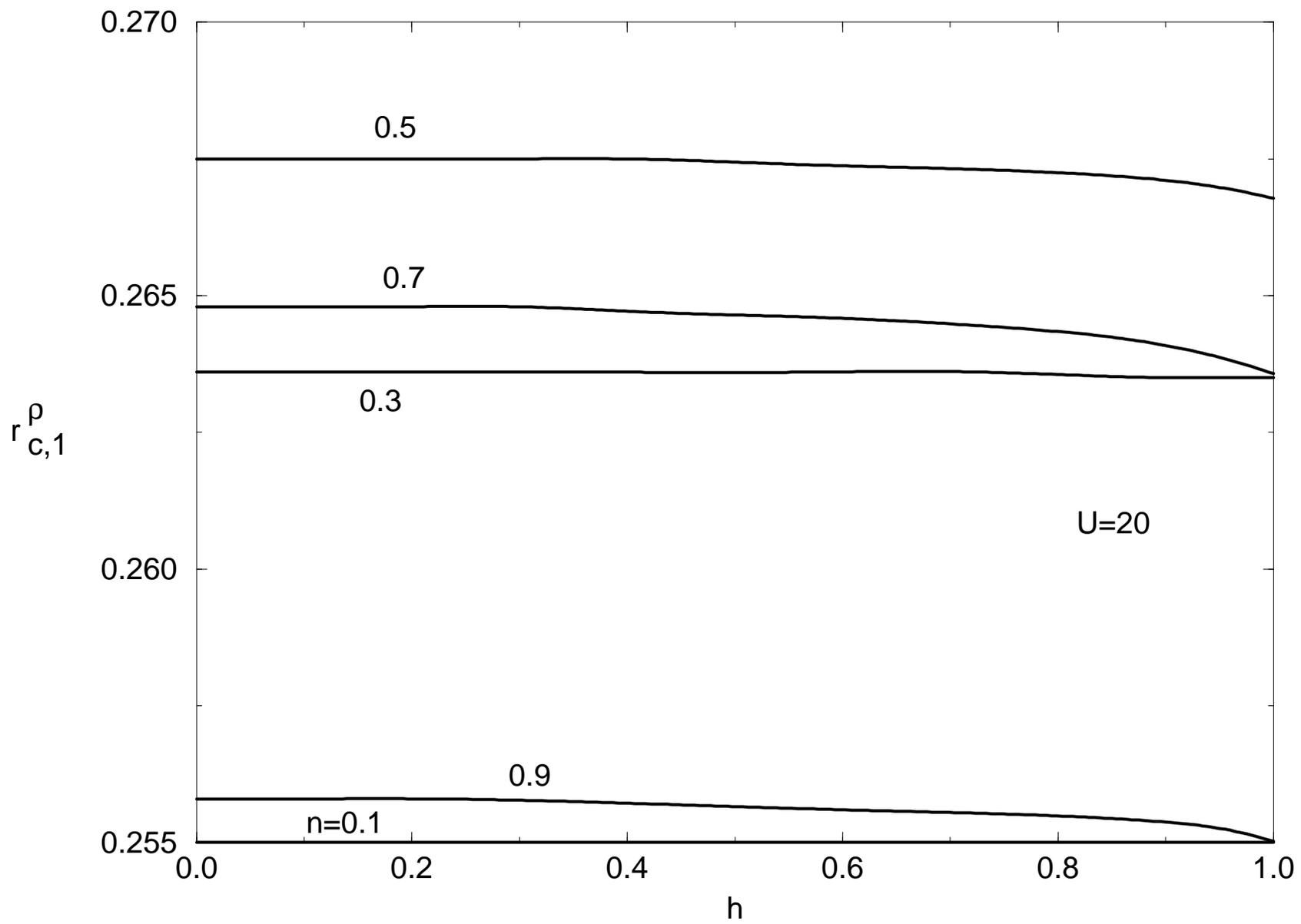
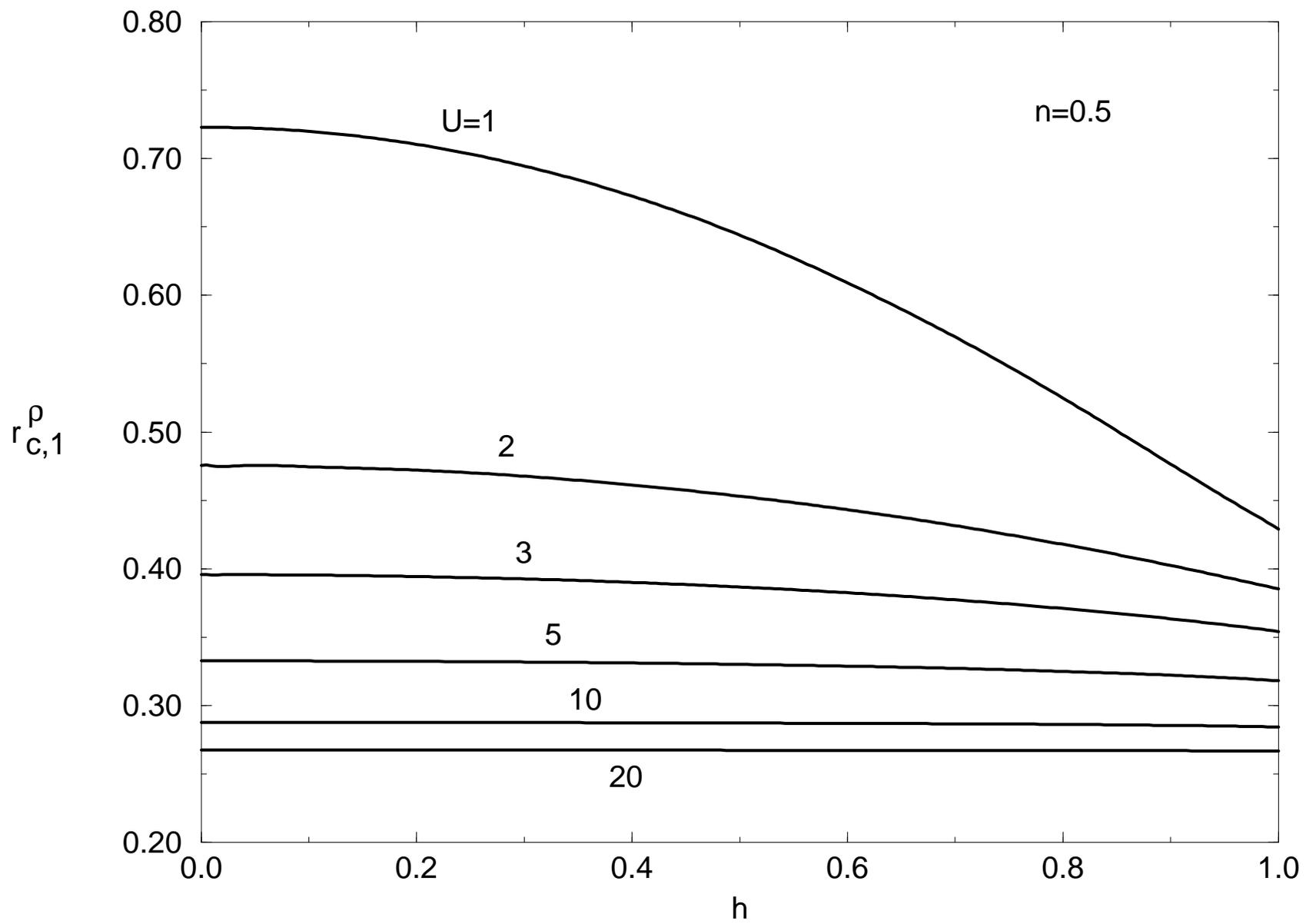
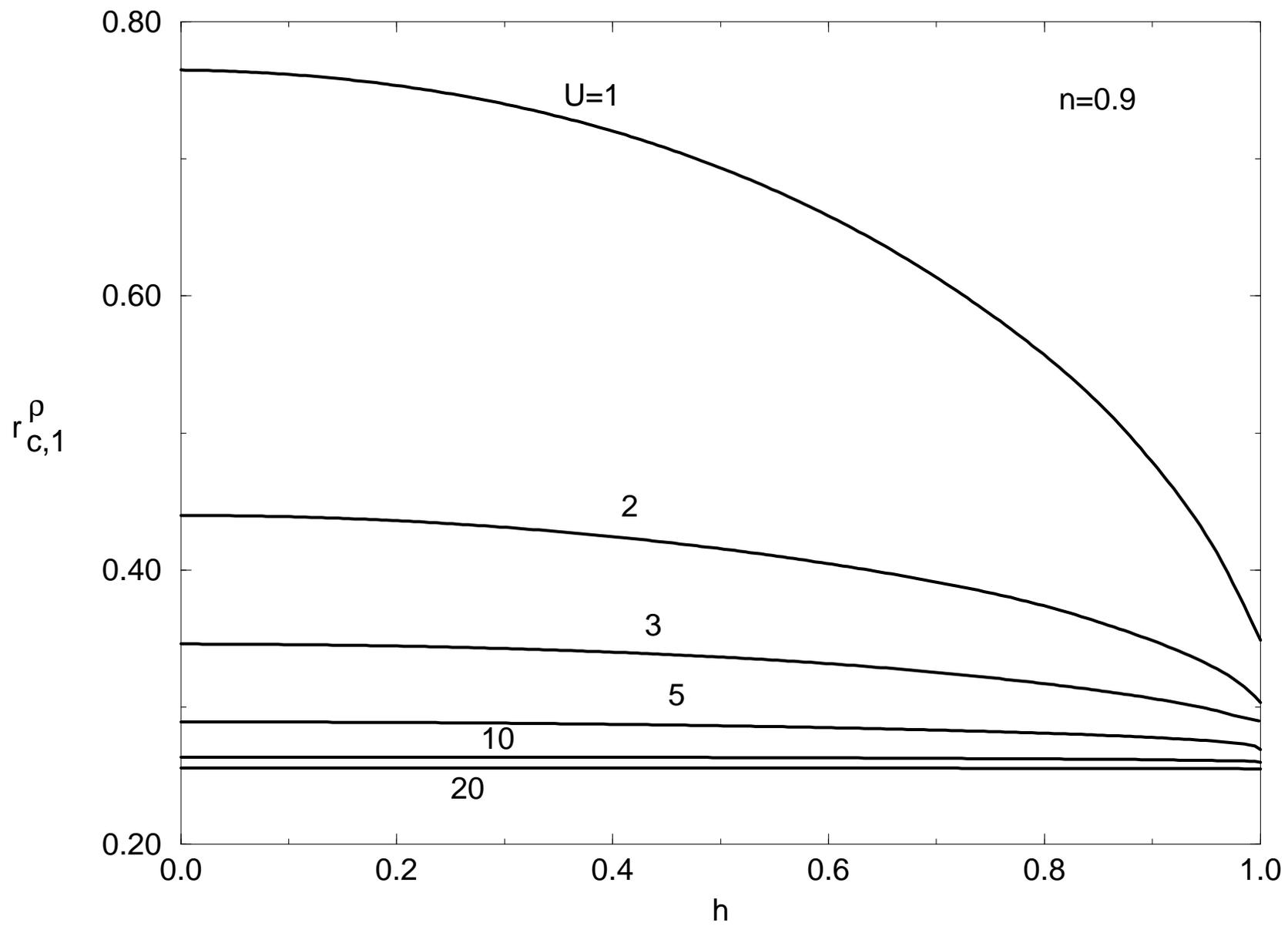
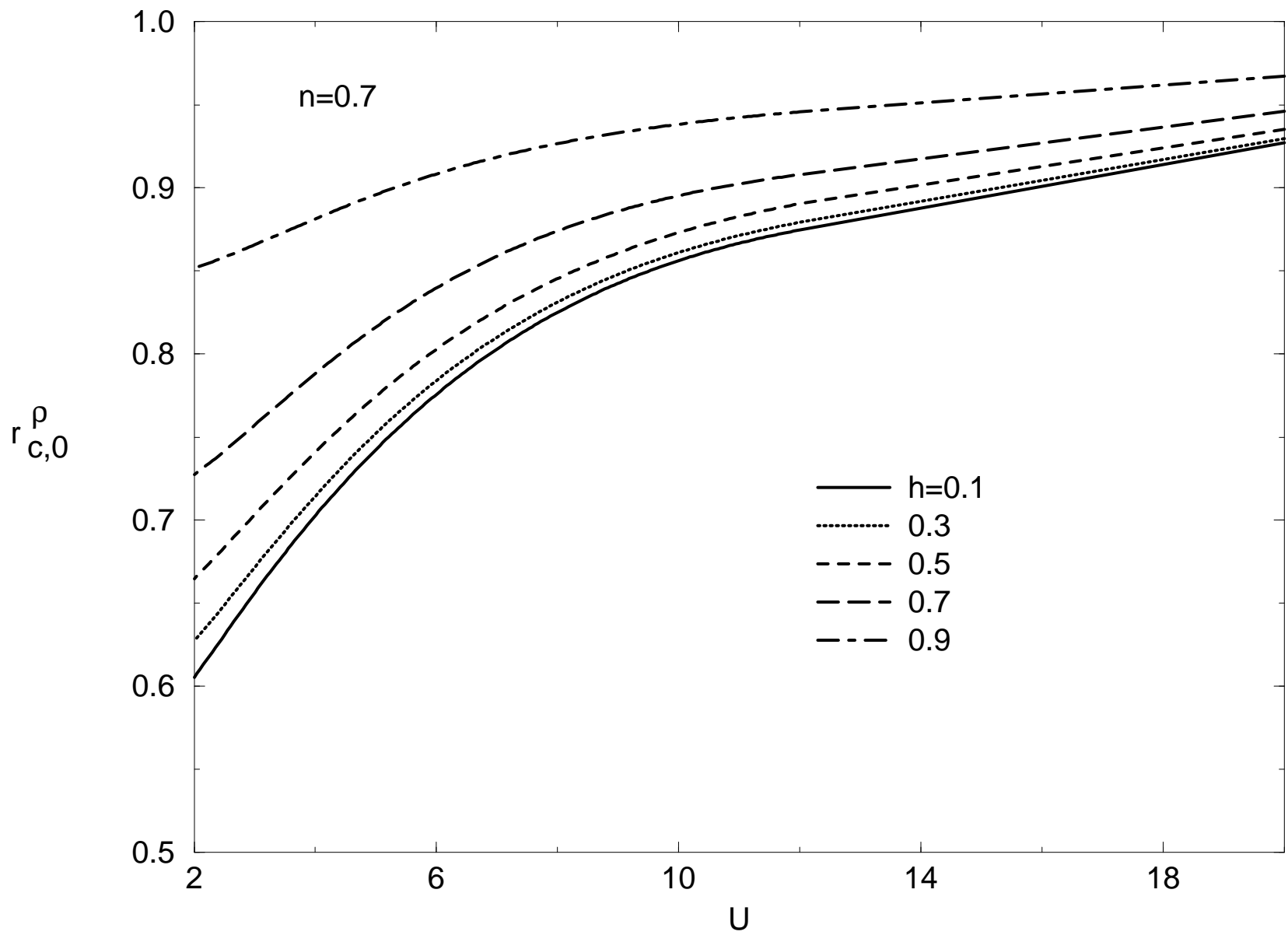
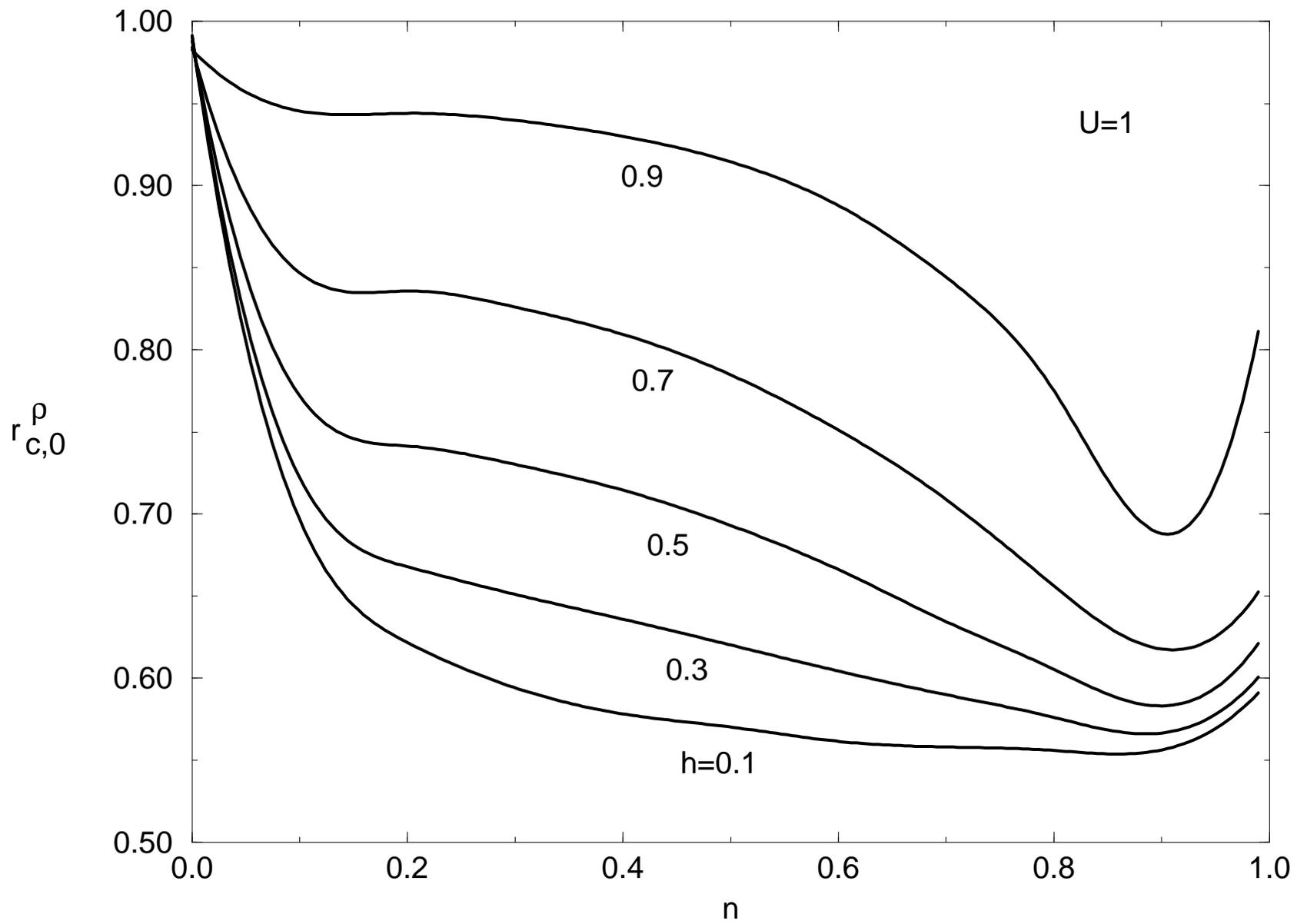
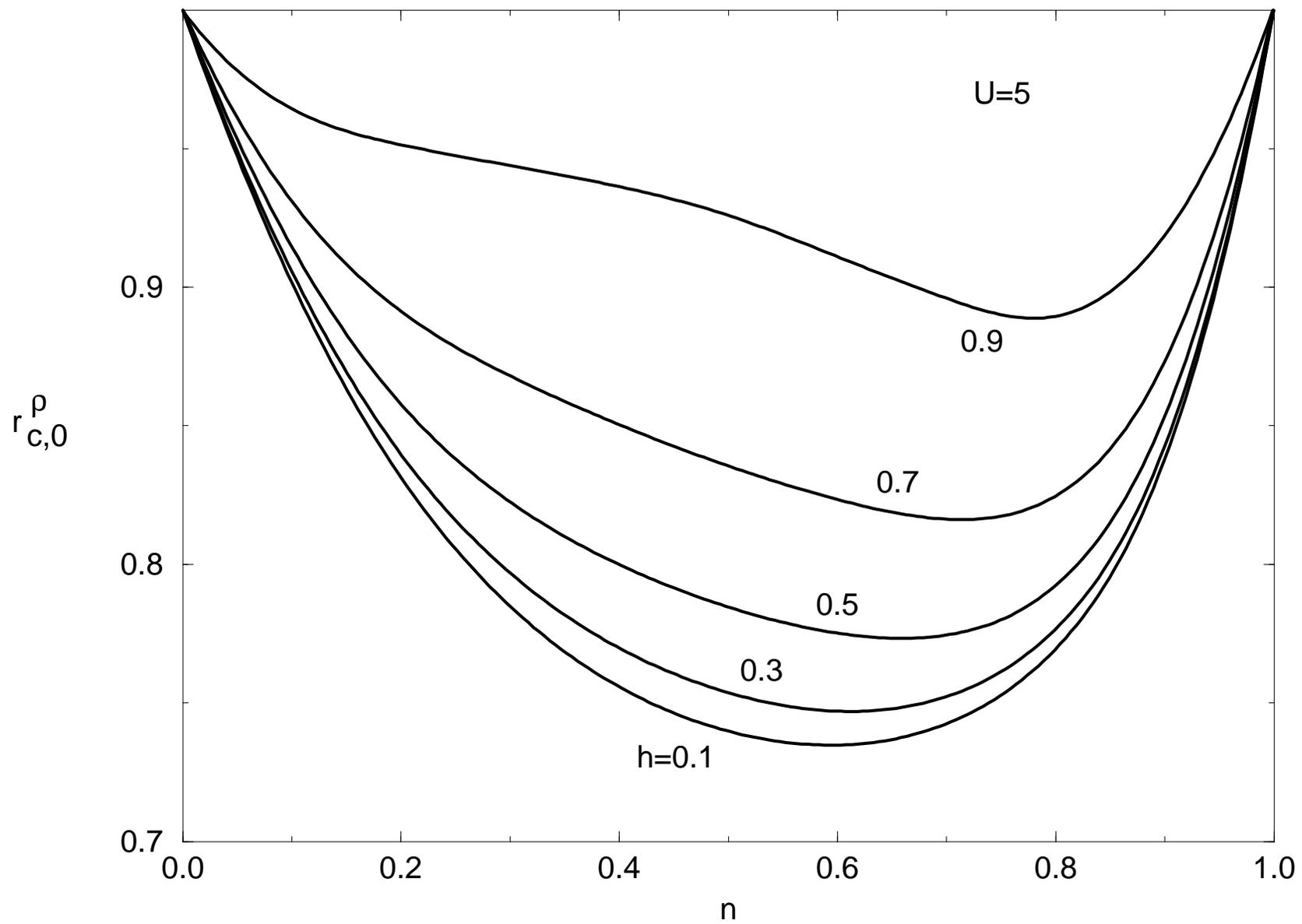
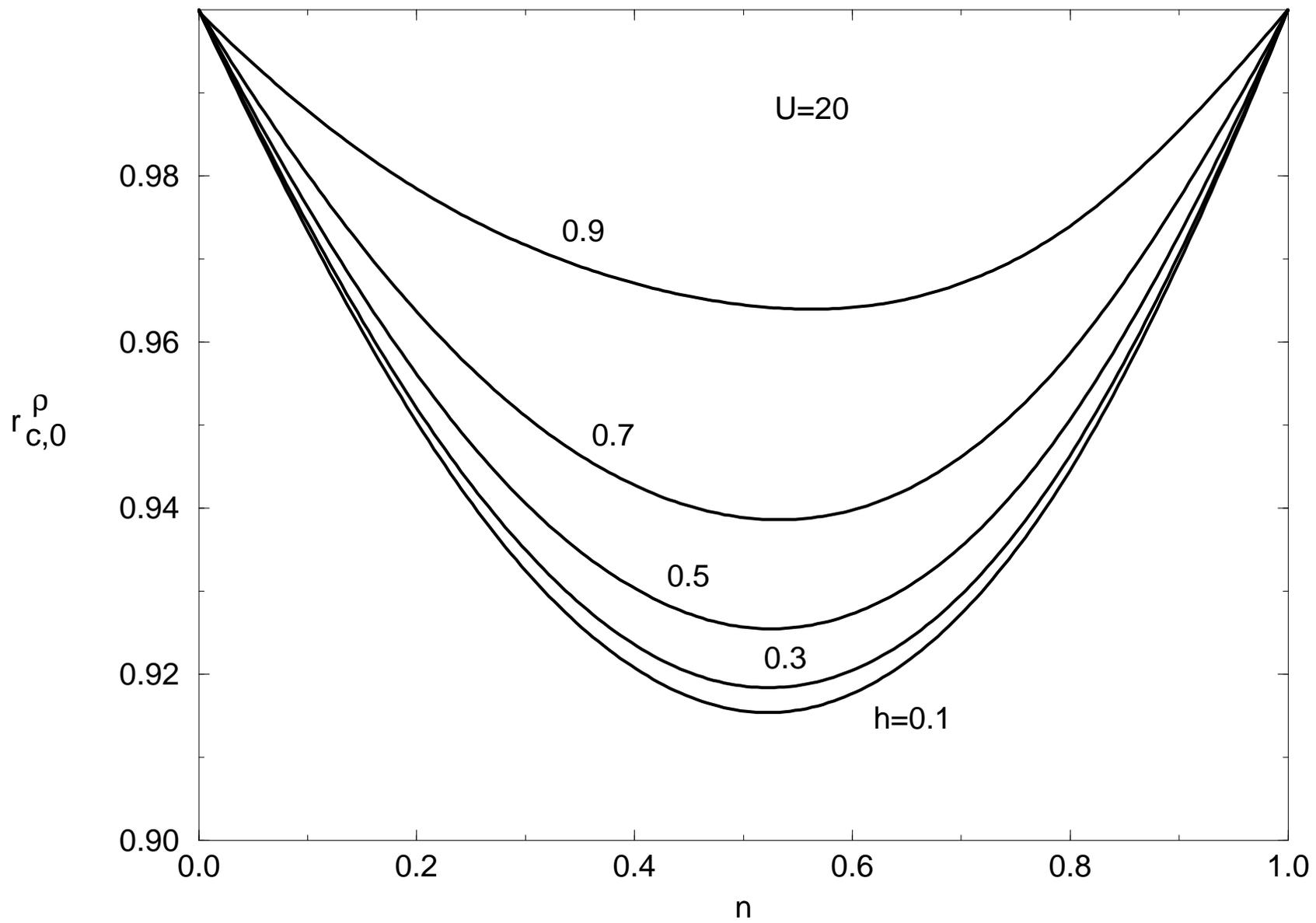
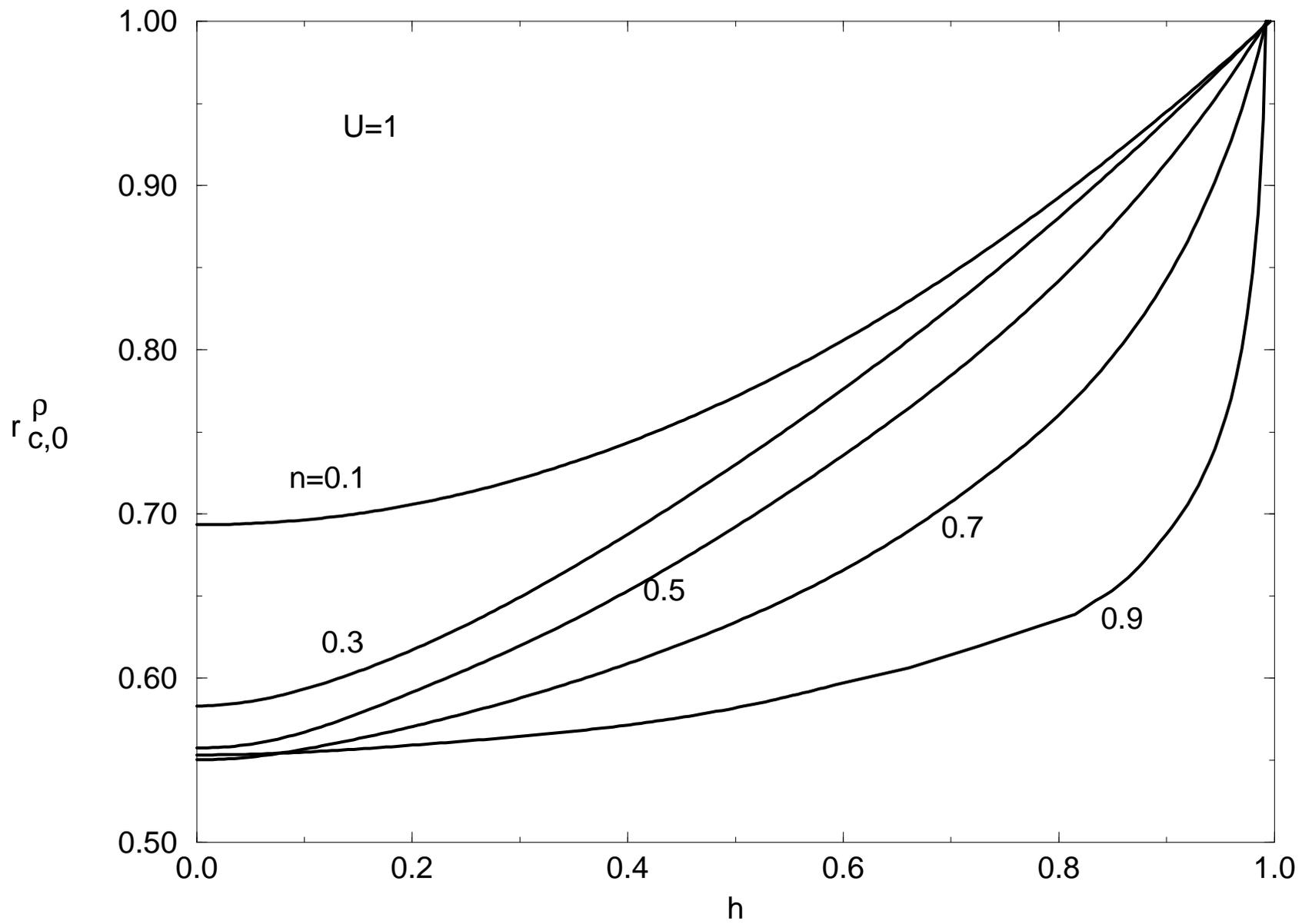
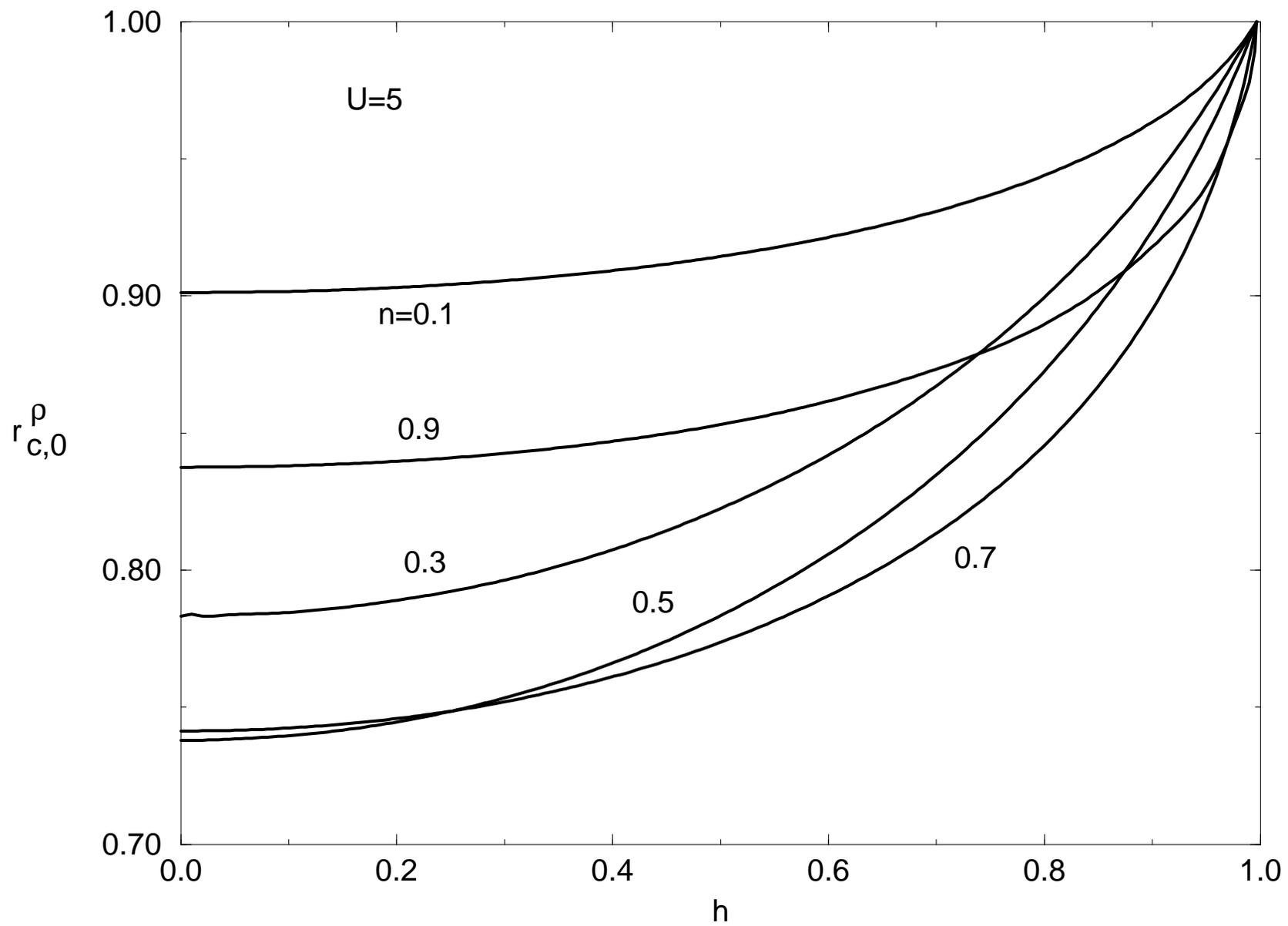
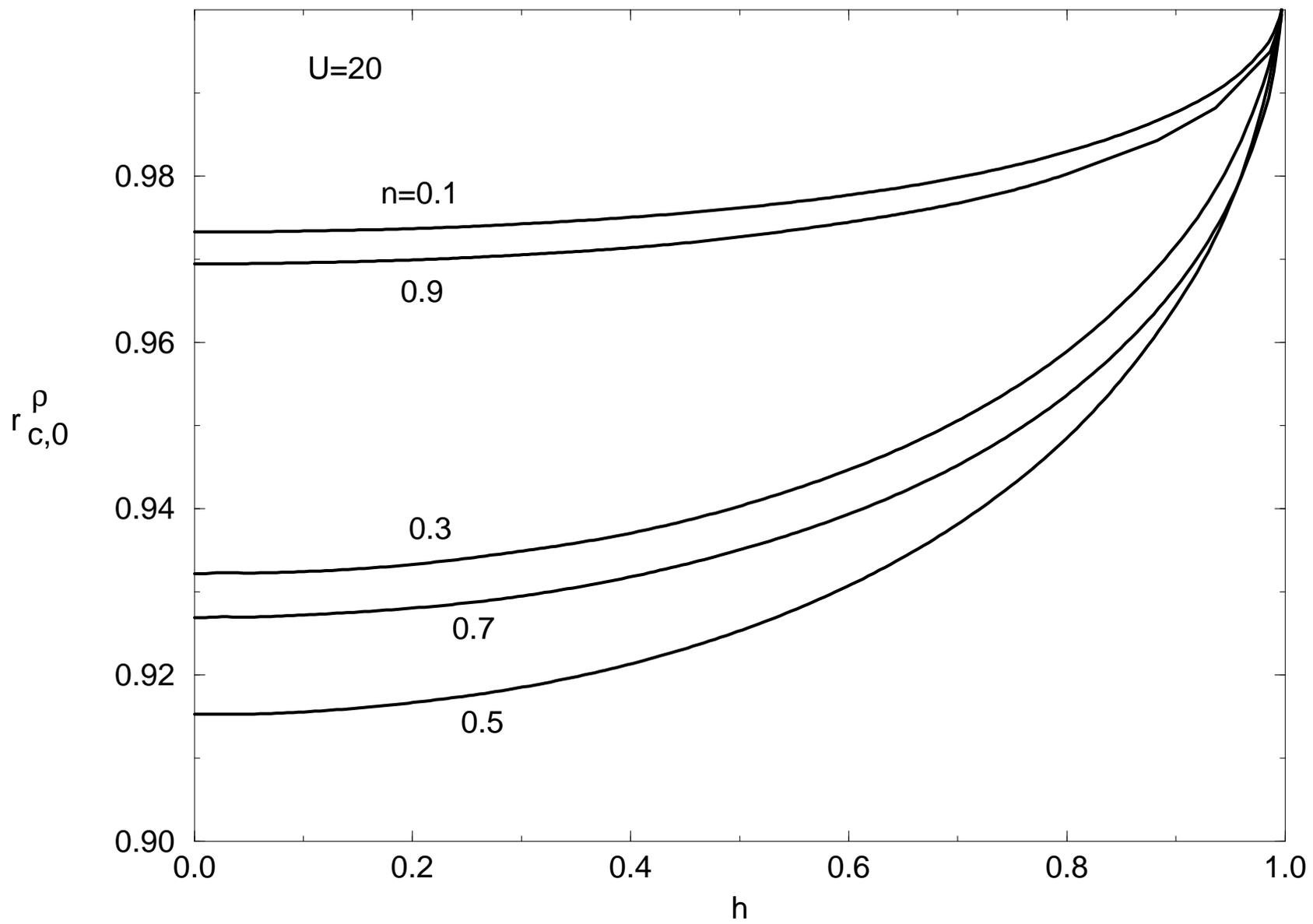
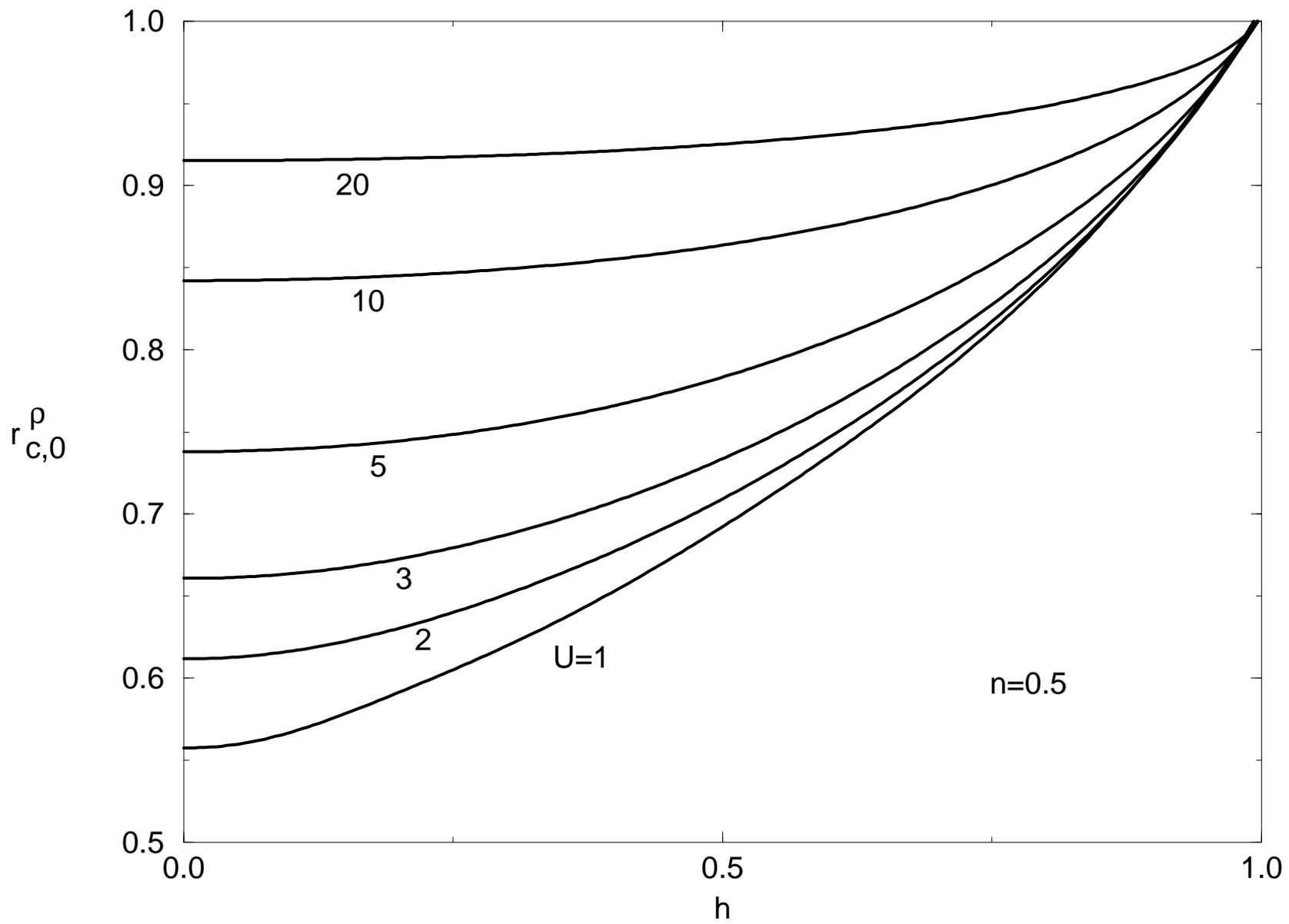
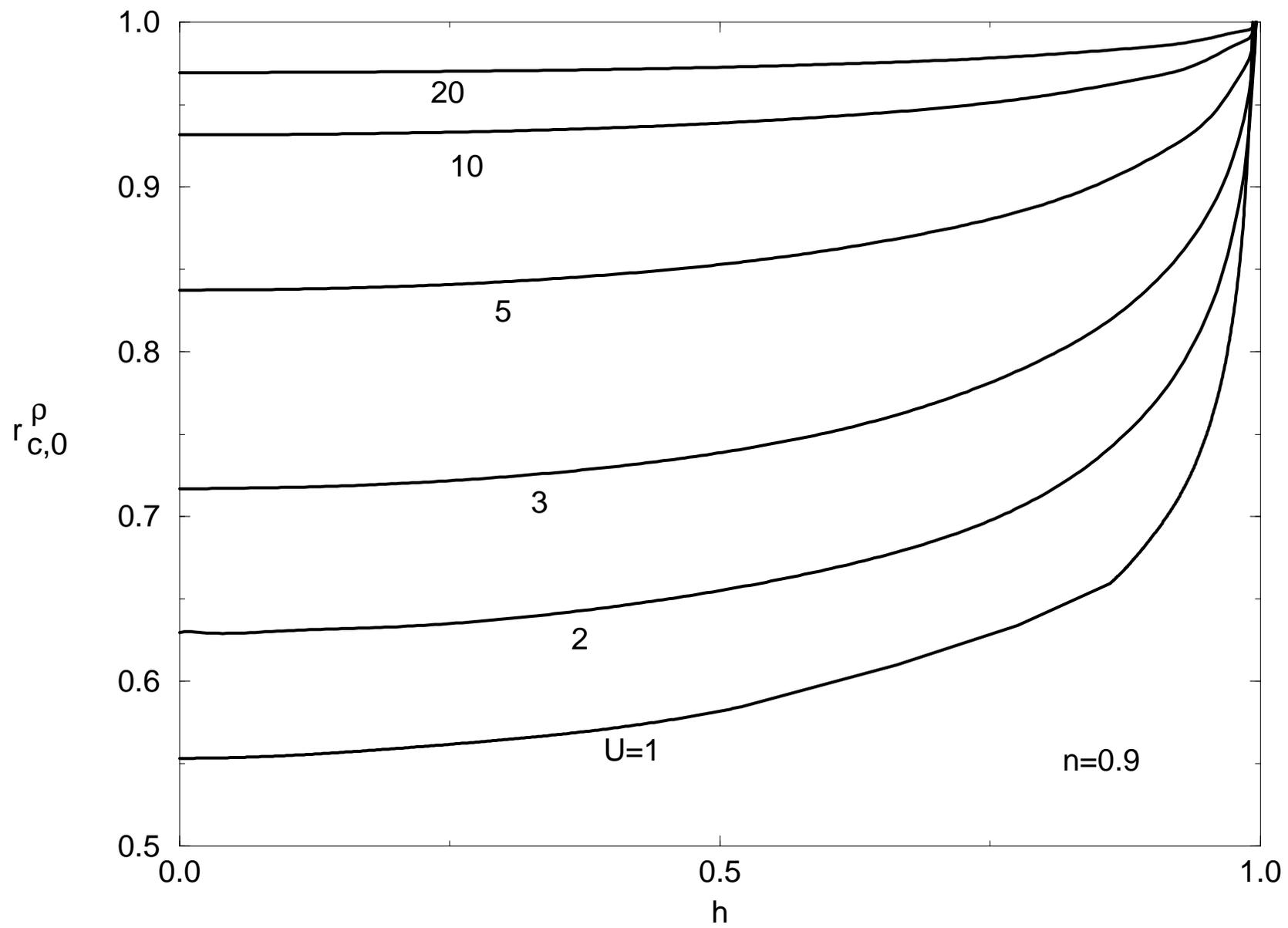
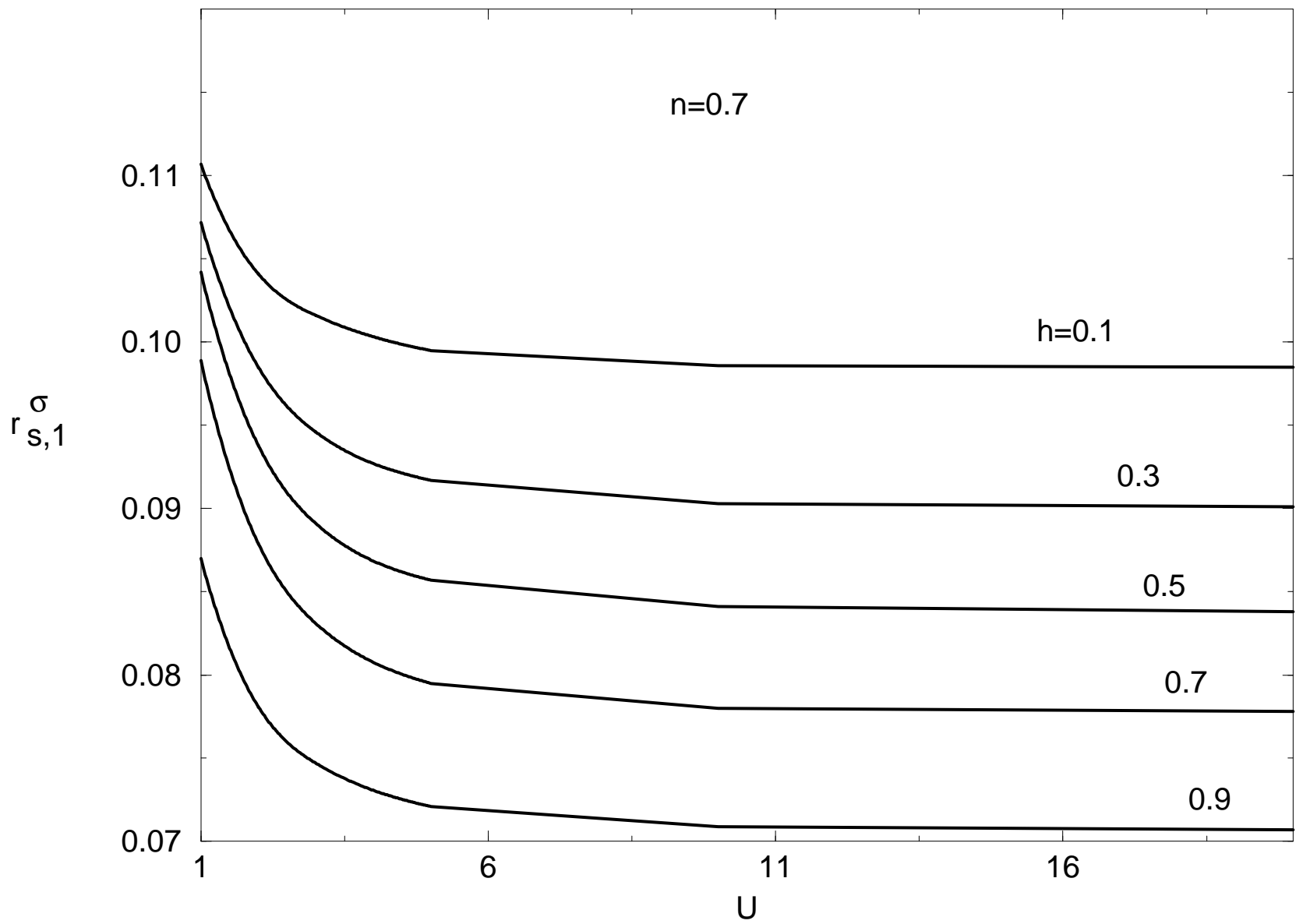
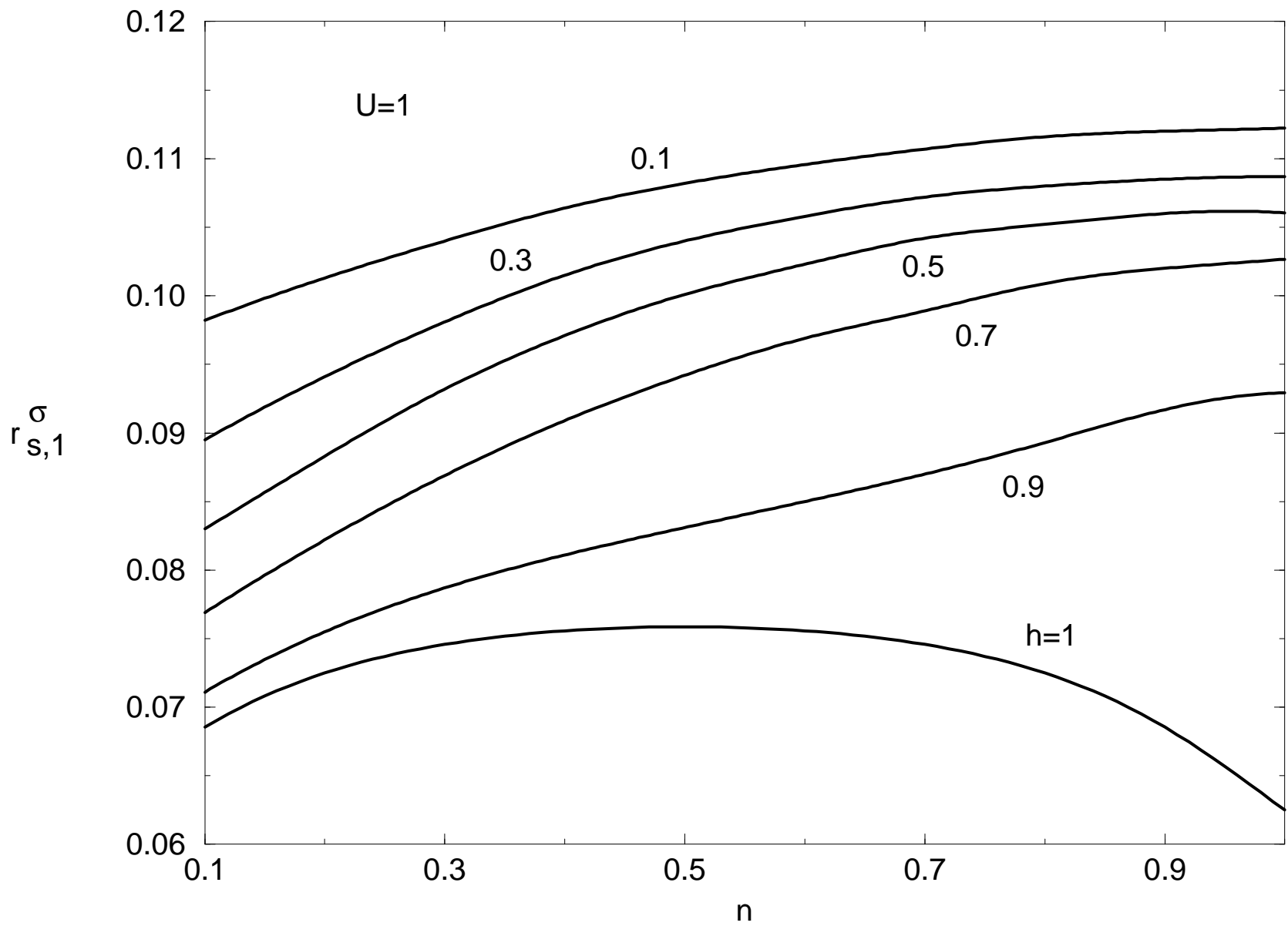
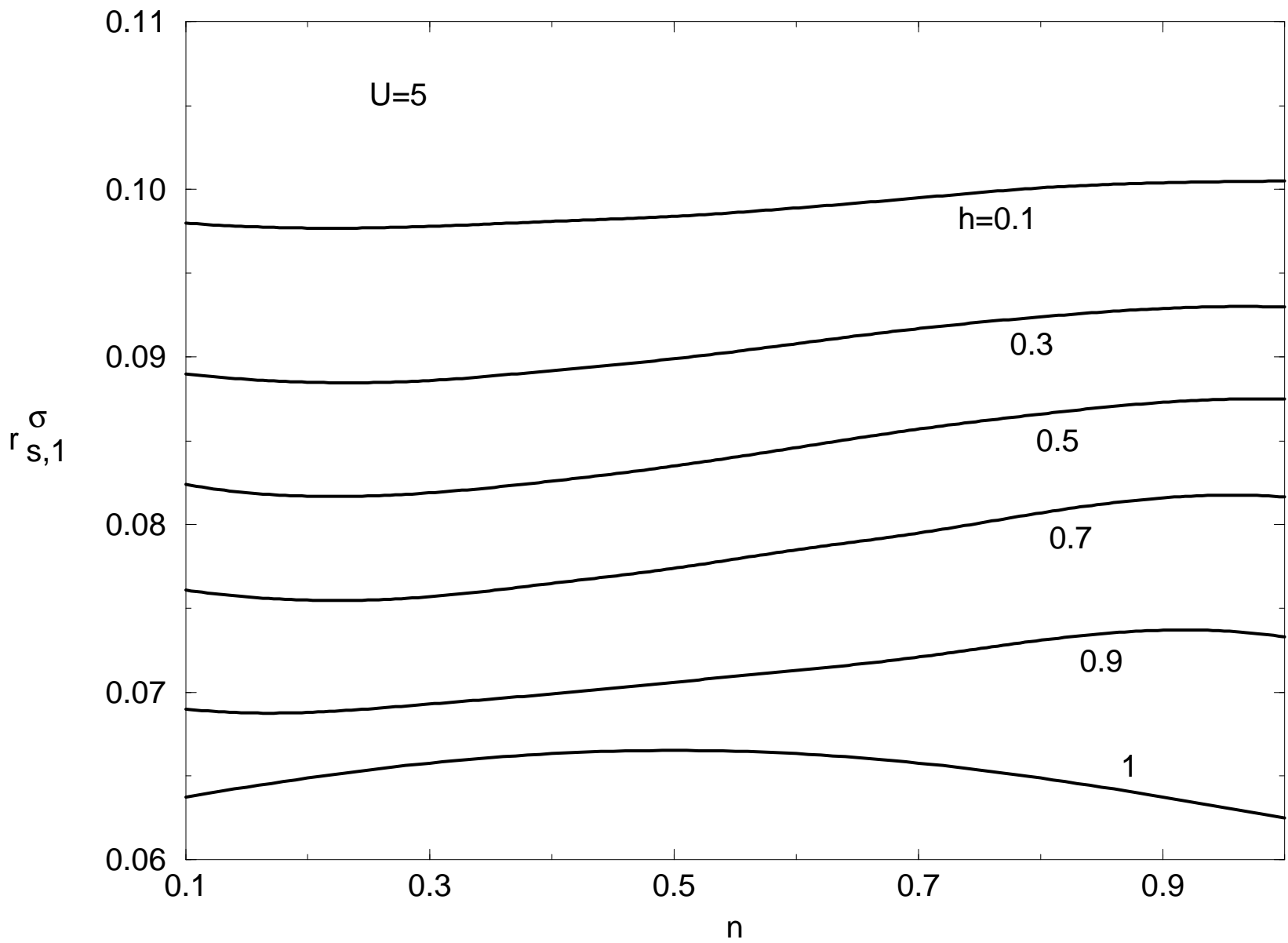
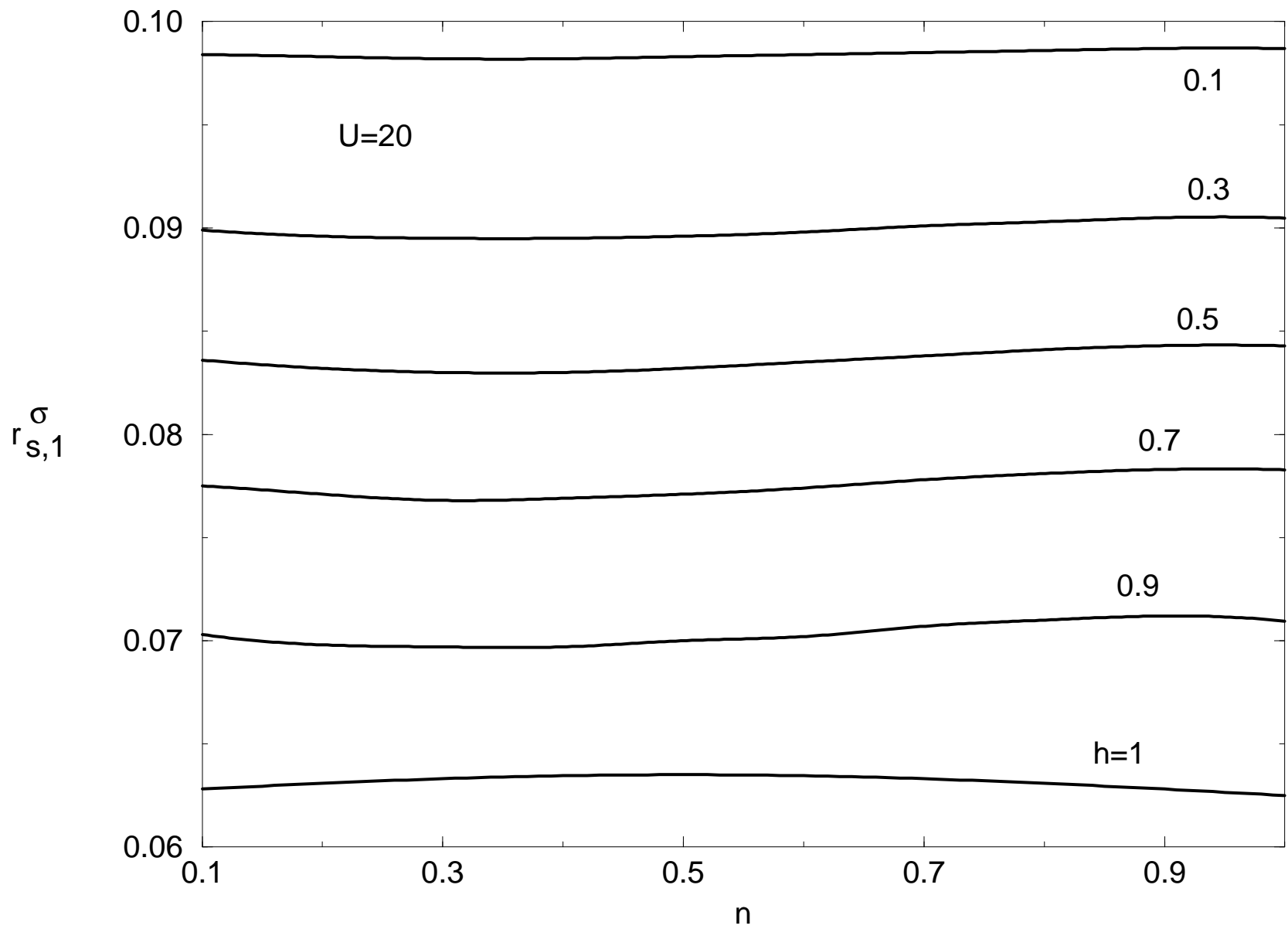
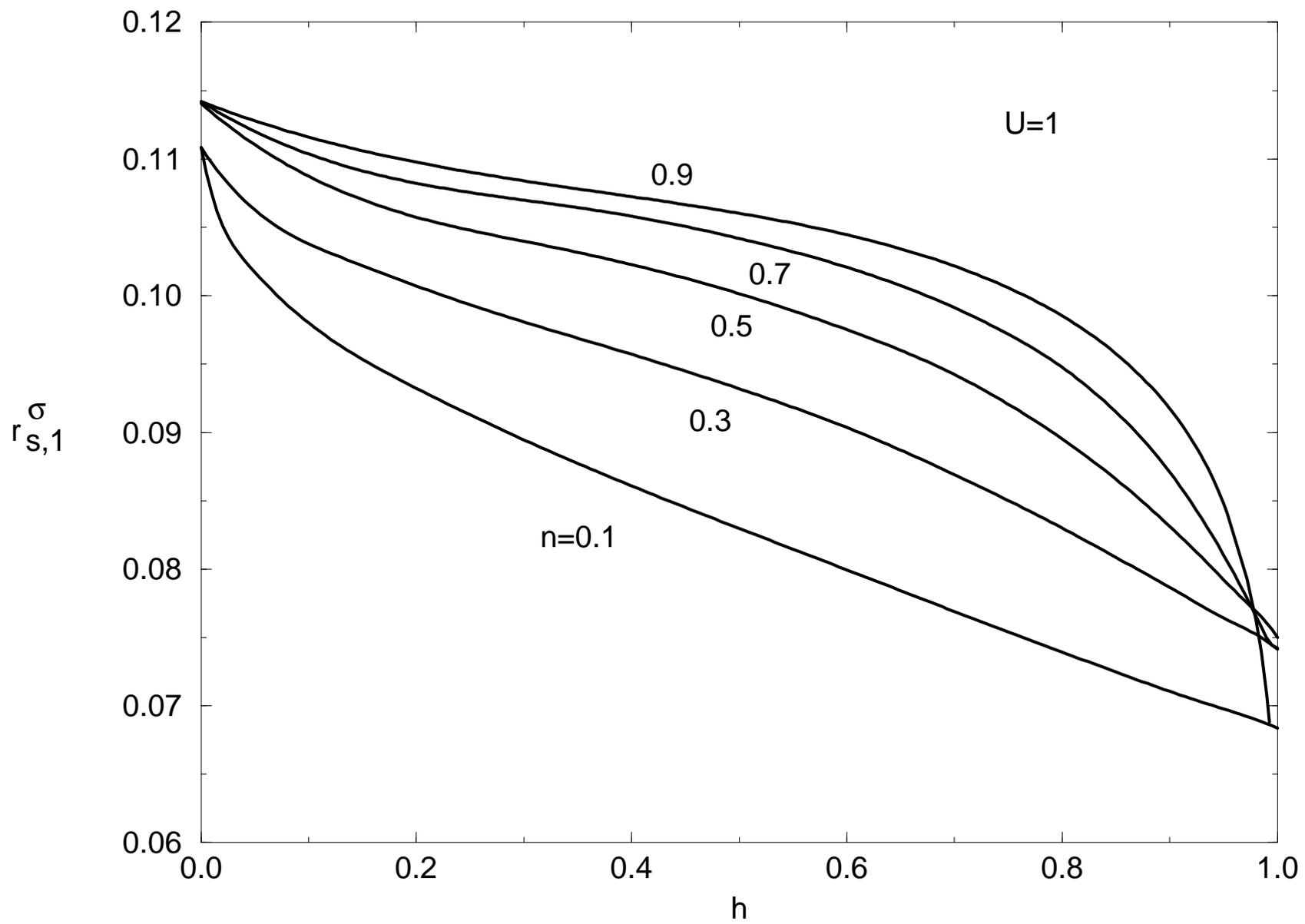
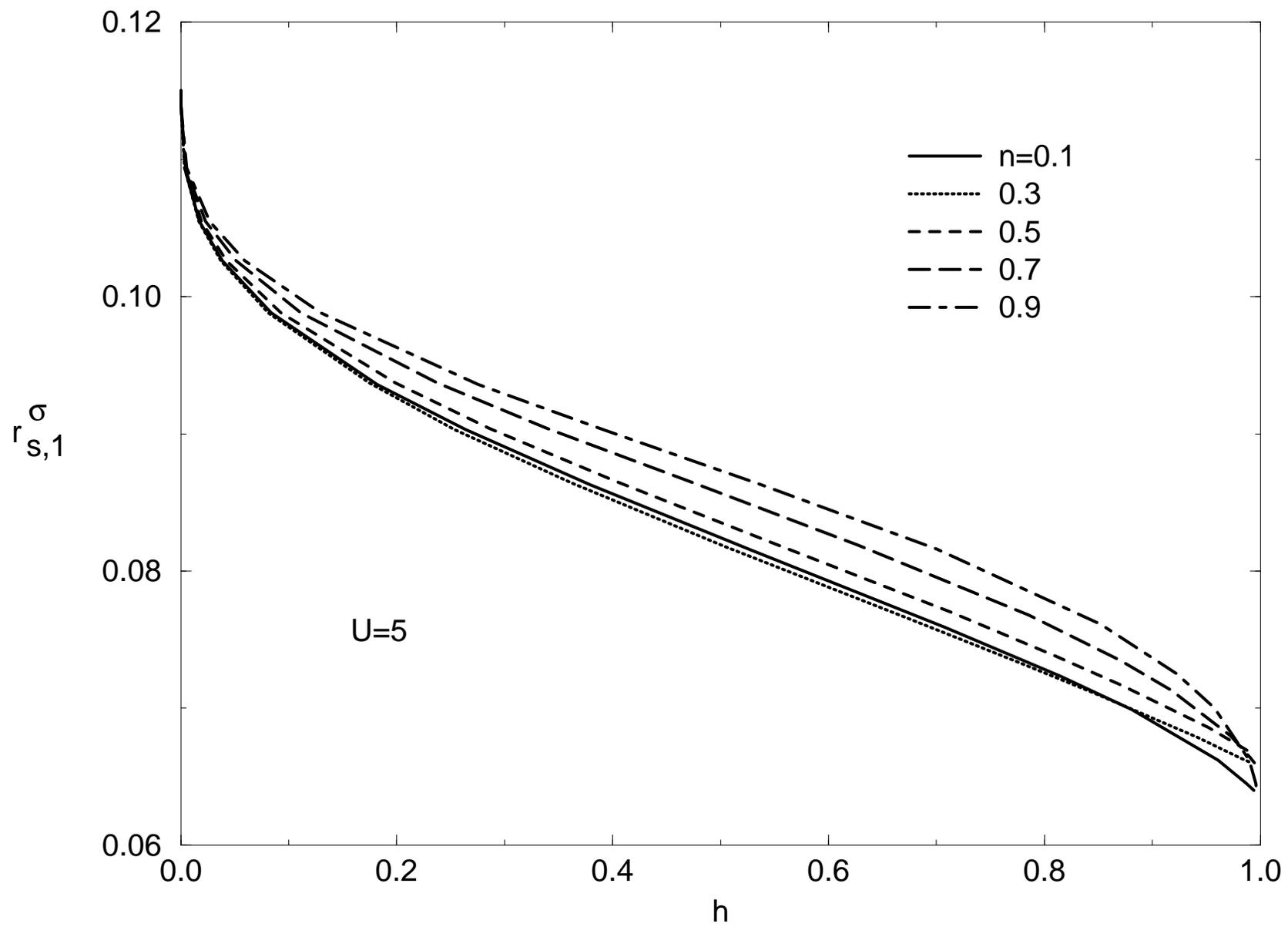
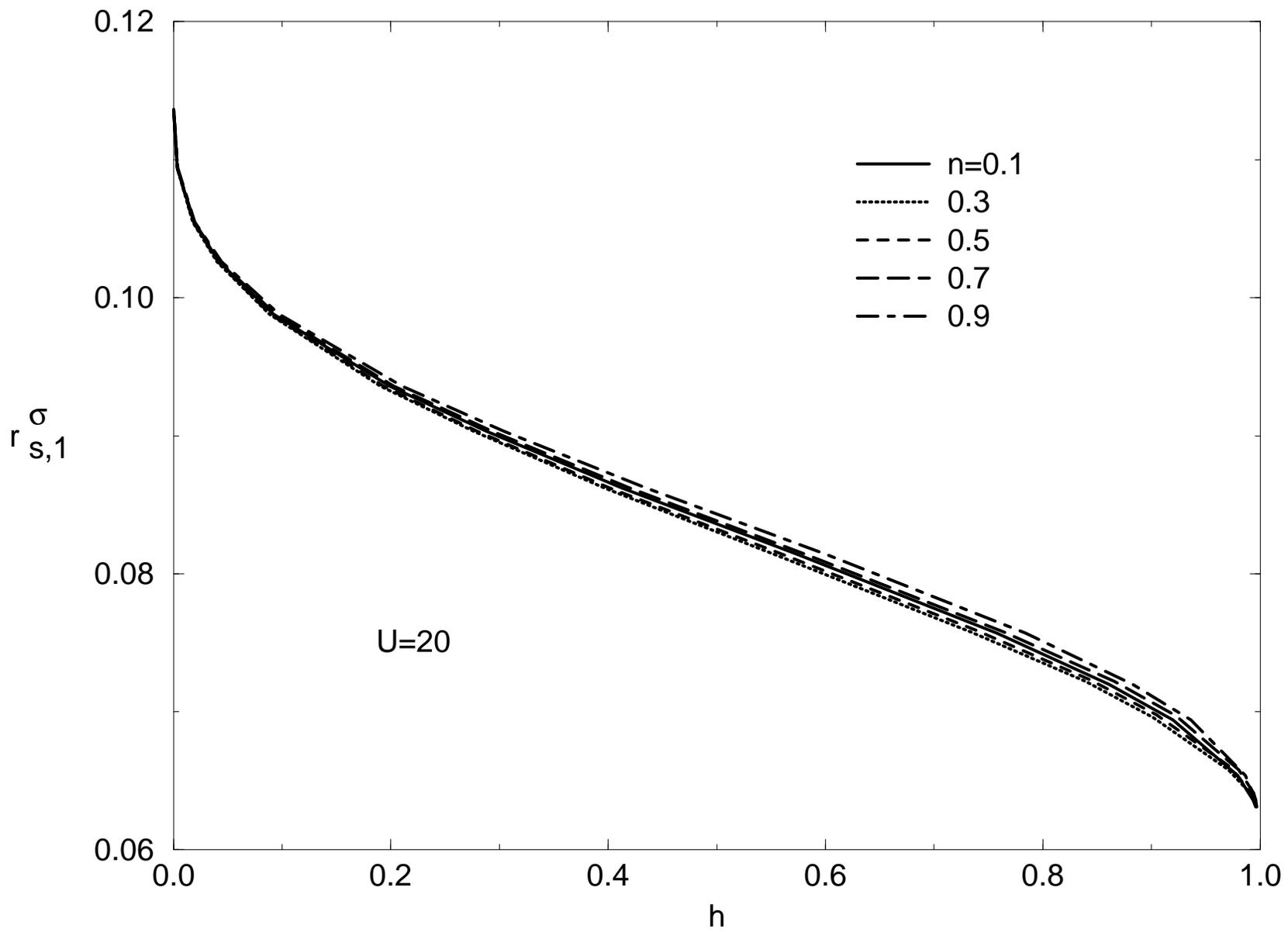
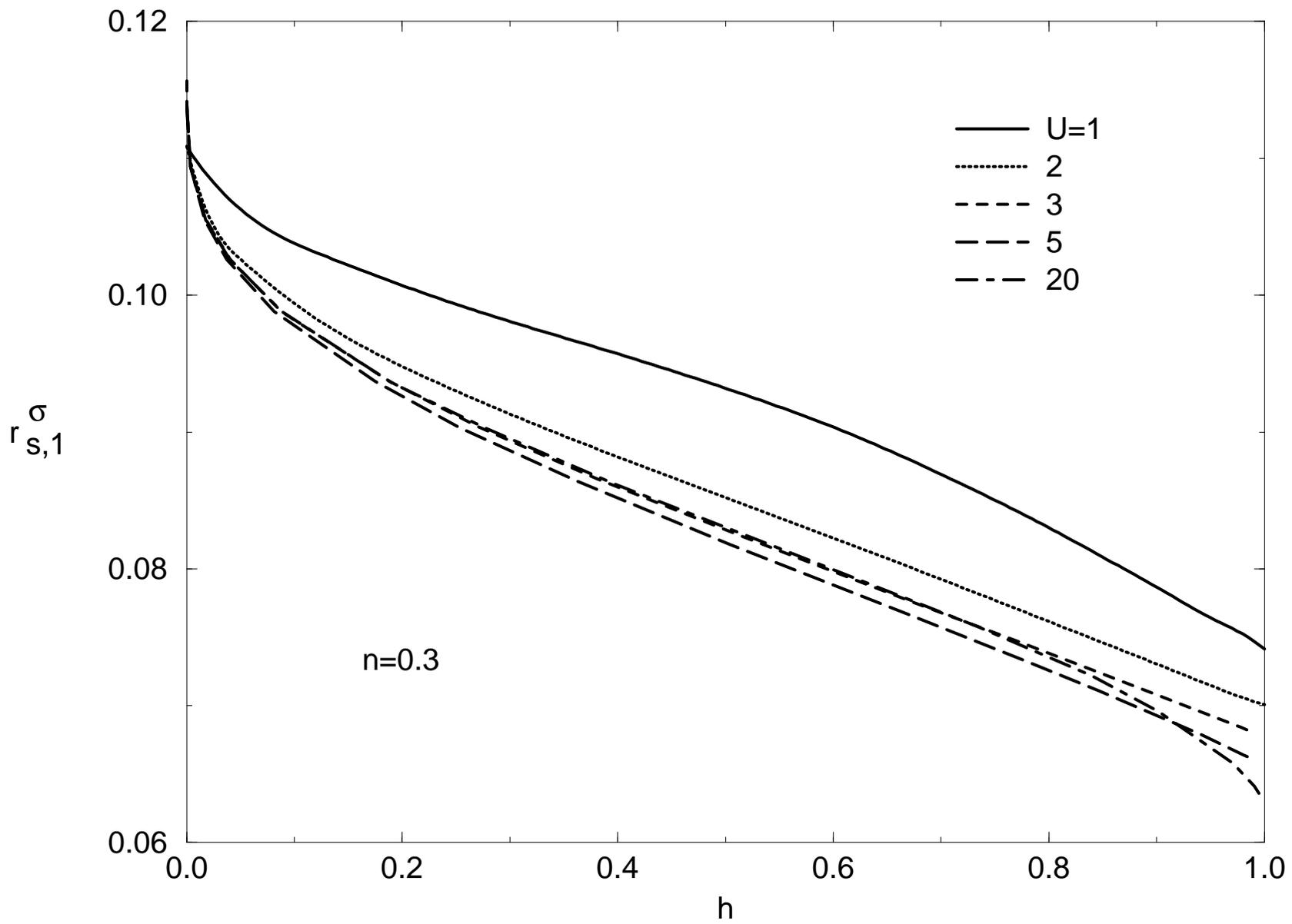
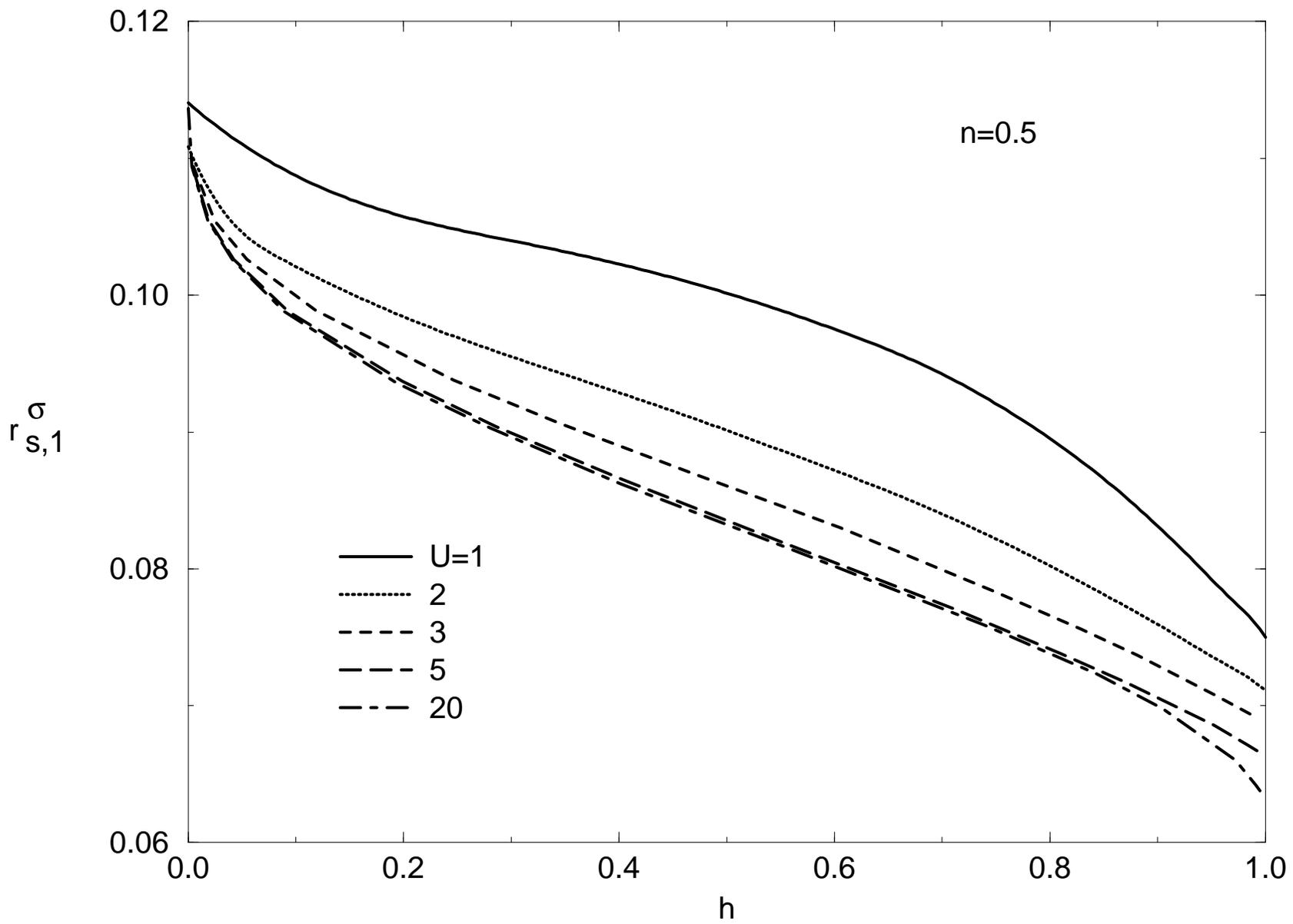
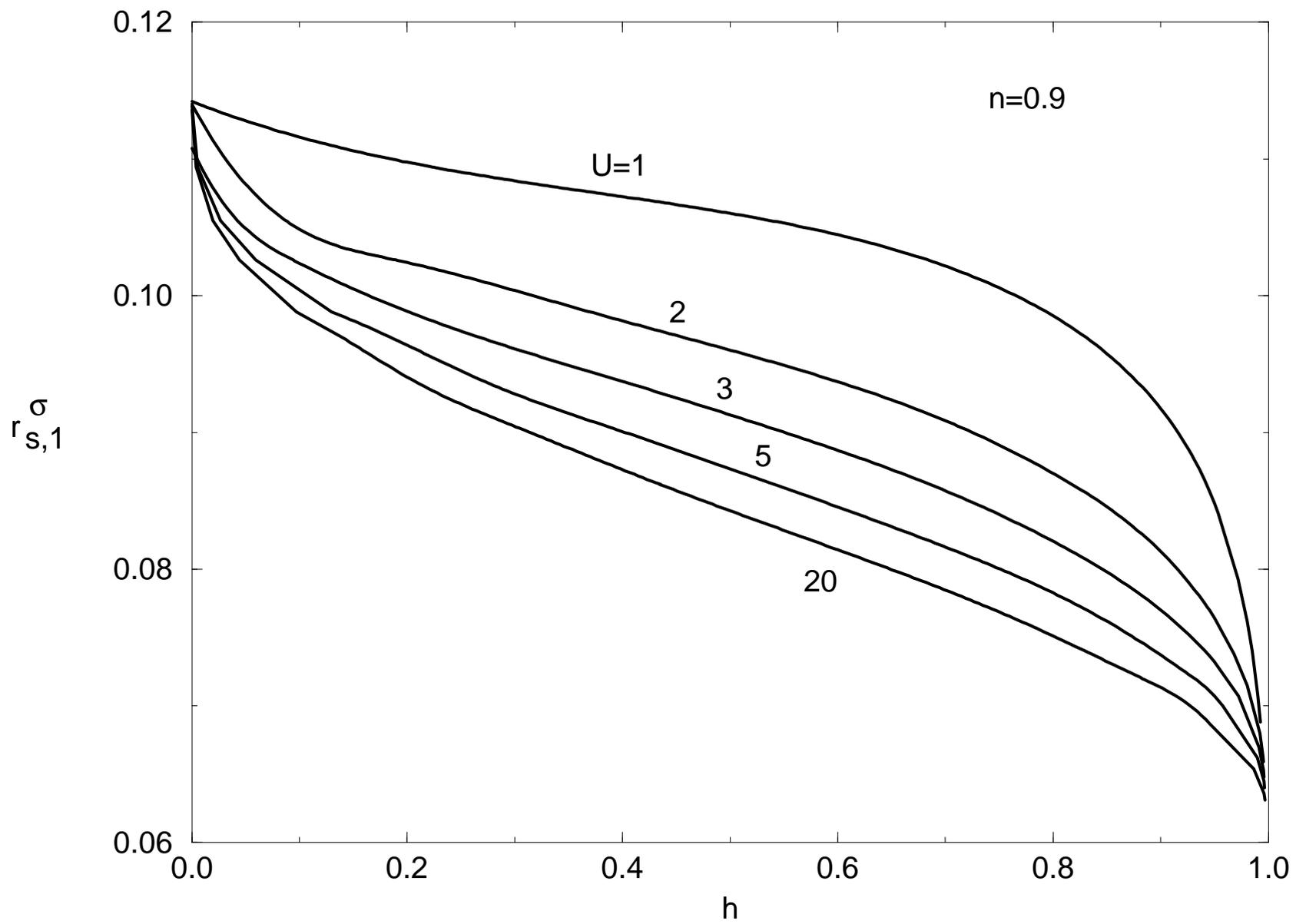
\end{document}